\RequirePackage{tikz}

\documentclass[pdflatex,sn-mathphys]{sn-jnl}

\usepackage[T1]{fontenc}
\usepackage[caption=false,font=normalsize,labelfon
t=sf,textfont=sf]{subfig}
\usepackage{siunitx}
\usepackage[disable]{todonotes}
\usepackage{float}
\usetikzlibrary{matrix,positioning,calc}
\usepackage{varwidth}
\usepackage{placeins}

\usepackage{xcolor, soul}

\hypersetup{allcolors=black }

\let\Oldsection\section
\renewcommand{\section}{\FloatBarrier\Oldsection}

\let\Oldsubsection\subsection
\renewcommand{\subsection}{\FloatBarrier\Oldsubsection}

\let\Oldsubsubsection\subsubsection
\renewcommand{\subsubsection}{\FloatBarrier\Oldsubsubsection}

\newcommand\figref[1]{Fig.~\ref{#1}}     

\jyear{2024}%

\theoremstyle{thmstyleone}%
\theoremstyle{thmstyletwo}%

\theoremstyle{thmstylethree}%

\begin{document}

\title[Utilizing Reinforcement Learning for \textit{de novo} Drug Design]{Utilizing Reinforcement Learning for \textit{de novo} Drug Design}


\author*[1,2]{\fnm{Hampus} \sur{Gummesson Svensson}}\email{hamsven@chalmers.se}

\author[3]{\fnm{Christian} \sur{Tyrchan}}\email{christian.tyrchan@astrazeneca.com}

\author[1,2]{\fnm{Ola} \sur{Engkvist}}\email{ola.engkvist@astrazeneca.com}

\author[1]{\fnm{Morteza} \sur{Haghir Chehreghani}}\email{morteza.chehreghani@chalmers.se}

\affil[1]{\orgdiv{Department of Computer Science and Engineering}, \orgname{Chalmers University of Technology and University of Gothenburg}, \orgaddress{\city{Gothenburg}, \country{Sweden}}}

\affil[2]{\orgdiv{Molecular AI, Discovery Sciences, R\&D}, \orgname{AstraZeneca}, \orgaddress{\city{Gothenburg}, \country{Sweden}}}

\affil[3]{\orgdiv{Medicinal Chemistry, Research and Early Development, Respiratory and Immunology (R\&I), BioPharmaceuticals R\&D}, \orgname{AstraZeneca}, \orgaddress{\city{Gothenburg}, \country{Sweden}}}


\abstract{Deep learning-based approaches for generating novel drug molecules with specific properties have gained a lot of interest in the last few years. Recent studies have demonstrated promising performance for string-based generation of novel molecules utilizing reinforcement learning. In this paper, we develop a unified framework for using reinforcement learning for de novo drug design, wherein we systematically study various on- and off-policy reinforcement learning algorithms and replay buffers to learn an RNN-based policy to generate novel molecules predicted to be active against the dopamine receptor DRD2. Our findings suggest that it is advantageous to use at least both top-scoring and low-scoring molecules for updating the policy when structural diversity is essential. Using all generated molecules at an iteration seems to enhance performance stability for on-policy algorithms. In addition, when replaying high, intermediate, and low-scoring molecules, off-policy algorithms display the potential of improving the structural diversity and number of active molecules generated, but possibly at the cost of a longer exploration phase. Our work provides an open-source framework enabling researchers to investigate various reinforcement learning methods for \textit{de novo} drug design.}




\maketitle

\section{Introduction}\label{sec:intro}
In recent years, there has been an increased interest in using machine learning for drug discovery. It has been applied to a large range of different tasks, including virtual screening, synthesis prediction, property prediction, and computer-assisted molecular design \cite{chen2018rise,yang2019concepts,vamathevan2019applications}.  Machine learning has obtained an important position in \textit{de novo} drug design \---- the design of novel chemical entities that fit certain constraints. \textit{De novo} drug design is an iterative optimization problem whose navigation in the optimization landscape relies on finding local optima of molecular structures, which does not necessarily lead to identifying the global optimum \cite{schneider2005computer}. Therefore, it is of interest to find a diverse set of local optima, meaning structurally different molecules with a high probability of being active against a desired target, i.e., with high activity. 

Numerous deep learning-based methods have been developed for \textit{de novo} drug design, including approaches based on reinforcement learning \cite{olivecrona2017molecular, zhou2019optimization,you2018graph,jin2020multi,yang2021hit,horwood2020molecular,gottipati2020learning,neil2018exploring} and variational autoencoders \cite{gomez2018automatic,maus2022local,jin2018junction,bradshaw2020barking}. These approaches use several different ways to encode molecules into something that the model can learn, such as fingerprint-, string- and graph-based encodings. The string-based simplified molecular-input line-entry system (SMILES) \cite{weininger1988smiles} is a popular way to encode the 2D structure of molecules. Previous work has demonstrated that graph-based and string-based generative models show equivalent performance in terms of chemical space coverage \cite{zhang2021comparative}. It is expected that the conclusions of this work are independent of how the molecules are represented. Moreover, recent evaluations of sample efficiency of \textit{de novo} molecular generation methods have concluded good performance when using reinforcement learning (RL) for learning a recurrent neural network (RNN) \cite{rumelhart1985learning} to generate SMILES strings representing high-scoring molecules \cite{gao2022sample,thomas2022re}. The objective is to learn a policy that can sample sequences of tokens to generate SMILES strings. Hence, policy optimization algorithms might have a significant impact on this task. 

To further improve the sample efficiency of RL, it has been proposed to combine RL with a Hill-climb algorithm, which learns on the $k$ top-scoring sequences \cite{thomas2022augmented,neil2018exploring,brown2019guacamol}. This method focuses the training on good samples from the current round of sequences. This can be interpreted as an off-policy algorithm with a replay buffer, filtering out low reward sequences and initializing the buffer memory between learning rounds.

The use of replay buffers is crucial in off-policy algorithms and is known to improve the sample efficiency of these algorithms \cite{fedus2020revisiting}. However, to our knowledge, no previous work in \textit{de novo} drug design has investigated off-policy algorithms with replay buffers utilizing past sequences. This work builds upon previous work in \textit{de novo} drug design using a reward-based replay mechanism. Many of the state-of-the-art replay mechanisms used in reinforcement learning are often based on the temporal difference (TD) error \cite{schaul2015prioritized}. However, since the TD error is not necessarily computable in a policy-based algorithm, a more general mechanism is needed for a fair comparison of algorithms. This paper focuses on reward-based replay mechanisms.

In this paper, we explore in a systematic way different on-policy and off-policy policy optimization reinforcement learning algorithms, in combination with several ways of replaying previous sequences or restricting the learning to a subset of the sequences sampled in the current episode. The objective is to investigate how large fraction of the generated molecules are predicted, with high probability, to be active to a desired target, and how structurally diverse these predicted active molecules are. Our work can be used as an open-source framework for researchers to investigate various RL methods for \textit{de novo} drug design\footnote{The source code of our framework is publicly available at: \url{https://github.com/MolecularAI/SMILES-RL}.}.  

\section{Problem Setup}
\begin{figure}[h]
    \centering
    \includegraphics[width=0.9\textwidth]{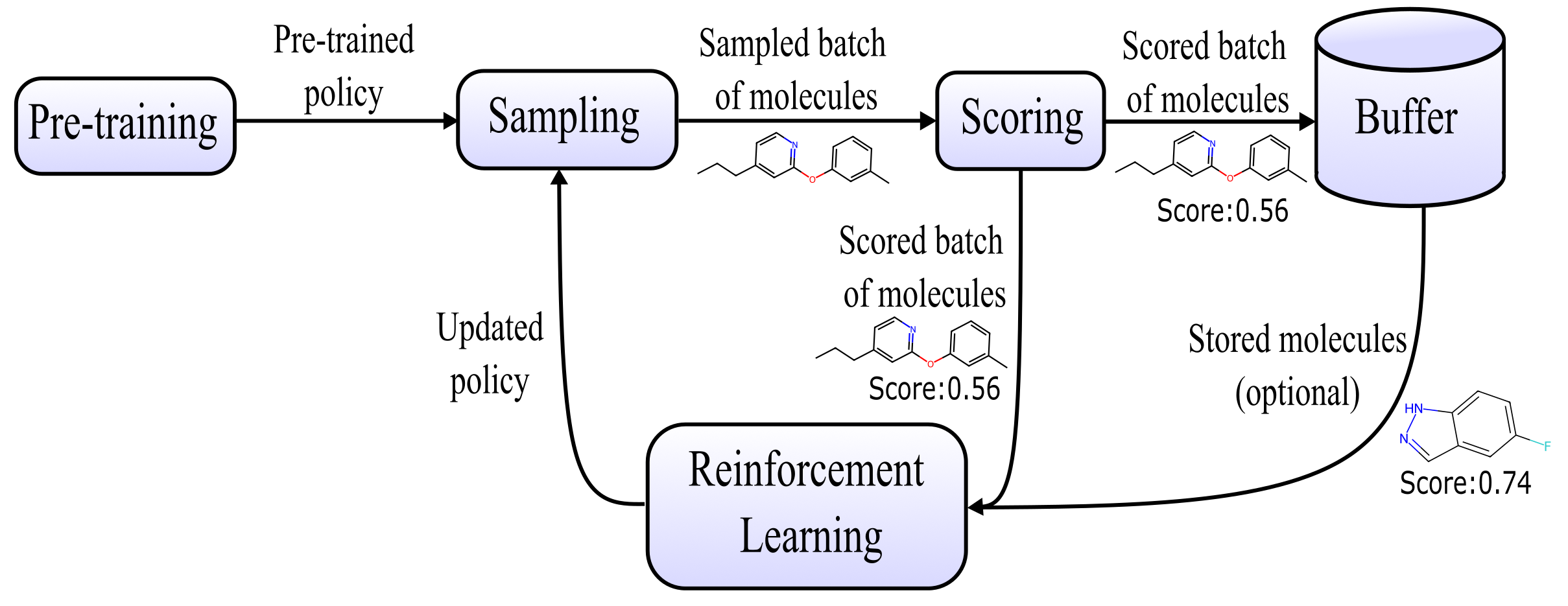}
    \caption{Schematic illustration of the \textit{de novo} drug design process using reinforcement learning (RL).}
    \label{fig:de_novo}
\end{figure}

The first step of \textit{de novo} drug design using RL involves training a pre-trained policy (and/or encoding certain structures into the policy), as illustrated in \figref{fig:de_novo}. Subsequently, a batch of molecules is sampled by the policy, e.g., by the policy choosing a sequence of characters in a SMILES string. In the next step, the sampled molecules are scored by an unknown ``black box'' objective function, i.e., the objective function can be evaluated at any point of its domain but its full expression is unknown. The molecules and corresponding scores are both stored for final inspection and replay (optional). The current molecules and corresponding scores are also fed into the RL algorithm, where the molecular sampling policy is updated. Depending on the use of the replay buffer, current and/or previous samples are provided for the learning step. Using the updated policy, a new batch of molecules is sampled. This continues until a stopping criterion has been reached, such as a pre-defined budget of samples.

\subsection{Problem Definition}

Molecular \textit{de novo} design using RNNs for optimizing molecules encoded as SMILES strings, can be formulated as an RL problem. The agent interacts with the environment over discrete time steps by adding tokens to a SMILES string. The environment is episodic, where a SMILES string is provided between a start and stop token, and the episode's length depends on the SMILES string's length, which is terminated when the stop token is added to the string. At time step $t=0$, the start token is added to the string and defines the first state $s_1$. At time step $t=1,\dots,T$ the agent observes a $n_s$-dimensional state vector $s_t \in \mathcal{S} \subseteq \mathbb{R}^{n_s}$, chooses an action $a_t \in \mathcal{A}$ according to a policy $\pi(a_t \vert s_t)$. The episode ends at time step $T+1$, for terminal state $s_{T+1}$, when the stop token is chosen as action $a_{T}$. Moreover, at the end of the episode, a reward signal $R(a_{1:T}) \in [0,1]$ is observed for a sequence of actions $a_{1:T}$. The scoring function provides this reward signal by scoring each valid SMILES string.

The state vector $s_t$ is given by the output states of the RNN at step $t-1$ and encodes information about the actions taken in previous steps. A discrete action space $\mathcal{A} = \{ 0,\dots,33\}$ which tokenizes the feasible characters in the SMILES string, including start and stop tokens, is considered. Under this setup, we explore various policy optimization RL methods, where the goal is to learn the policy directly parameterized by $\theta$, $\pi_\theta(a_t \vert s_t)$. The output gates of the RNN are fed into a fully connected layer to provide either the probability (utilizing a softmax layer) or values of each action.

\section{Policy Optimization Algorithms for \textit{de novo} Drug Design} \label{sec:rl}
\begin{figure}[h]
    \centering
    \includegraphics[width=0.9\textwidth]{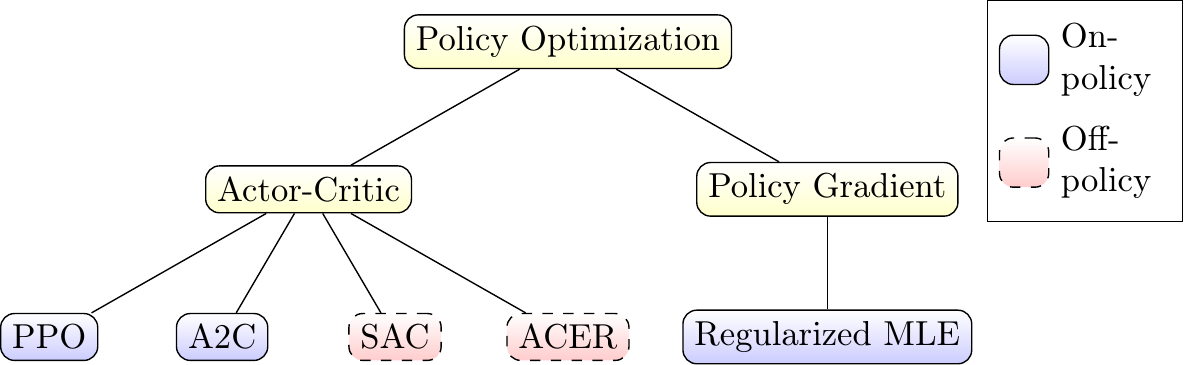}
    \caption{Taxonomy of the reinforcement learning (RL) algorithms explored in this work.}
    \label{fig:rl_taxonomy}
\end{figure}

In this paper, we explore the following policy optimization algorithms for \textit{de novo} drug design, to generate diverse molecules with high scores: (1) Regularized maximum likelihood estimation (Reg. MLE); (2) Advantage Actor-Critic (A2C); (3) Proximal Policy Optimization (PPO); (4) Actor-Critic with Experience Replay (ACER); (5) Soft Actor-Critic (SAC). \figref{fig:rl_taxonomy} illustrates the taxonomy of these algorithms. These are the major on- and off-policy policy optimization algorithms.

\subsection{Regularized Maximum Likelihood Estimation}
The regularized maximum likelihood estimation (Reg. MLE) algorithm is currently used in REINVENT \cite{blaschke2020reinvent}. Recent evaluations by both \cite{gao2022sample} and \cite{thomas2022re} have concluded good performance compared to both RL-based and non-RL-based approaches for \textit{de novo} drug design. The idea is that the likelihood of the agent's policy should stay close to that of the pre-trained policy (See Sec.~\ref{sec:sampling})  while still focusing on the high-scoring sequences. It minimizes the following policy loss

\begin{equation}
    L^{\mathrm{Reg.~MLE}}(\theta) = \left( \log \pi_{\textrm{prior}}(a_{1:T}) + \sigma R(a_{1:T}) - \log \pi_\theta (a_{1:T})  \right)^2,
\end{equation}
where $\pi_{\textrm{prior}}(a_{1:T}) = \pi_{\textrm{prior}}(a_1\vert s_1) \cdots \pi_{\textrm{prior}}(a_{T} \vert s_{T})$ is the likelihood of the pre-trained policy for a sequence of length $T$ (excluding start token), and $\pi_{\theta}(a_{1:T}) = \pi_{\theta}(a_1\vert s_1) \cdots \pi_{\theta}(a_{T} \vert s_{T})$ is the corresponding likelihood of the policy that is optimized. The policy has the same network architecture and is initialized as the pre-trained policy. $\sigma$ is a hyperparameter \cite{blaschke2020reinvent} that determines the importance of the reward signal.

Note that it uses a margin guard, which resets the agent to the prior and adjusts sigma if the margin between the augmented likelihood and the agent likelihood $\log \pi_{\textrm{prior}}(a_{1:T}) + \sigma R(a_{1:T}) - \log \pi_{\theta}(a_{1:T})$ is below a pre-defined threshold. Table \ref{table:reg_mle} in Appendix \ref{app:hyper} displays a list of hyperparameters and the values employed in this paper.

\subsection{Proximal Policy Optimization}\label{sec:ppo}
Proximal Policy Optimization (PPO) \cite{schulman2017proximal} uses a clipping loss function defined by

\begin{equation}
    \label{eq:clip_loss}
    L^{\mathrm{CLIP}} \left(\theta \right) = \hat{\mathbb{E}}_t \left[\mathrm{min}\left(r(\theta)\hat{A}(s_t), \mathrm{clip}\left(r_t(\theta), 1-\epsilon, 1+\epsilon \right) \hat{A}(s_t) \right)\right],
\end{equation}
where $r(\theta) = \frac{\pi_{\theta}(a_t \vert s_t)}{\pi_{ \theta_{\mathrm{old}}}(a_t \vert s_t)}$ is the probability ratio and advantage $\hat{A}(s_t) = \gamma^{T-t} R(a_{1:T}) - V_{\phi}(s_t)$ is used, where $V_{\phi}(s_t)$ is the value function. This corresponds to Monte-Carlo (MC) samples where the reward $R(a_{1:T})$ is only given at time step $T$.  $\gamma$ is the discount factor and $\epsilon$ is a hyperparameter determining the clipping range. We adapt PPO to the setting where the state vectors are represented by the RNN outputs of the latest recurrent node. The actor is a neural network with parameters $\theta$, providing probabilities $\pi_{\theta}(a \vert s)$ for an action $a$ at state vector $s$ using a softmax layer. It is initialized as the pre-trained policy (see \ref{sec:sampling}) and the parameters are updated using the loss function in Eq.~\eqref{eq:clip_loss}. The value function $V_{\phi}(s_t)$ is a neural network with (non-shared) parameters $\phi$. It has the same network architecture as the actor, but the output layer only consists of one output, i.e., the value, and uses no softmax layer. The output layer is reset at initialization, while the initial embedding and long short-term memory (LSTM) \cite{hochreiter1997long} layer are the same as the pre-trained policy network. The value network is trained by minimizing the following mean squared error loss

\begin{equation}
    L^{\mathrm{MSE}} (\phi) = \mathbb{E}_t \left[\frac{1}{2}\left( \gamma^{T-t} R(a_{1:T}) - V_{\phi}(s_t)\right)^2\right].
\end{equation}
For each batch of sequences, the actor and critic loss are minimized over 4 epochs, each doing minibatch updates where sequences are shuffled into 4 mini-batches. 

It is possible for the actor to diverge from the pre-trained actor (policy), which has learned how to sample a valid SMILES string. This rarely happens but should be properly handled when happening. Therefore, if the fraction of valid SMILES strings (out of the 128 sampled in each episode) is less than 0.8 for more than 10 consecutive episodes, the parameters of the algorithm will be reset to that of the pre-trained model. Table \ref{table:ppo} in Appendix \ref{app:hyper} displays a list of hyperparameters and the values employed in this paper.

\subsection{Advantage Actor-Critic}\label{sec:a2c}
Advantage Actor-Critic (A2C) is a synchronous version of the A3C algorithm \cite{mnih16a3c}, wherein the following definition of the advantage is used
\begin{equation}
    \hat{A}(s_t) = \gamma^{T-t} R(a_{1:T}) - V_{\phi}(s_t),
\end{equation}
where $V_{\phi}(s_t)$ is the value network with (non-shared) parameters $\phi$ and discount factor $\gamma$. We adapt A2C to the setting where the state vectors are represented by the RNN outputs of the latest recurrent node. The actor is a neural network with parameters $\theta$, providing probabilities $\pi_{\theta}(a \vert s)$ for an action $a$ and state vector $s$ using a softmax layer. It is initialized as the pre-trained policy (see \ref{sec:sampling}) and the parameters are updated using the policy gradient with the advantage. A discount factor slightly smaller than 1 is used to slightly favor small molecules. The value network has the same network architecture as the actor, but the output layer only consists of one output, i.e., the value, and uses no softmax layer. The output layer is reset at initialization, while the initial embedding and LSTM layer are the same as the pre-trained policy network. The value network is trained by minimizing the following mean squared error loss

\begin{equation}
    L^{\mathrm{MSE}} (\phi) = \mathbb{E}_t \left[ \frac{1}{2}\left(\gamma^{T-t} R(a_{1:T}) - V_{\phi}(s_t)\right)^2\right].
\end{equation}

If the actor generates a large fraction of invalid SMILES strings, the algorithm is reset in the same way as for PPO (see Sec.~\ref{sec:ppo}). Table \ref{table:a2c} in Appendix \ref{app:hyper} displays a list of hyperparameters and the values employed in this paper.

\subsection{Actor-Critic with Experience Replay}\label{sec:acer}
Actor-Critic with Experience Replay (ACER) \cite{wang2016sample} is an off-policy actor-critic algorithm with experience replay. ACER is the off-policy counterpart of the A3C algorithm \cite{mnih16a3c} where the aim is to stabilize the off-policy estimator, e.g, by applying a trust region policy optimization (TRPO) method. The algorithm performs one on-policy update and $r \sim \mathrm{Pois}(\lambda)$ off-policy updates using replay, where each replay samples $128$ sequences.  We adapt ACER to the setting where the state vectors are represented by the RNN outputs of the latest recurrent node.  It uses a shared network utilizing the same architecture as the pre-trained policy network  (see Sec.~\ref{sec:sampling}) but with an additional value head, i.e., a parallel fully connected layer with output dimension 1. The value head is randomly initialized, while the other weights are the same as the pre-trained policy network, including the policy head.
We add an entropy term to the loss with weight $0.001$, slightly favoring sequences with larger cumulative entropy.

We use retrace Q-value estimation, as proposed in the original algorithm \cite{wang2016sample}. Using retrace Q-value estimations, instead of Monte-Carlo samples, does slightly improve the stability. For each sequence, $a_{1:T}$, reward $R(a_{1:T})$ is only given at time step $T$ for action $a_{T}$, when the stop token is chosen as action. A reward signal of $\num{-1}$ is given to invalid SMILES strings, which are by default given a reward of $0$ for the on-policy algorithms. The penalty for invalid SMILES bias the algorithm strongly toward valid SMILES. This seems to be crucial, especially when performing many off-policy updates using the replay memory. Furthermore, 10 initial steps without updating the policy are performed, only using the pre-trained policy to store initial sequences in the replay memory. Table \ref{table:acer} in Appendix \ref{app:hyper} displays a list of hyperparameters and the values employed in this paper.

\subsection{Soft Actor-Critic}

\begin{algorithm}
\caption{Discrete Soft Actor-Critic for \textit{de novo} drug design}\label{alg:sac}
\begin{algorithmic}[1]
\Statex \textbf{Input:} $\phi$, $\theta$, initial episodes $K_{\mathrm{init}}$, total budget of episodes $K_{\mathrm{E}}$,
\Statex \textbf{Init:} $\phi' \gets \phi$, $\theta' \gets \theta$, $\mathcal{D} \gets \emptyset$
\For{each initial episode $1,\dots,K_{\mathrm{init}}$}
\State Sample a batch $\mathcal{T}$ of $M$ sequences using pre-trained policy $\pi_\theta$
\State Score each sequence in $\mathcal{T}$
\State Add unique, valid sequences to replay memory $\mathcal{D}$
\EndFor
\For{each episode $K_{\mathrm{init}}+1,\dots,K_{\mathrm{E}}$}
\State Sample a batch $\mathcal{T}$ of $M$ sequences using current policy $\pi_\theta$
\State Score each sequence in $\mathcal{T}$
\State Add unique, valid sequences to replay memory $\mathcal{D}$
\State $\phi \gets \phi - \lambda_Q \hat{\nabla}_\phi J_Q (\phi \vert \mathcal{T})$ \Comment{On-policy update of Q-function parameters}
\State $\theta \gets \theta - \lambda_\pi \hat{\nabla}_\theta J_\pi (\theta\vert \mathcal{T})$ \Comment{On-policy update of policy parameters}
\State $\alpha \gets \alpha - \lambda_\alpha \hat{\nabla}_\alpha J_\alpha (\alpha \vert \mathcal{T})$ \Comment{On-policy update of temperature}
\State $\phi' \gets \tau \phi' + (1-\tau) \phi$  \Comment{Update target parameters}
\State $\theta' \gets \tau \theta' + (1-\tau) \theta$  \Comment{Update average policy parameters}
\For{each off-policy update}
\State $\phi \gets \phi - \lambda_Q \hat{\nabla}_\phi J_Q (\phi \vert \mathcal{D})$
\State $\theta \gets \theta - \lambda_\pi \hat{\nabla}_\theta J_\pi (\theta\vert \mathcal{D})$
\State $\alpha \gets \alpha - \lambda_\alpha \hat{\nabla}_\alpha J_a (\alpha \vert \mathcal{D})$
\State $\phi' \gets \tau \phi' + (1-\tau) \phi$
\State $\theta' \gets \tau \theta' + (1-\tau) \theta$
\EndFor
\EndFor
\end{algorithmic}
\end{algorithm}

Soft Actor-Critic (SAC) \cite{haarnoja2018bsoft} is an off-policy algorithm that incorporates the entropy of the policy into the reward signal to encourage a stochastic policy with more randomness, while still fulfilling the task. It is based on the maximum entropy objective, with the aim of optimal policy $\pi^*$ to maximize both its reward and entropy at each visited state.

It uses an automatic entropy adjustment to control the temperature parameter $\alpha$ that determines the relative importance of the entropy term. The original algorithm is formulated for a continuous action space, but \cite{christodoulou2019soft} has extended it to discrete action spaces. The soft actor-critic algorithm for discrete action spaces is utilized in this work, with some adaptions discussed below. 
 We adapt SAC to the setting where the state vectors are represented by the RNN outputs of the latest recurrent node. The actor is a neural network with parameters $\theta$, providing probabilities $\pi_{\theta}(a \vert s)$ for an action $a$ and state vector $s$ using a softmax layer. It is initialized as the pre-trained policy (see \ref{sec:sampling}) and the parameters are updated by minimizing the loss in Eq.~\eqref{eq:sac_actor_loss}. The value function $Q_{\phi}(a,s)$ is given by a neural network that has the same network architecture as the actor, where each output corresponds to the action-state value of an action $a$ at current state $s$ but uses no softmax layer. We initialize the parameters $\phi$ to that of the parameters of the pre-trained policy network. We perform no updates of parameters during the $K_{\mathrm{init}} = 10$ first episodes, where only experiences are sampled to the replay buffer. Since a budget of 2000 episodes is considered in the following experiments, there are 1990 episodes left for learning to generate high-scoring molecules. When updating parameters, we utilize a reward of $-1$ for invalid SMILES. Moreover, we perform one on-policy update, using all current sequences, before doing any off-policy update. Four off-policy updates are performed, where 64 sequences are sampled from the replay memory, for each episode. We observed no significant difference in performance when using the larger replay size of 128 as for ACER (seer Sec.~\ref{sec:acer}).

The following entropy-augmented reward is defined
\begin{equation}
    r_\pi (s_t,a_t) \triangleq r_{(s_t,a_t)} + \mathbb{E}_{s_{t+1}\sim p} \left[ \alpha \mathcal{H} \left( \pi_{\theta} \left(\cdot \vert s_{t+1}\right) \right) \right],
\end{equation}
where $p$ is the state transition probability of the environment. Full sequences are used for updating, i.e., Monte-Carlo (MC) samples, instead of one-step update in \cite{haarnoja2018bsoft,christodoulou2019soft}.  The reward signal $R(a_{1:T})$ is given at the end of the episode, at time step $T$, and uses no discount, i.e., discount factor $\gamma=1$ is utilized. This gives the following target for each soft Q-value update
\begin{equation}
    y_{s_i} = R(a_{1:T}) + \sum^{T-2}_{l=i} \mathbb{E}_{s_{l+1}\sim p} \left[ \alpha \mathcal{H} \left( \pi_{\theta} \left(\cdot \vert s_{l+1}\right) \right) \right],
\end{equation}
where  $y_{s_{T+1}} = 0$,  and $y_{s_{T-1}} = y_{s_{T}} = R(a_{1:T})$, since 
 it is defined that $\mathcal{H} \left( \pi_{\theta} \left(\cdot \vert s_{T+1}\right) \right) = \mathcal{H} \left( \pi_{\theta} \left(\cdot \vert s_{T} \right) \right)= 0$ to keep the pre-trained high probabilities of stop tokens at certain states. This gives each action an equal contribution from the reward signal of the full sequence and an additional cumulative entropy term that favors actions where future states have high entropy.  

To improve the stability, \cite{zhou2022revisiting} has proposed to include more regularization in the actor and critic losses. The proposed clipping of the critic loss with target critic(s) is used, which is found to improve the stability, giving the following loss of the critic network
\begin{multline}
    J_Q(\phi \vert \Pi) = \mathbb{E}_{\mathcal{T} \sim \Pi} \left[ \mathbb{E}_{(a_t,s_t)\sim \mathcal{T}} \left[ \mathrm{max} \left( \left(Q_{\phi}(a_t,s_t) - y_{s_t}  \right)^2, \right. \right.  \right. \\ \left. \left. \left.  \left( Q_{\phi'} (a_t,s_t) + \mathrm{clip} \left( Q_{\phi} (a_t, s_t)  - Q_{\phi'} (a_t, s_t), -c, c\right) - y_{s_t} \right)^2 \right)\right] \right],
\end{multline}
where $\Pi$ is a set of sequences (either current sequences or replay memory), $\mathcal{T}$ is a sequence, $Q_{\phi}$ is the estimate of the critic network, and $Q_{\phi'}$ is the estimate of the target-critic network. When $\Pi$ contains the current sequences, all sequences are sampled; otherwise, 64 sequences are sampled when using a full replay memory. The weights of the target-critic are updated using the moving average between the current weights of the target-critic and critic,

\begin{equation}
\label{eq:moving_average}
    \phi' \gets \tau \phi' + (1-\tau) \phi,
\end{equation}
where $\tau$ is the smoothing coefficient, determining how much of the updated critic network that will be transferred to the target-critic network. We only use one critic and target-critic network, since we observe that using the pre-trained model as the initial critic network is advantageous, instead of using randomly initialized weights. This yields more stable learning in terms of the validity of generated SMILES strings and biases the learning towards what the pre-trained model knows. It could possibly be useful to use two critic networks if their updates utilize different replay experiences.

For regularizing the actor loss, we include the Kullback-Leibler (KL) divergence term between the (current) actor and the average policy network. The average policy network $\pi_{\theta'}$ is initialized as the actor and whose parameters $\theta'$ is updated using moving average (as in Eq.~\eqref{eq:moving_average}) with $\tau = 0.99$. Adding this KL divergence term yields the following actor loss

\begin{multline}
    \label{eq:sac_actor_loss}
    J_{\pi} (\theta \vert \Pi) = \mathbb{E}_{\mathcal{T} \sim \Pi} \left[ \mathbb{E}_{s_t \sim \mathcal{T}} \left[\pi_\theta (\cdot \vert s_t) \left( \alpha \log \left(\pi_{\theta}\left( \cdot \vert s_t\right) \right) - Q_\phi (\cdot, s_t)\right) \right. \right. \\ \left.\left. + D_{\mathrm{KL}} \left( \pi_{\theta}(\cdot\vert s_t) \|  \pi_{\theta'}(\cdot\vert s_t) \right)\right] \right] .
\end{multline}

The temperature $\alpha$ is updated by minimizing the following objective, extending the objective proposed by \cite{christodoulou2019soft},

\begin{equation}
    J_\alpha (\alpha \vert \Pi) = \mathbb{E}_{\mathcal{T} \sim \Pi} \left[ \mathbb{E}_{(a_t,s_t) \sim \mathcal{T}} \left[-\alpha \left(\log \pi_{\theta} (a_t \vert s_t) +\bar{H}\right) \right] \right],
\end{equation}
where $\bar{H}$ is the target entropy. Algorithm \ref{alg:sac} illustrates all steps in the (discrete) soft actor-critic algorithm used for \textit{de novo} design. Table \ref{table:sac} in Appendix \ref{app:hyper} displays a list of hyperparameters and the values employed in this paper.

\section{\textit{de novo} drug design using RNNs for generating SMILES generation}\label{sec:method}
This section describes the experimental setup, including the replay buffers used to investigate the on- and off-policy policy optimization algorithms.
A batch of $M=128$ sequences is sampled in each episode, generating 128 SMILES strings in each episode. Any duplicates of SMILES strings are removed afterward, possibly yielding a list of fewer unique molecules. At the end of each episode, using on-policy or/and off-policy batches, the policies are updated by the full roll-out of each sequence, as described in Sec.~\ref{sec:rl}. Technical details of the experiments are discussed in Appendix \ref{app:tech}.

\subsection{Molecular and Topological Scaffolds} 
\begin{figure}[h]
    \centering
    \includegraphics[width=0.7\textwidth]{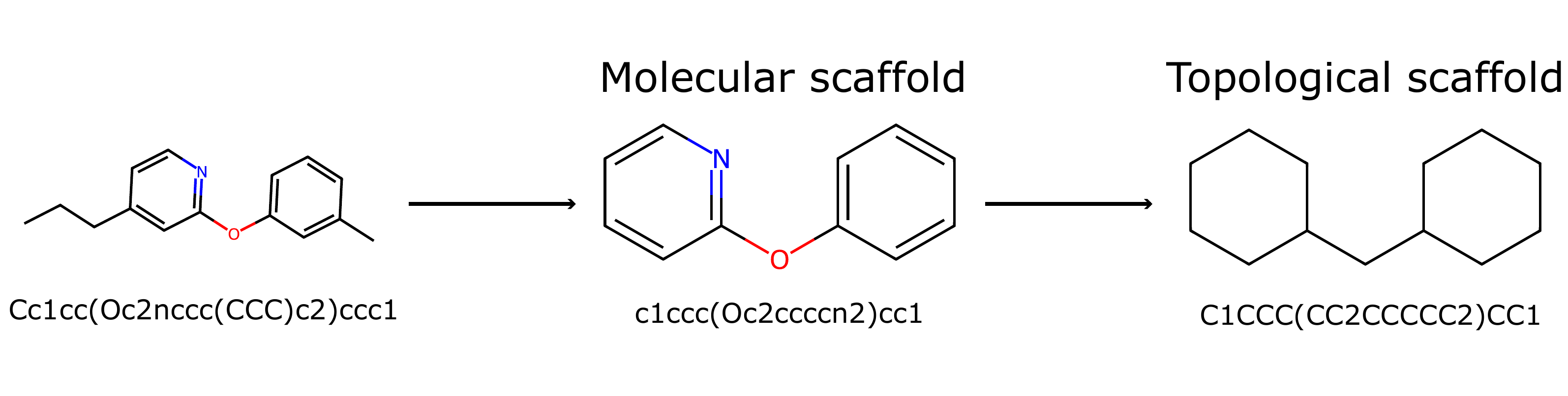}
    \caption{The structural formula and SMILES strings for an arbitrary molecule, and its corresponding molecular and topological scaffold.}
    \label{fig:scaffold}
\end{figure}

The scaffold of a molecule is defined as its core structure. This is a common structure characterizing a group of molecules. This provides a basis for a systematic investigation of molecular cores and building blocks. This assists in finding structurally distinct molecules having similar activity, providing several structural alternatives when optimizing the properties of 
potential drug candidates \cite{hu2016computational}. Hence, scaffolds provide a diversity measure of the identified active molecules.

The molecular scaffold defined by \cite{bemis1996properties}, also known as the Bemis-Murcko scaffold, is used in this work to generate scaffolds. The topological scaffold is defined as the generic molecular scaffold where all atom types are converted into carbon atoms and all bonds are converted into single bonds, as illustrated in \figref{fig:scaffold}. All scaffolds in this work are generated using \cite{landrum2006rdkit}.

\subsection{Sampling}\label{sec:sampling}
 We use the pre-trained policy network provided by \cite{blaschke2020reinvent}. The model is trained on a data set derived from the ChEMBL databse\cite{gaulton2012chembl} and is capable of generating molecules in terms of SMILES strings. The parameters of the policy network are sequentially updated using reinforcement learning (see Sec. \ref{sec:rl}). The policy consists of an embedding layer, an LSTM layer, and a fully connected output layer. The embedding layer consists of 256-dimensional embedding vectors. The LSTM layer has an input size of 256 and an output size of 512 and consists of 3 recurrent layers, i.e., three LSTMs stacked together. The LSTM output is fed to an output layer of output dimension 34. Each output entity corresponds to a token in the vocabulary, including start and stop tokens, defined by \cite{blaschke2020reinvent}. 

When sampling a sequence, the output layer is fed through a softmax function to obtain estimates of the probabilities of each token (action). Multimodal sampling, using the estimated probabilities from the softmax function, is performed to select the next action in a sequence. Small molecules are of interest and, therefore, the length of a sequence is limited to 256. If a sequence reaches this length, the sampling is stopped, returning the sequence sampled so far.

\subsection{Scoring}\label{sec:scoring}
A scoring function provides the rewards signal $R(a_{1:T})$ of each sequence. The scoring function is given by a random forest model with 1300 trees and a maximum depth of 300 for each tree. Class weights, which determine the sample probability during bootstrapping, are inversely proportional to the class frequencies. The random forest model is trained to predict the binary activity of a molecule to the Dopamine receptor D2 (DRD2), using the activity data in ExCAPE-DB \cite{sun2017excape}. 2048-bit Morgan-like fingerprints, computed by RDKit \cite{landrum2006rdkit}, with radius 2, utilizing features and counts are used as feature vectors. Each SMILES string is encoded into a such feature vector for scoring. Class probabilities, of the binary activity, are given by the fraction of trees predicting the corresponding class. The reward of a sequence is defined as the probability of predicting a positive label for the corresponding sequence. A sequence (or SMILES string) is defined to be valid if the corresponding SMILES string can be constructed into a Mol object by RDKit \cite{landrum2006rdkit}, which is done when computing the fingerprints for scoring. When constructing a Mol object, RDKit first performs a grammar check and then applies basic chemical rules. An invalid SMILES string is given a reward of $0$ and $-1$ for the on- and off-policy algorithms, respectively. The score can also be modified by the diversity filter, see Sec.~\ref{sec:diversity_filter}. 

A molecule is defined to be active if the corresponding sequence has a reward greater than or equal to 0.7, and a scaffold is defined as active if it contains at least one active molecule. These definitions are used to compare the policy optimization algorithm utilizing different replay buffers. 

\subsection{Diversity Filter}\label{sec:diversity_filter}
A diversity filter is a memory-assisted approach to improve the diversity of generated molecules. It keeps track of the molecules with similar structures, e.g., scaffolds. We use the diversity filter based on identical molecular scaffolds, proposed by \cite{blaschke2020memory}. This diversity filter consists of a scaffold memory that stores molecules and their corresponding molecular scaffold. A molecule is saved into the scaffold memory if the corresponding sequence reaches at least a reward of 0.4. No molecules with the same canonical SMILES are allowed in the scaffold memory, hence only storing unique molecules. If the number of saved molecules with the same molecular scaffold reaches 25, all future molecules with the same molecular scaffold are given a reward of $0$, and consequently not saved in the scaffold memory. This changes the reward function used for learning when a certain number of molecules with the same molecular scaffold have been generated. 

\subsection{Replay Buffers}
In light of the current use of the Hill-climb algorithm \cite{thomas2022augmented,neil2018exploring,brown2019guacamol} for training, we study different approaches using both current and previous sequences. All of these approaches are collected under the term \emph{replay buffers}.

In each episode, a batch of $M = 128$ sequences are sampled. For on-policy algorithms with replay buffers considering historical data, the current batch of sequences plus $k=64$ sequences from history, not including the current sequences, are used for learning. Moreover, for replay buffers only using current data, $k=64$ sequences from the current batch of sequences are used for learning, except for \textit{All current} where the entire current batch is used for training. For the off-policy algorithms, only replay buffers using historical data are considered, since they are defined to always do one on-policy update with the current sequences. For each off-policy update, SAC and ACER replay $k=64$ and $k=128$ sequences, respectively, from the buffer. Opposite of what is done for the on-policy algorithms, the sequences of the current episode are immediately stored in the replay memory, i.e., sequences from the current batch can be sampled from the replay memory when performing off-policy updates in the current episode. Below follows descriptions of each replay buffer investigated in this paper, seven in total.

\paragraph{All current (AC)}
For the \textit{All current} (AC) replay buffer, the entire batch of sampled sequences in the current episode is used during learning. No sequences from previous episodes are utilized. In practice, this corresponds to performing a full on-policy update. 

\paragraph{Bin history (BH)}
The \textit{Bin history} (BH) replay buffer sorts sequences into bins with respect to their reward. It consists of the following fixed binds with respect to rewards: $[0,0.1],(0.1,0.2],\dots, (0.9,1]$. For the off-policy algorithms, the buffer also includes a bin storing invalid SMILES strings, i.e., SMILES string with a score $-1$. Each bin has a maximum size of 1000 sequences. First in, first out (FIFO) is applied if the bin is full.  To the extent possible, it does, without replacement, sample an equal number of sequences from each bin and otherwise uniformly samples, without replacement, from the bins with elements that have not been sampled until $k$ sequences have been acquired. 

\paragraph{Bin current (BC)}
\textit{Bin current} (BC) replay buffer sorts the current batch of sampled sequences into bins with respect to their rewards. It consists of the following fixed binds with respect to rewards: $[0,0.1],(0.1,0.2],\dots, (0.9,1]$. It does, to the extent possible, sample an equal number of sequences from each bin without replacement. If $k$ sequences have not been acquired after this, it uniformly samples from the bins with unsampled sequences (i.e., bins whose sequences are not yet in the set used for update) until $k$ sequences have been sampled in total. 

\paragraph{Top-Bottom history (TBH)}
When using \textit{Top-Bottom history} (TBH) replay buffer with an on-policy algorithm, the previous $k/2$ highest and $k/2$ lowest rewarding sequences, and the current sequences, are used for updating the policy. It prioritizes storing the newest sequence(s) if several sequences have an equal reward. No duplicates with the same canonical SMILES string are allowed, keeping the newest sequence with the lowest reward.

For off-policy algorithms, it consists of three sub-buffers, each with a memory size of 1000 sequences. It consists of one sub-buffer with the highest rewarding sequences and two sub-buffers with low rewarding sequences. One of these low-reward sub-buffers stores only sequences with $-1$ reward, i.e., invalid SMILES strings, and the other one stores the lowest-scoring sequences that correspond to a valid molecule (having a reward of at least zero). It uniformly samples, without replacement, from a buffer consisting of all three sub-buffers. FIFO is utilized for each sub-buffer, where the oldest sequences are removed when a sub-buffer is full. No duplicates with the same canonical SMILES string are allowed, where the newest sequence with the lowest score is kept.

\paragraph{Top-Bottom current (TBC)}
For the \textit{Top-Bottom current} (TBC) replay buffer, $k/2$ highest and $k/2$ lowest rewarding sequences of the current batch are used for the update.

\paragraph{Top history (TH)}
When using the \textit{Top history} (TH) replay buffer with an on-policy algorithm, the top-$k$ rewarding sequences from previous episodes and sequences from the current episode are used for updating the actor and critic. Hence, it only needs to store the top-$k$ sequences from previous episodes. Any duplicate with the same canonical SMILES string as another stored sequence is removed.

For off-policy algorithms, it consists of a buffer with the highest rewarding sequences. It has a memory size of 1000 sequences and does not allow any sequences with the same canonical SMILES string, keeping the newest sequence with the lowest score. It uniformly samples $k$ sequences, without replacement, from the buffer to use for the update of the parameters. 

\paragraph{Top current (TC)}
For the \textit{Top current} (TC) replay buffer, the top-$k$ rewarding sequences of the current batch is utilized for updating. This is similar to what is utilized in the Hill-climb algorithm.

\section{Results}\label{sec:results}
\begin{figure}[h]
    \centering
    \includegraphics[width=0.9\textwidth]{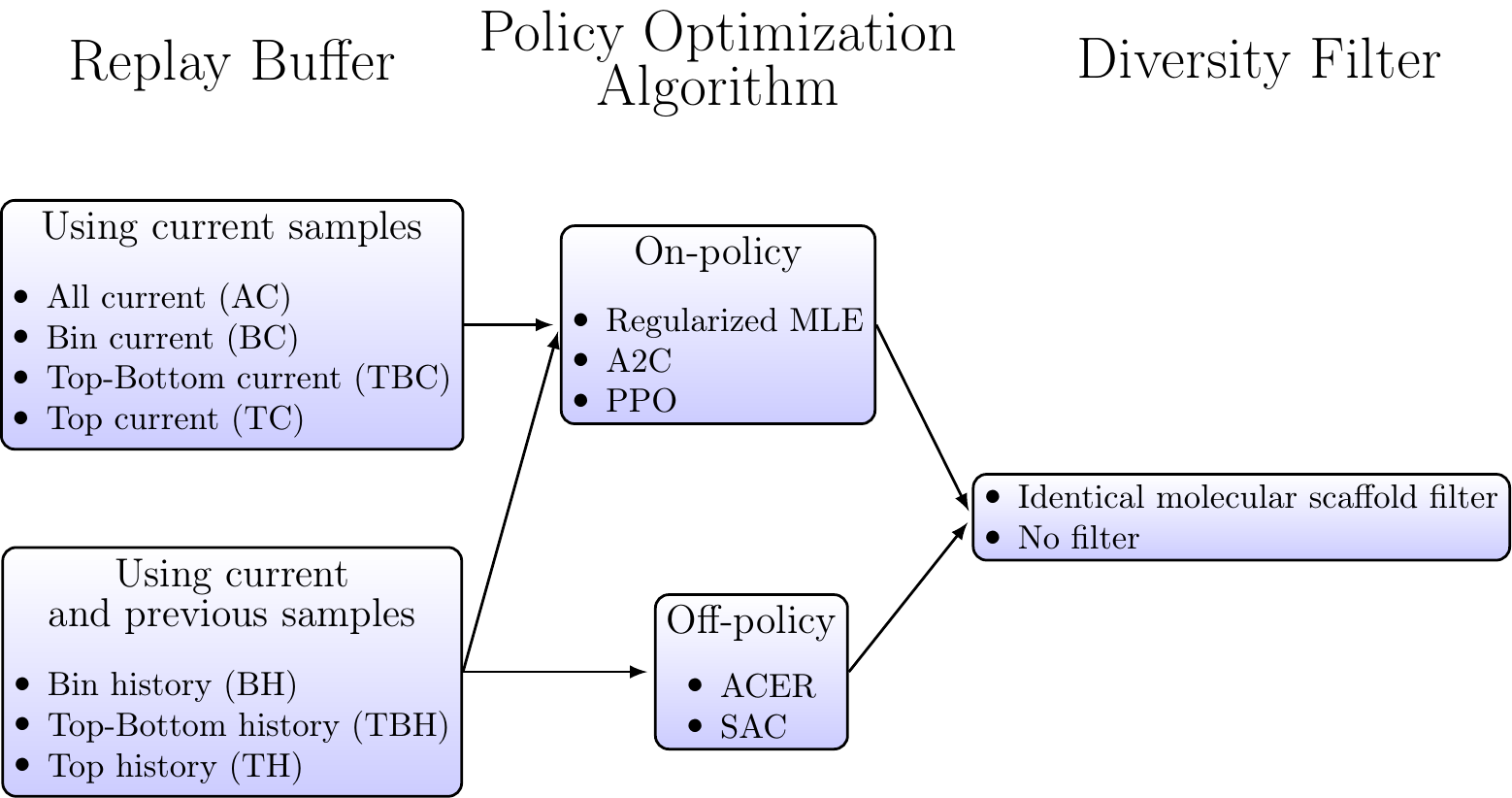}
    \caption{Illustration of the different combinations of replay buffer, policy optimization algorithm and diversity filter that are investigated in this paper.}
    \label{fig:experiments}
\end{figure}

In this section, we investigate various policy optimization algorithms, described in Sec.~\ref{sec:rl}, for the \textit{de novo} drug design setup defined in Sec.~\ref{sec:method}. To compare all algorithms under the same budget constraint, the generation is limited to 2000 episodes, giving a budget of \num{256000} possible SMILES strings in total. We investigate both the use of the diversity filter based on identical molecular scaffolds (see Sec. \ref{sec:diversity_filter}) and the use of no diversity filter. The results are divided into on- and off-policy algorithms. \figref{fig:experiments} displays the different combinations of replay buffer, policy optimization algorithm and diversity filter that are investigated. Figs. \ref{fig:smiles_a2c} to \ref{fig:smiles_sac_nofilter} in Appendix \ref{app:smiles} display the ten top-scoring molecules, each from a unique topological scaffold, of one representative run for each combination. The generated molecules look like what has been published in previous work targeting DRD2 \cite{olivecrona2017molecular,thomas2022augmented}.

\subsection{On-policy Algorithms}
Firstly, we study \textit{de novo} drug design utilizing on-policy algorithms. We investigate the algorithms A2C, Regularized MLE and PPO, when either using the identical molecular scaffold diversity filter or no diversity filter.

\subsubsection{With Diversity Filter}\label{sec:onpolicy_filter}
\begin{figure}[h]
     \centering
     \subfloat[Number of active molecules]{
    \includegraphics[width=0.49\textwidth]{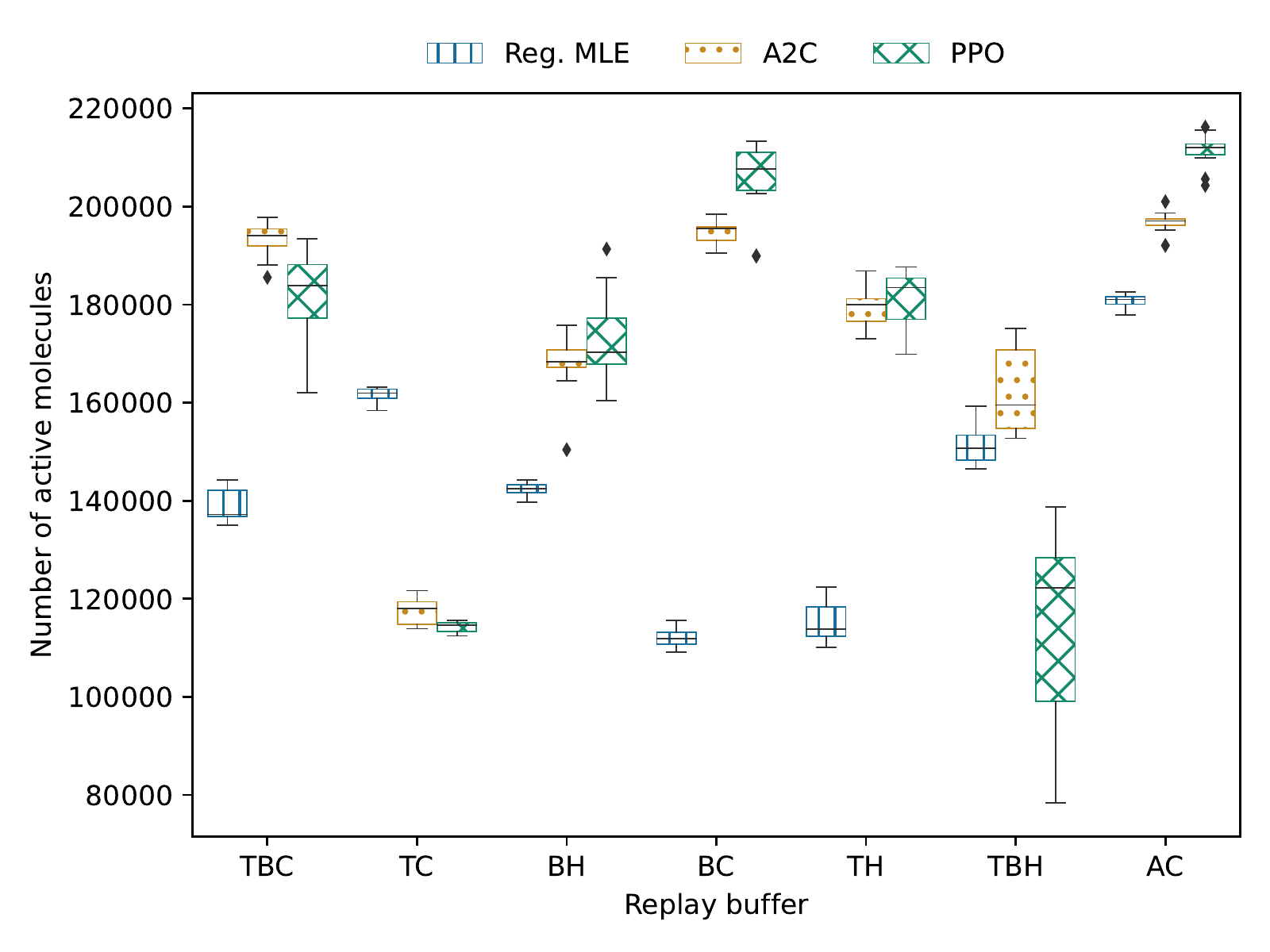}\label{fig:actives_filter}}
     \hfill
    \subfloat[Number of active molecular scaffolds]{\includegraphics[width=0.49\textwidth]{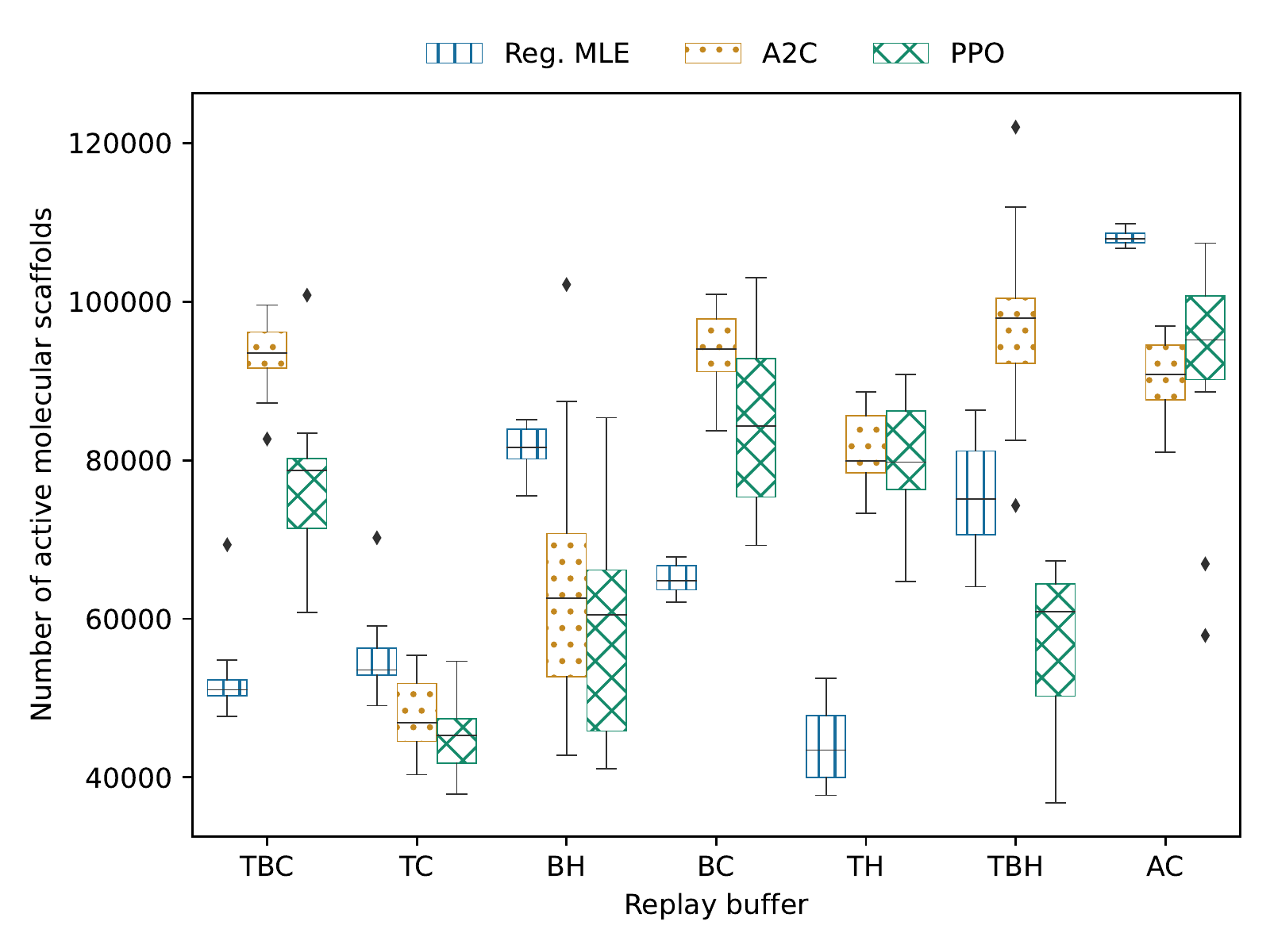}\label{fig:mol_scaff_filter}}
     \hfill
     \subfloat[Number of active topological scaffolds]{\includegraphics[width=0.49\textwidth]{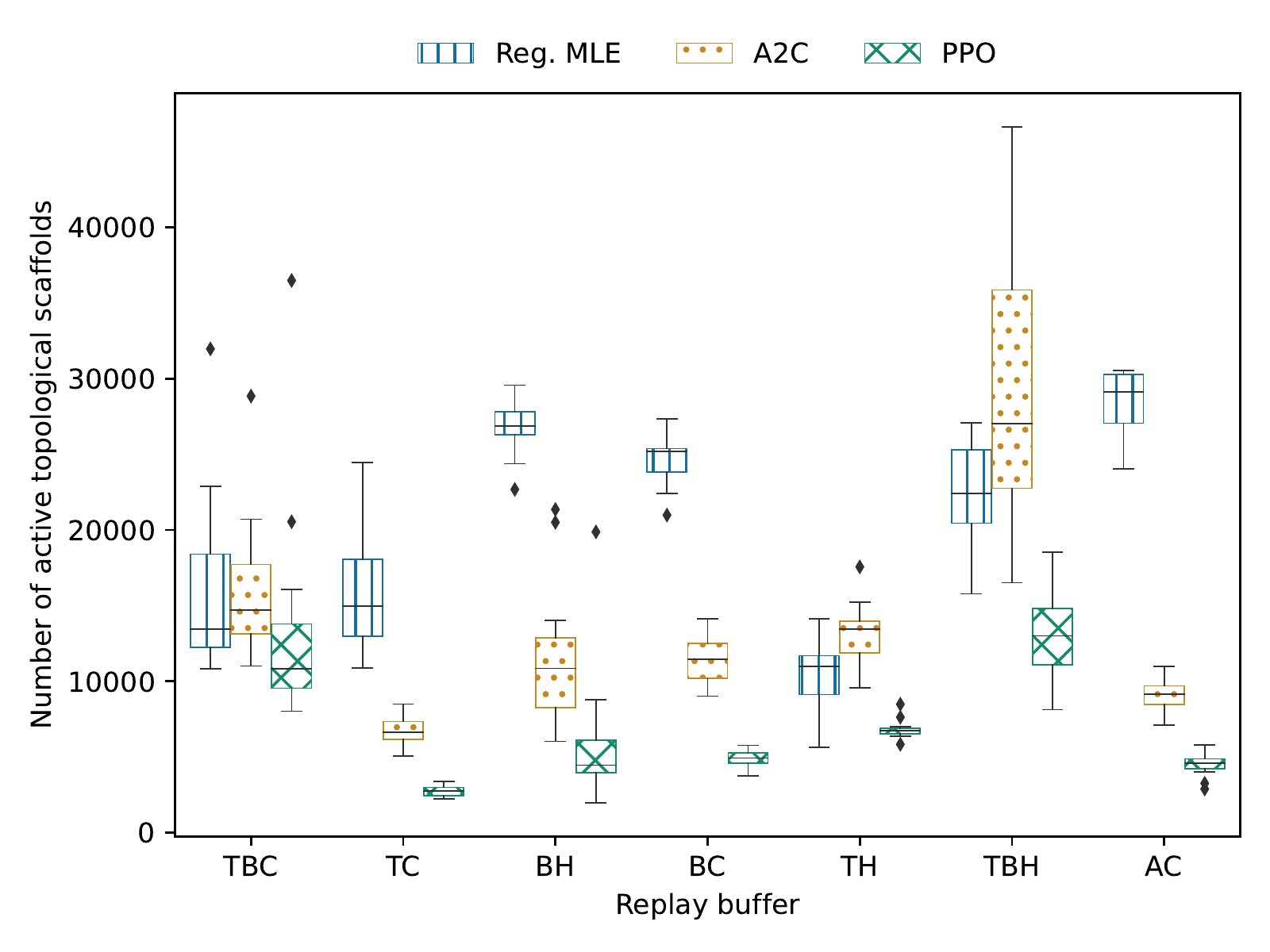}\label{fig:topo_scaff_filter} }
    \caption{Box plots of the number of unique active molecules and active scaffolds for the on-policy algorithms A2C, Regularized MLE, and PPO (higher is better) when utilizing identical molecular scaffold filter.  It shows mean and standard deviation over 11 runs for each policy and replay buffer. 128 SMILES strings are sampled in each episode, with a budget of 2000 episodes in total.}
    \label{fig:results_filter}
\end{figure}

\figref{fig:results_filter} shows box plots of the number of (unique) active molecules and active scaffolds over 11 runs for the on-policy algorithms utilizing different replay buffers, and identical molecular scaffolds diversity filter. As illustrated in \figref{fig:actives_filter}, for most replay buffers, both A2C and PPO sample a significantly larger number of active molecules compared to Regularized MLE; whereas there is relatively little difference between A2C and PPO in terms of the number of active molecules generated. Regularized MLE only generates notably more unique active molecules when using the \textit{top current} replay buffer, compared to A2C and PPO using the same replay buffer. When using PPO and A2C, the replay buffers \textit{All current} and \textit{Bin current} yield the largest number of active molecules. Moreover, when using Regularized MLE, utilizing \textit{All current} generates the largest number, whereas \textit{Top current} is the second-best replay buffer.

As seen in \figref{fig:mol_scaff_filter}, Regularized MLE utilizing \textit{All current} show the largest median of (unique) active scaffolds with low variability, compared to A2C and PPO. This combination also shows the largest median of generated (unique) active topological scaffolds, as seen in \figref{fig:topo_scaff_filter}. Regularized MLE generates a significantly larger number of active topological scaffolds with most replay buffers, compared to A2C and PPO. However, A2C with \textit{Top-Bottom history} performs on par, in terms of active scaffolds, with Regularized MLE but has a larger variability. In fact, A2C with \textit{Top-Bottom history} displays the best runs in terms of the number of active scaffolds. Moreover, in most cases, A2C generates a significantly larger number of topological scaffolds than PPO; while comparable numbers of active molecular scaffolds are often generated. 

\begin{figure}[h]
     \centering
     \subfloat[PPO]{
    \includegraphics[width=0.49\textwidth]{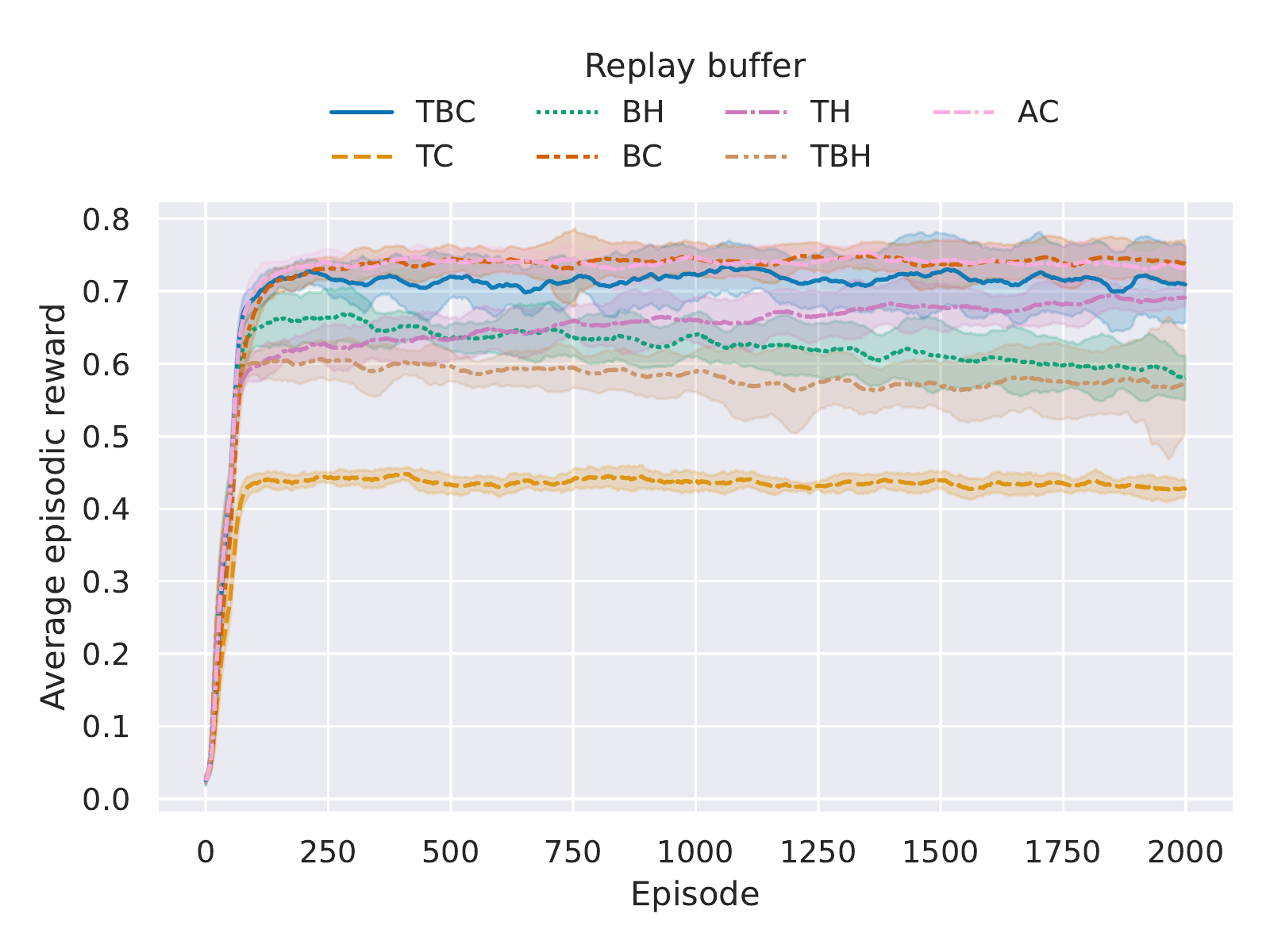}\label{fig:reward_ppo_filter}}
     \hfill
    \subfloat[A2C]{\includegraphics[width=0.49\textwidth]{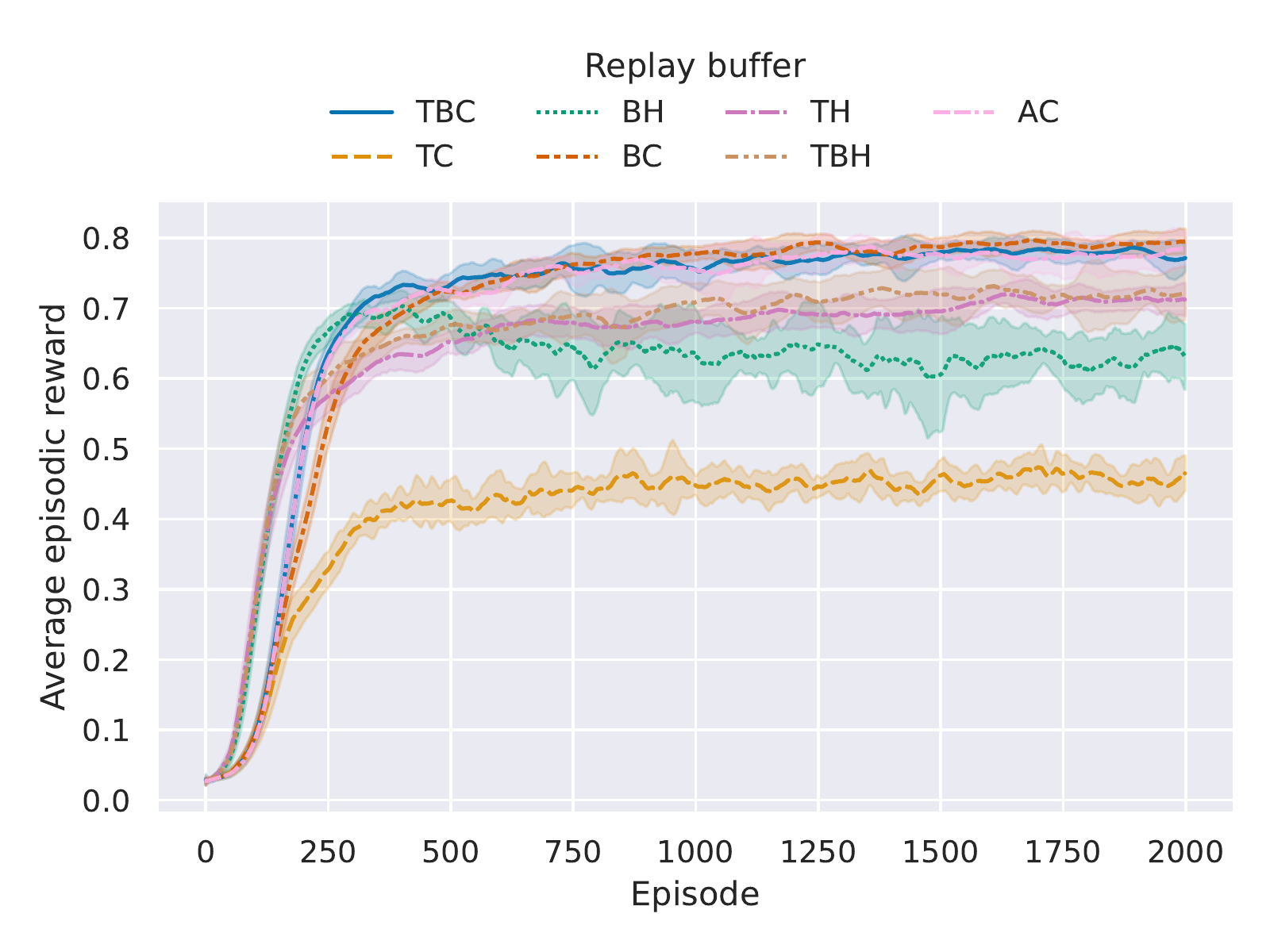}\label{fig:reward_a2c_filter}}
     \hfill
     \subfloat[Reg. MLE]{\includegraphics[width=0.49\textwidth]{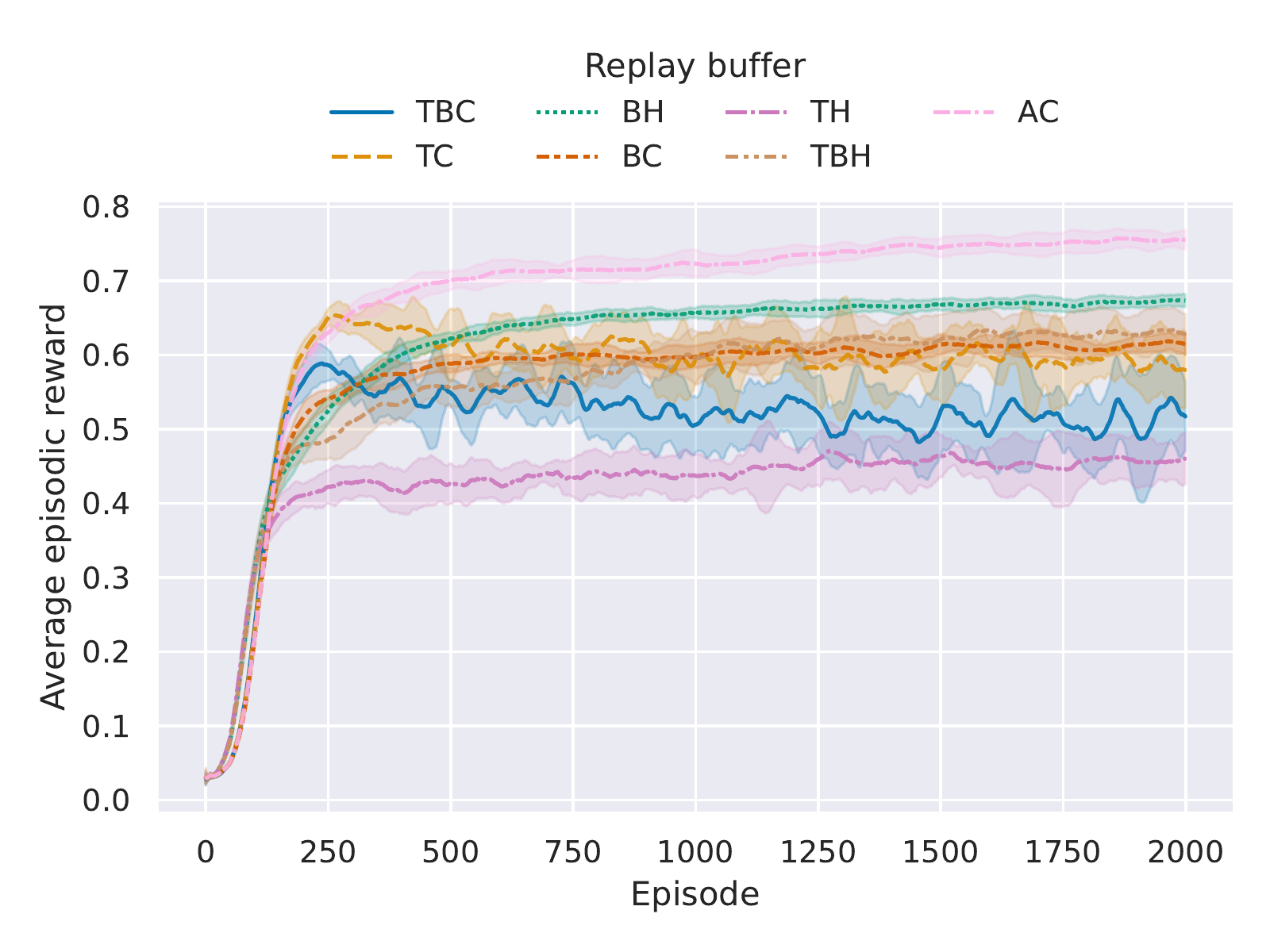}\label{fig:rew_mle_filter} }
    \caption{Average episodic reward computed over a batch of sequences, for the on-policy algorithms A2C, Regularized MLE and PPO (higher is better) when utilizing identical molecular scaffold filter. It shows mean and standard deviation of moving average, of window size 50, over 11 runs for each policy and replay buffer. 128 SMILES strings are sampled in each episode, with a budget of 2000 episodes in total.}
    \label{fig:reward_filter}
\end{figure}

\figref{fig:reward_filter} displays means and standard deviations, over 11 repeated runs, of the average episodic rewards of sampled molecules for all combinations of on-policy algorithms and replay buffers over 2000 episodes of batch size 128. The average episodic rewards are displayed using a moving average with a window size of 50. All runs use the identical molecular scaffold filter. For both PPO and A2C, \textit{Top current} gives the lowest average episodic reward, as shown in \figref{fig:reward_ppo_filter} and \figref{fig:reward_a2c_filter}, respectively. Furthermore, PPO utilizing \textit{Bin current} and \textit{All current} performs on par, giving moving averages between 0.8 and 0.7 after roughly 125 episodes. It seems that most replay buffers converge after around 125 episodes. Moreover, for A2C, \textit{All current}, \textit{Top-Bottom current} and  \textit{Bin current} perform among the best in terms of the average episodic reward, when using a diversity filter. Compared to PPO, it takes slightly more episodes for the rewards to converge. For Regularized MLE, utilizing \textit{All current} gives the largest average episodic reward, converging to a reward between 0.7 and 0.8 with low variance. It is the only replay buffer for Regularized MLE that can reach an average episodic reward above 0.7. Utilizing \textit{Top history} yields the lowest episodic reward.

\subsubsection{Without Diversity filter} \label{sec:onpolicy_nofilter}
\begin{figure}[h]
     \centering
     \subfloat[Number of active molecules]{
    \includegraphics[width=0.49\textwidth]{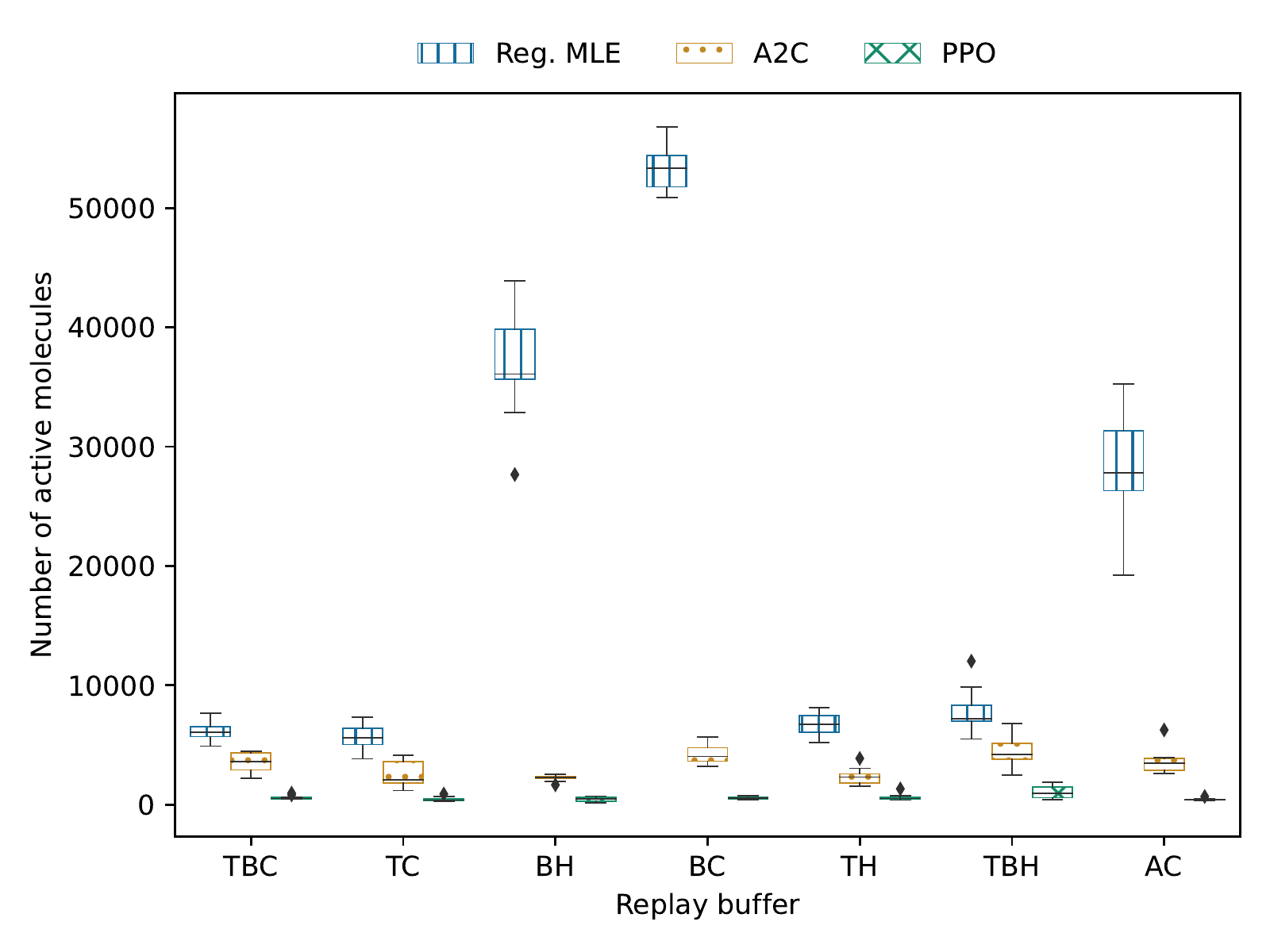}\label{fig:actives_nofilter}}
     \hfill
    \subfloat[Number of active molecular scaffolds]{\includegraphics[width=0.49\textwidth]{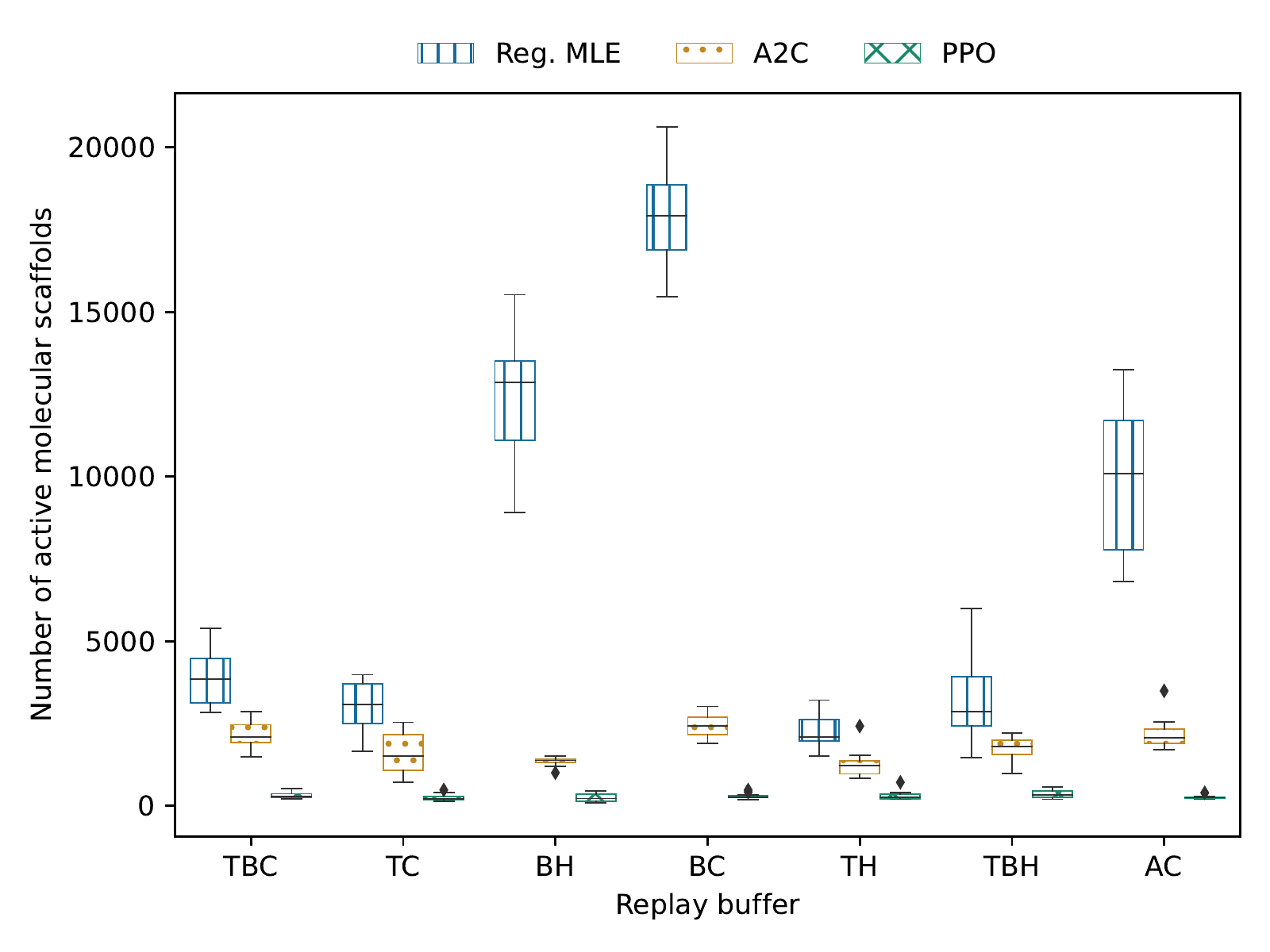}\label{fig:mol_scaff_nofilter}}
     \hfill
     \subfloat[Number of active topological scaffolds]{\includegraphics[width=0.49\textwidth]{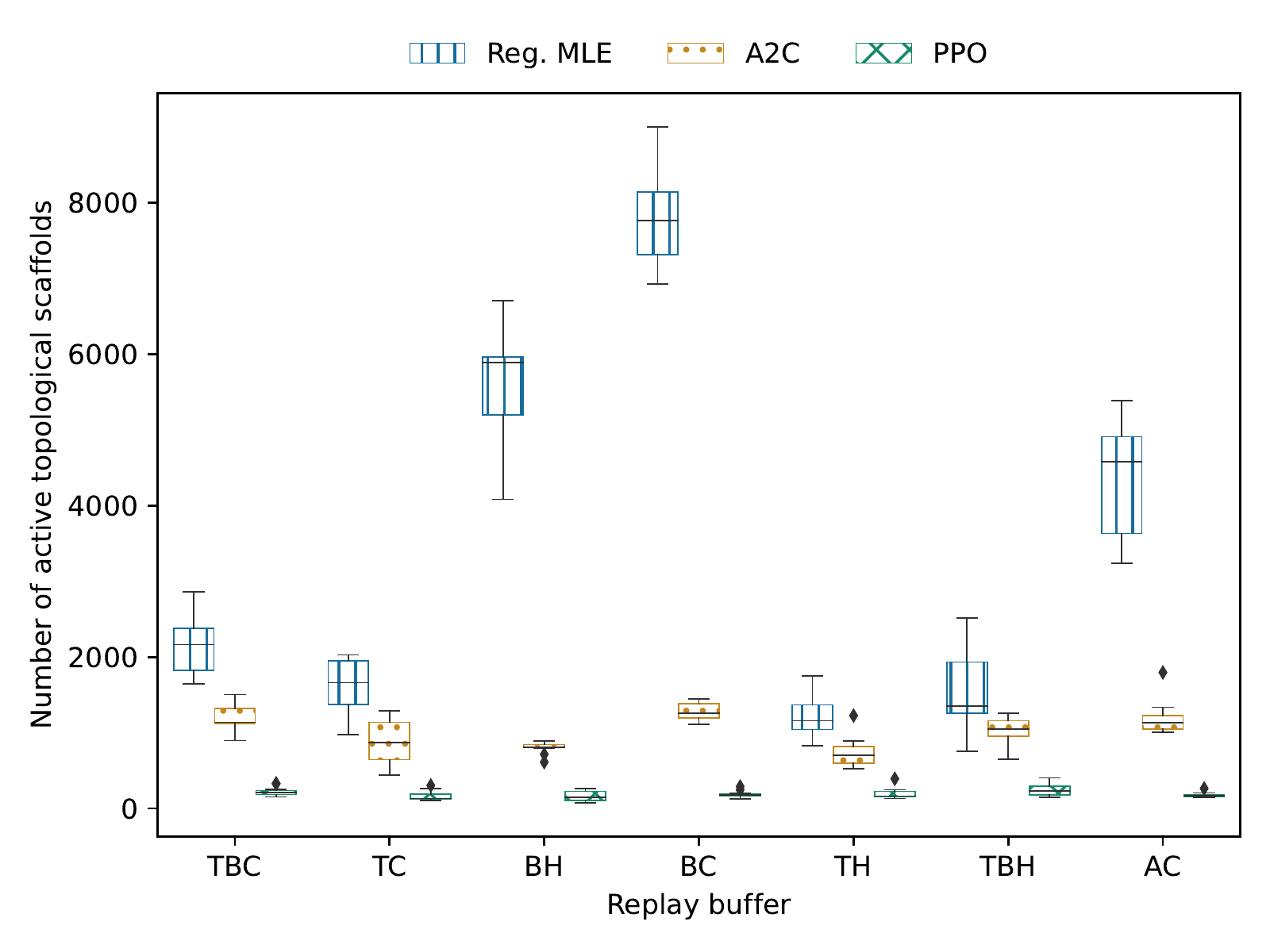}\label{fig:topo_scaff_nofilter} }
    \caption{Box plots of the number of unique active molecules and active scaffolds for the on-policy algorithms A2C, Regularized MLE and PPO (higher is better) when utilizing no diversity filter. 128 SMILES strings are sampled in each episode, with a budget of 2000 episodes in total.}
    \label{fig:results_nofilter}
\end{figure}

\figref{fig:results_nofilter} shows boxplots of the number of (unique) active molecules and active scaffolds over 11 runs for the on-policy algorithms utilizing different replay buffers. No diversity filter is used.
In this setting, Regularized MLE using either \textit{Bin history}, \textit{Bin current} or \textit{All current} consistently generates a larger number of active molecules, molecular scaffolds and topological scaffolds, compared to all other combinations of on-policy algorithm and replay buffer. \textit{Bin current} yields the largest number for all these metrics. For all replay buffers, Regularized MLE gives the largest number of active molecules and scaffolds, while A2C gives a higher number than PPO.

\begin{figure}[h]
     \centering
     \subfloat[PPO]{
    \includegraphics[width=0.49\textwidth]{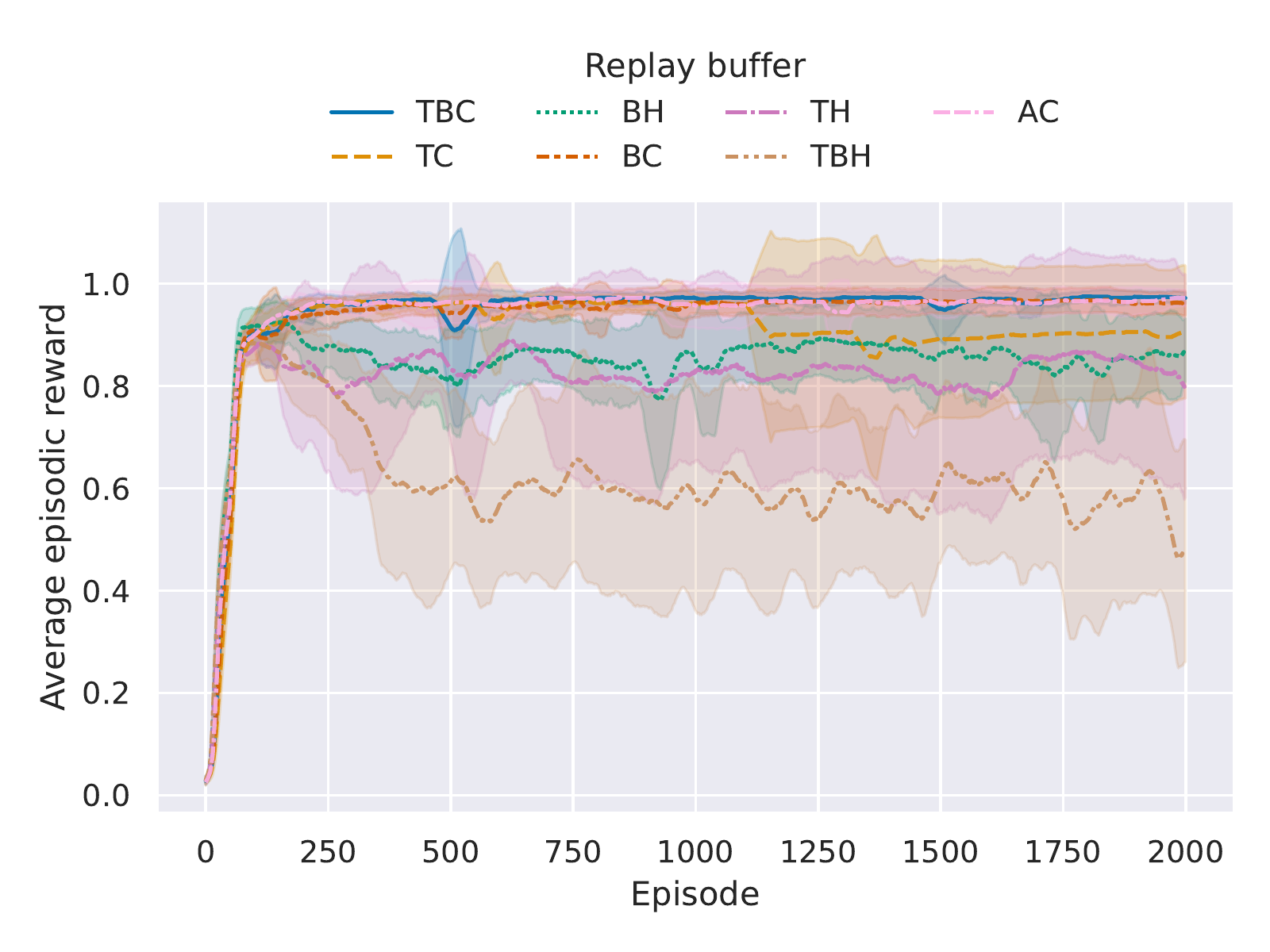}\label{fig:reward_ppo_nofilter}}
     \hfill
    \subfloat[A2C]{\includegraphics[width=0.49\textwidth]{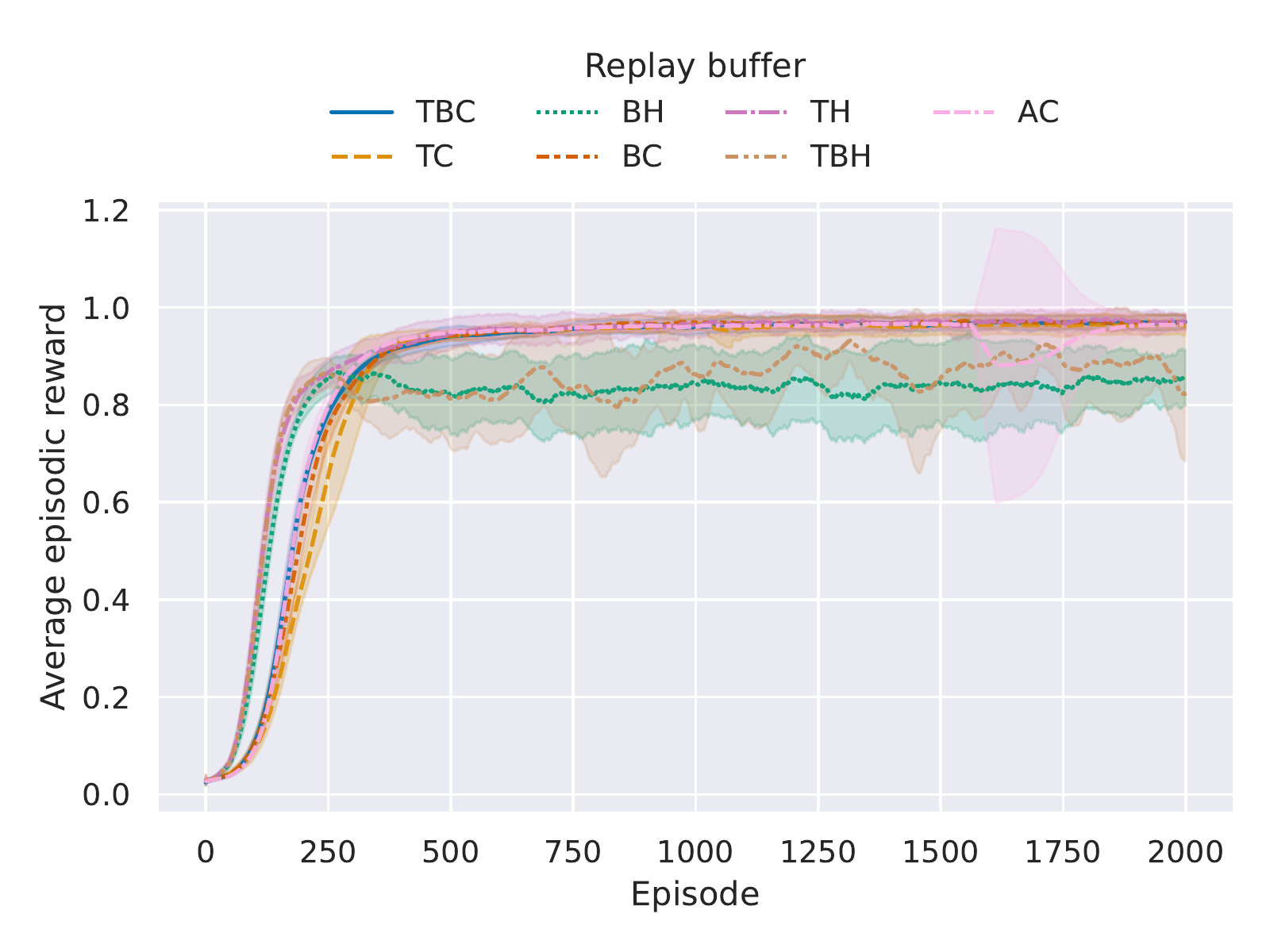}\label{fig:reward_a2c_nofilter}}
     \hfill
     \subfloat[Reg. MLE]{\includegraphics[width=0.49\textwidth]{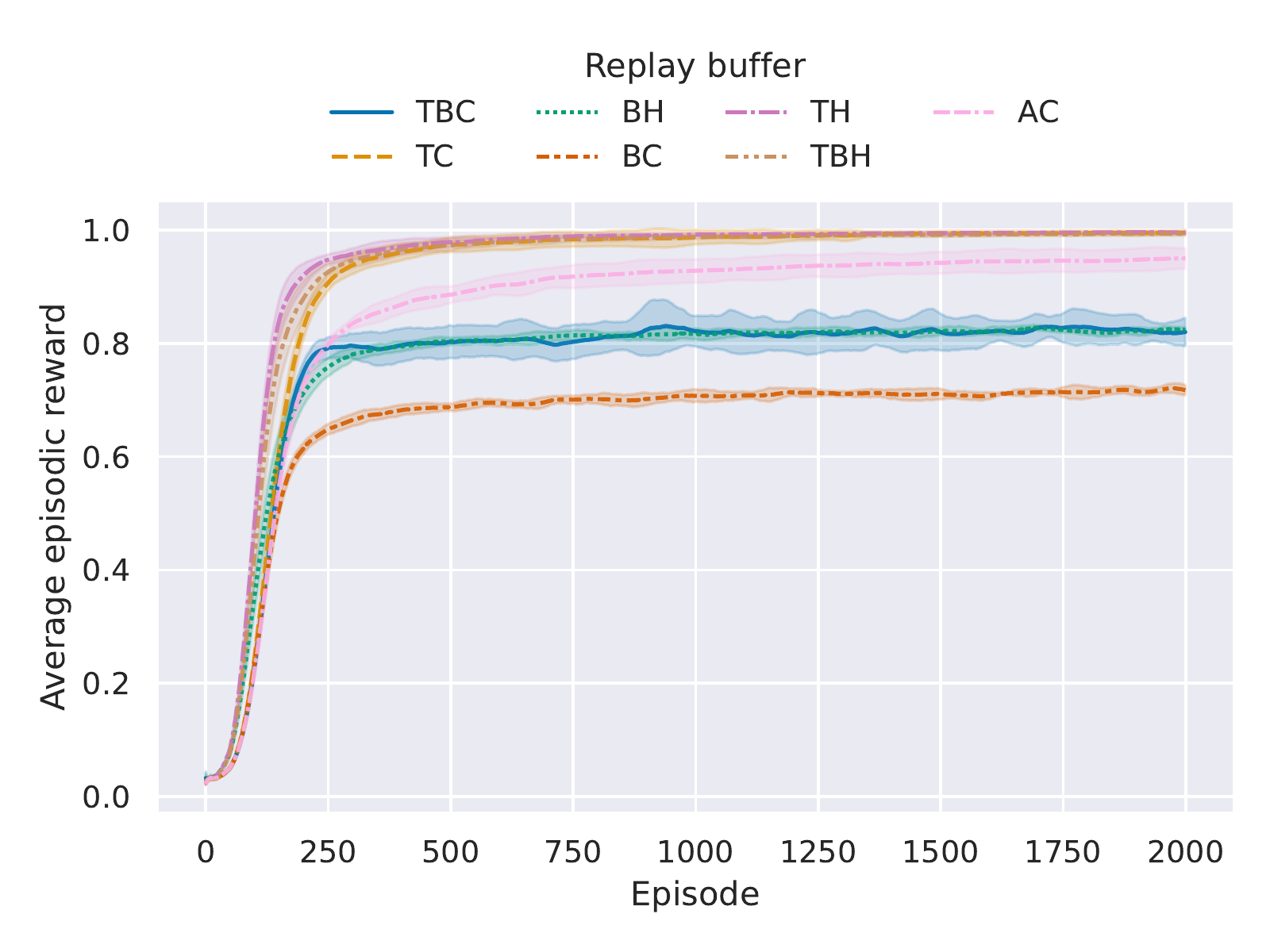}\label{fig:rew_mle_nofilter} }
    \caption{Average episodic reward computed over a batch of sequences, for the on-policy algorithms A2C, Regularized MLE and PPO (higher is better) when utilizing no diversity filter. It shows mean and standard deviation of moving average, of window size 50, over 11 runs for each policy and replay buffer. 128 SMILES strings are sampled in each episode, with a budget of 2000 episodes in total.}
    \label{fig:reward_nofilter}
\end{figure}

\figref{fig:reward_nofilter} displays means and standard deviations, over 11 repeated runs, of the average episodic rewards of sampled molecules for all combinations of on-policy algorithms and replay buffers using no diversity filter. A budget of 2000 episodes of batch size 128 is investigated. For visualization purposes, the average episodic rewards correspond to moving averages using a window size of 50. PPO seems to require the least number of episodes to converge but, on the other hand, shows a higher variance. 

As seen in \figref{fig:reward_ppo_nofilter}, PPO with \textit{Top-Bottom current}, \textit{All current} and \textit{Top-Bottom history} reaches an average episodic reward 1 approximately. Using \textit{Top-bottom history} shows a substantially lower episodic reward but has a larger variance. 

For A2C, displayed in \figref{fig:reward_a2c_nofilter}, using replay buffers with data samples from previous episodes, except for \textit{Top current}, gives slightly lower episodic reward compared to using only data sampled in the current episode. All replay buffers only using immediate samples give an average episodic reward close to 1. One should note that the short drop of reward for \textit{All current} occurs due to the resetting of the parameters of one run to that of the pre-trained model, since the policy samples less than 80\% valid molecules for more than 10 consecutive episodes. After restarting, it quickly gets back on track. This run is kept to highlight that it is possible for the networks to diverge from the pre-trained model and forget how to generate valid SMILES strings. When this happens, it can quickly find its way back by restarting from the pre-trained model.

For Regularized MLE, there is generally a lower variance for the episodic reward compared to PPO and A2C. When using either \textit{Top-Bottom history}, \textit{Top current} or \textit{Top history}, the average episodic reward converges to 1, as illustrated in \figref{fig:rew_mle_nofilter}. Note that neither PPO nor A2C consistently reaches such high episodic reward. Also, for regularized MLE, \textit{All current} is not among the best ones, which is the case for both PPO and A2C.

\subsection{Off-policy Algorithms}
To further investigate the benefits of a replay buffer, two off-policy algorithms have been explored: (1) Soft Actor-Critic (SAC); (2) Actor-Critic with Experience replay (ACER).

The off-policy algorithms perform one step of on-policy update (using all sampled data in the current episode) and several off-policy updates using sequences from both current and previous episodes. The off-policy updates use replays from either \textit{Bin history}, \textit{Top history} or \textit{Top-Bottom history}. Opposite to the on-policy algorithms, the sequences of the current episode are stored in the replay memory before use in the current episode, i.e., it is possible to utilize sequences from the current batch for off-policy updates in the current episode.

\subsubsection{With Diversity Filter}
\begin{figure}[h]
     \centering
     \subfloat[Number of active molecules]{
    \includegraphics[width=0.49\textwidth]{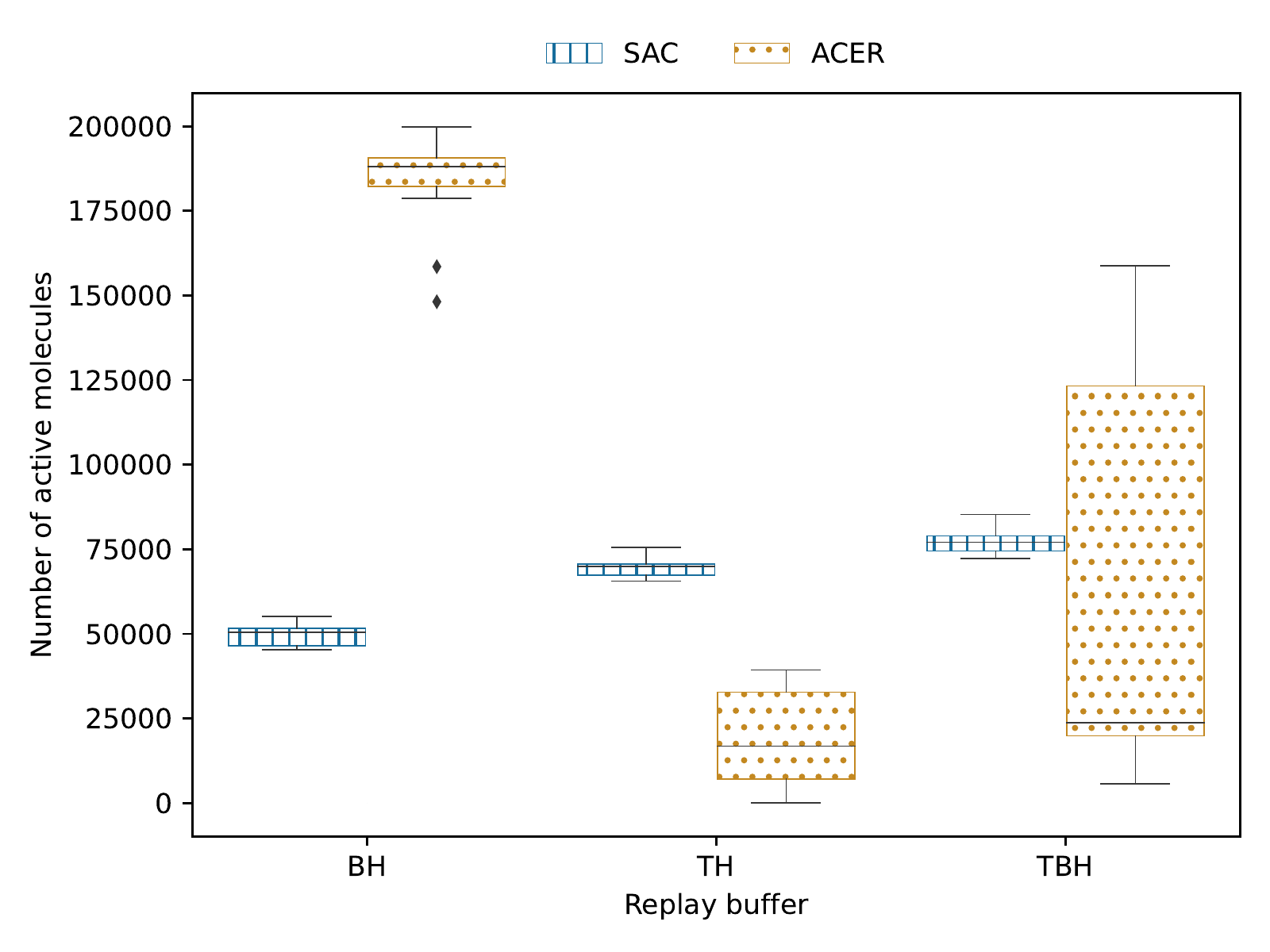}\label{fig:actives_filter_offpolicy}}
     \hfill
    \subfloat[Number of active molecular scaffolds]{\includegraphics[width=0.49\textwidth]{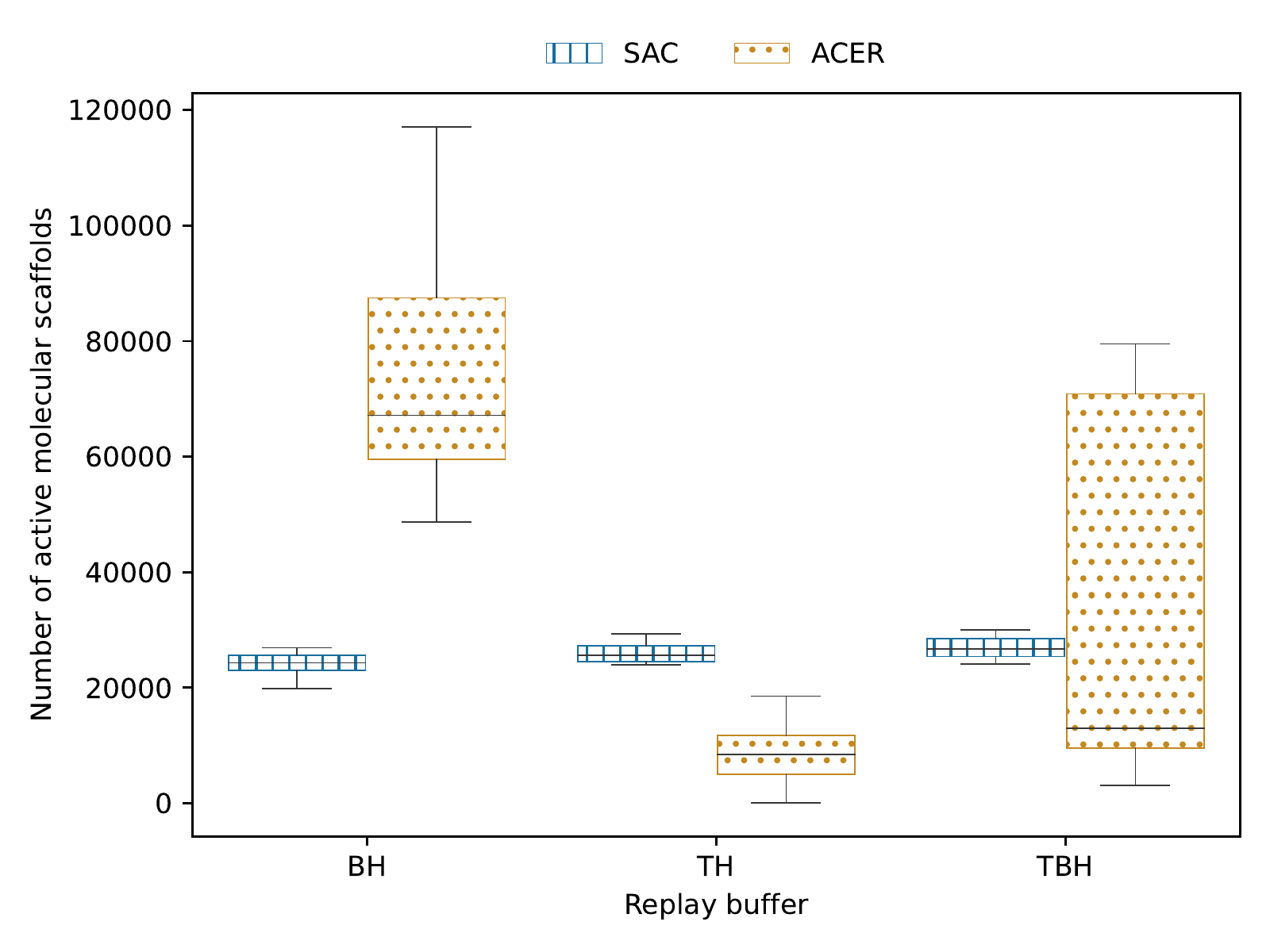}\label{fig:mol_scaff_filter_offpolicy}}
    \hfill
     \subfloat[Number of active topological scaffolds]{\includegraphics[width=0.49\textwidth]{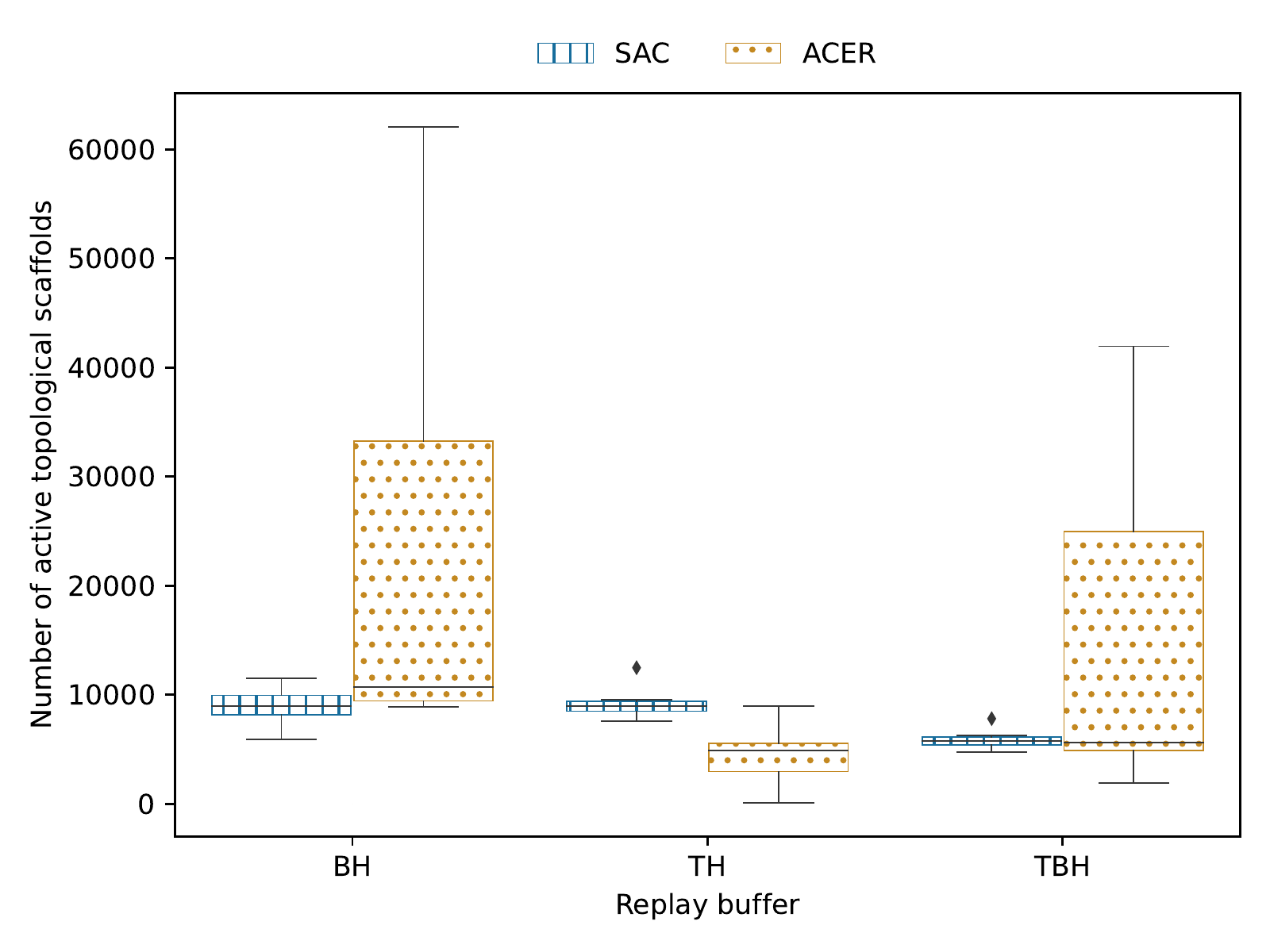}\label{fig:topo_scaff_filter_offpolicy}}
    \caption{Box plots of the number of unique active molecules and active scaffolds for the off-policy algorithms ACER and SAC (higher is better) when utilizing identical molecular scaffold filter. The box plot is computed over 11 repeated runs for each combination of policy and replay buffer. 128 SMILES strings are sampled in each episode, with a budget of 2000 episodes in total.}
    \label{fig:results_filter_offpolicy}
\end{figure}

\figref{fig:results_filter_offpolicy} displays box plots of the number of active molecules and scaffolds for 11 repeated runs of each combination of off-policy algorithm and replay buffer. 128 molecules are generated in each episode, not necessarily valid and/or unique molecules, with a total budget of 2000 episodes. ACER using \textit{Bin history} generates the largest number of active molecules and scaffold. It is able to yield numbers close to or better than the best on-policy results in Sec. \ref{sec:onpolicy_filter}, which the other off-policy combinations are not able to. However, in general, ACER shows a larger variability compared to SAC, where ACER using \textit{Top-Bottom history} shows the largest variability over the repeated runs.

\begin{figure}[h]
     \centering
     \subfloat[SAC]{
    \includegraphics[width=0.49\textwidth]{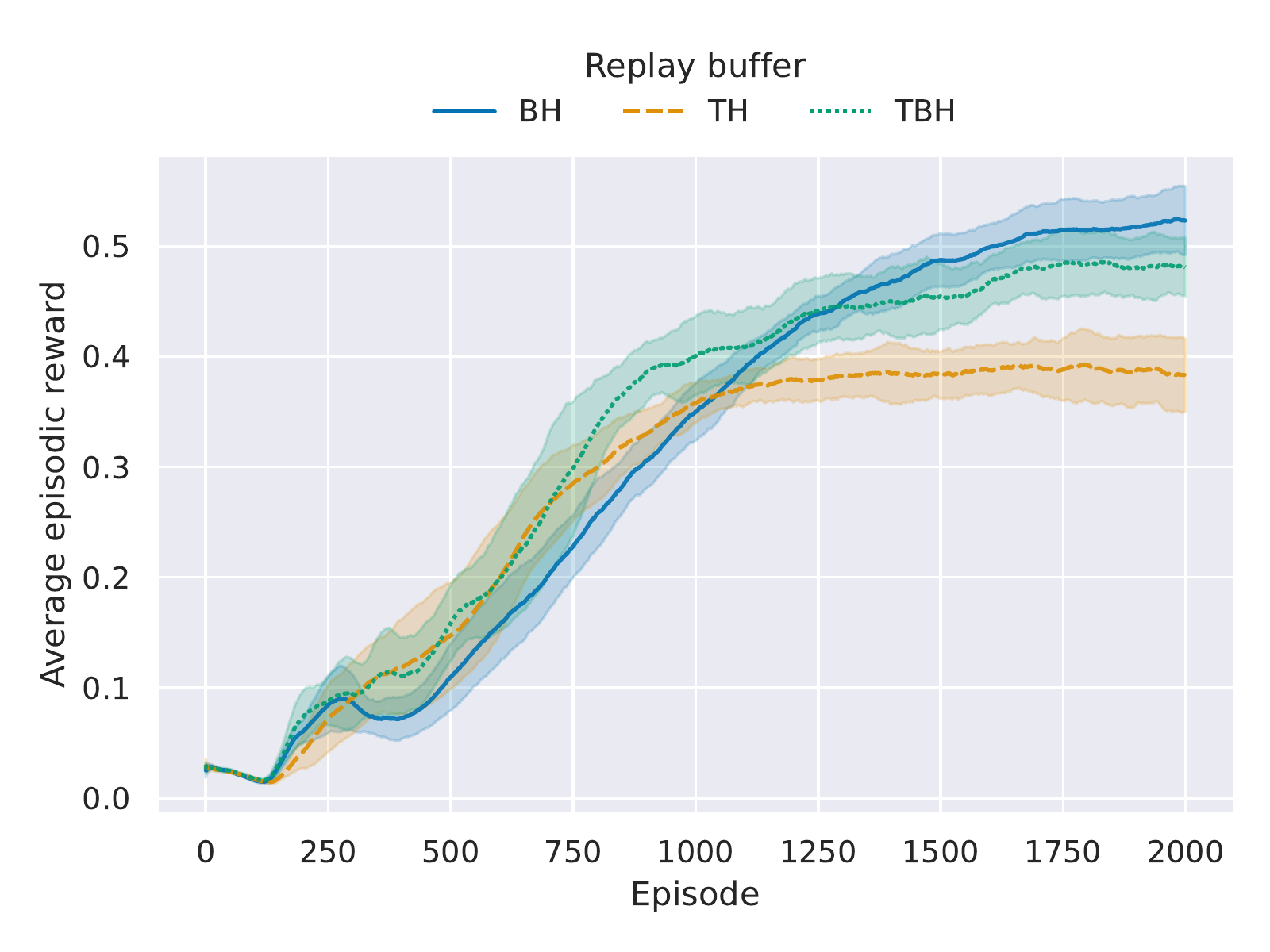}\label{fig:reward_sac_filter}}
     \hfill
    \subfloat[ACER]{\includegraphics[width=0.49\textwidth]{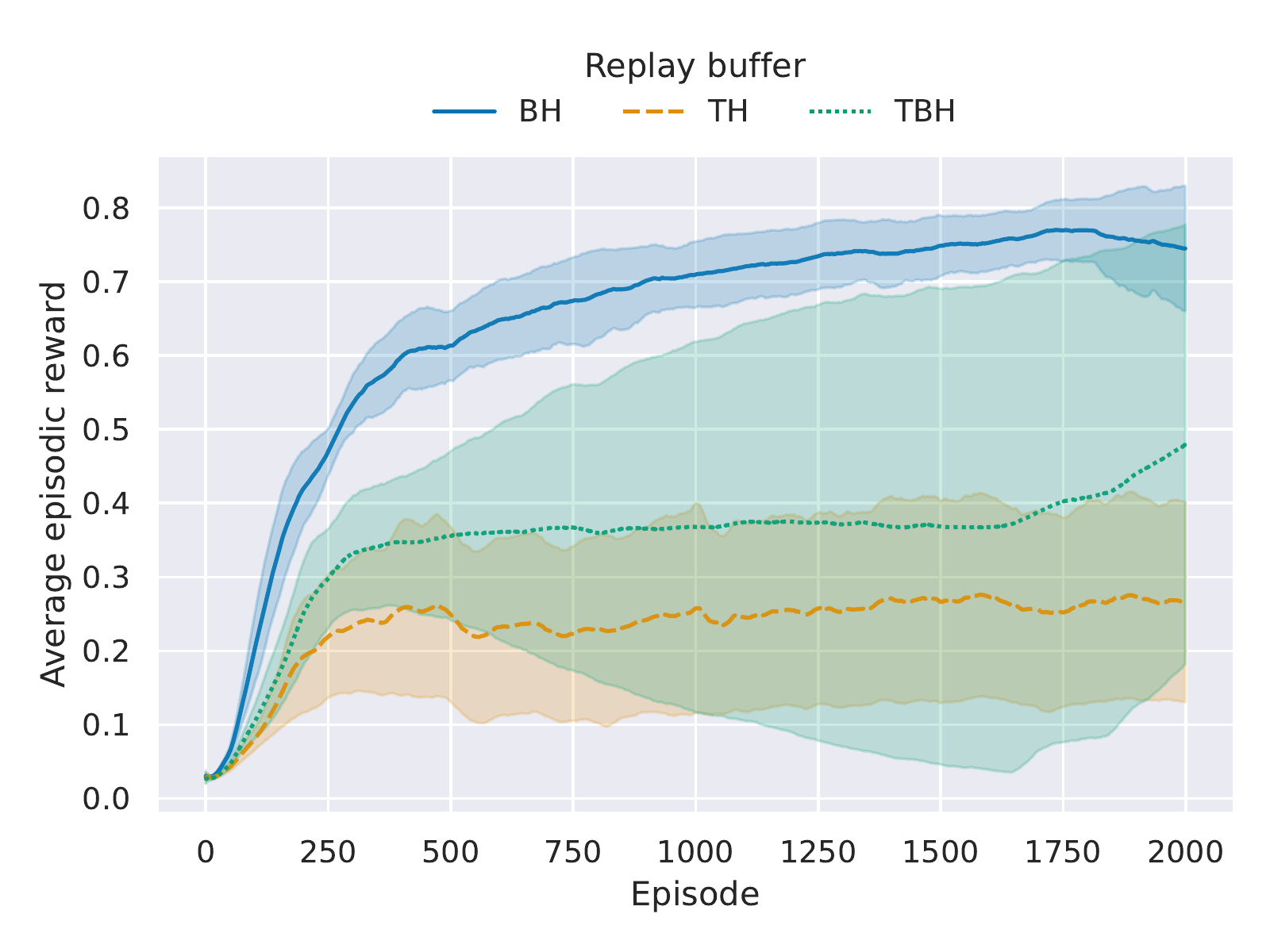}\label{fig:reward_acer_filter}}
    \caption{Average episodic reward computed over a batch of sequences, for the off-policy algorithms SAC and ACER (higher is better) when utilizing identical molecular scaffold filter. It shows mean and standard deviation of moving average, of window size 50, over 11 runs for each policy and replay buffer. 128 SMILES strings are sampled in each episode, with a budget of 2000 episodes in total. Note that invalid SMILES are displayed with a reward of $0$ in this figure, but are given a reward of $-1$ during training.}
    \label{fig:reward_filter_offpolicy}
\end{figure}

\figref{fig:reward_filter_offpolicy} shows means and standard deviations, over 11 repeated runs, of the average episodic rewards of sampled molecules for all combinations of off-policy algorithms and replay buffers using the identical molecular scaffold filter. A budget of 2000 episodes, each sampling a batch of 128 molecules, is investigated. For visualization purposes, each average episodic reward corresponds to the moving average using a window size of 50. Note that, in this figure, invalid SMILES are displayed with a reward of $0$ but are given a reward of $-1$ during training.

When using SAC, both \textit{Bin history} and \textit{Top-Bottom history} reach an average episodic reward of 0.5 approximately; while \textit{Top history} reaches an average episodic reward of 0.4 approximately, which is the minimum reward for a sequence to be saved in the diversity filter. For ACER, \textit{Bin history} reaches an episodic reward of around 0.75; while using the other two replay buffers yields substantially lower episodic rewards and higher variances, in particular for \textit{Top-Bottom history}. \textit{Top history} reaches an episodic reward below 0.3, not reaching above the diversity filter threshold. ACER with \textit{Bin history} is the only combination that is able to obtain an average episodic reward comparable to the best on-policy algorithms, but with a slower increase in average episodic reward. SAC displays a significantly slower increase in average episodic reward than both ACER and the on-policy algorithms. 

\subsubsection{Without Diversity filter}
\begin{figure}[h]
     \centering
     \subfloat[Number of active molecules]{
    \includegraphics[width=0.49\textwidth]{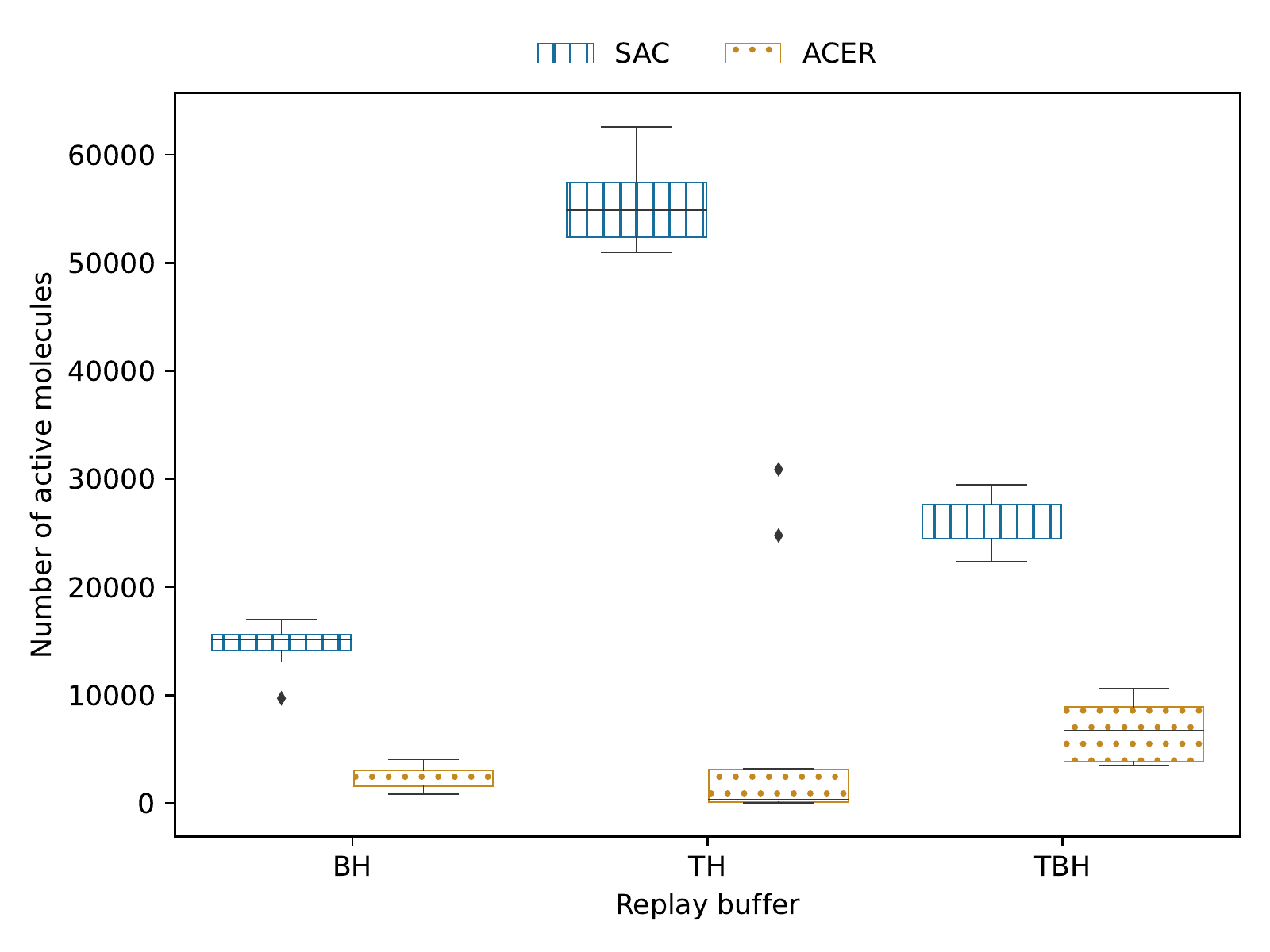}\label{fig:actives_nofilter_offpolicy}}
     \hfill
    \subfloat[Number of active molecular scaffolds]{\includegraphics[width=0.49\textwidth]{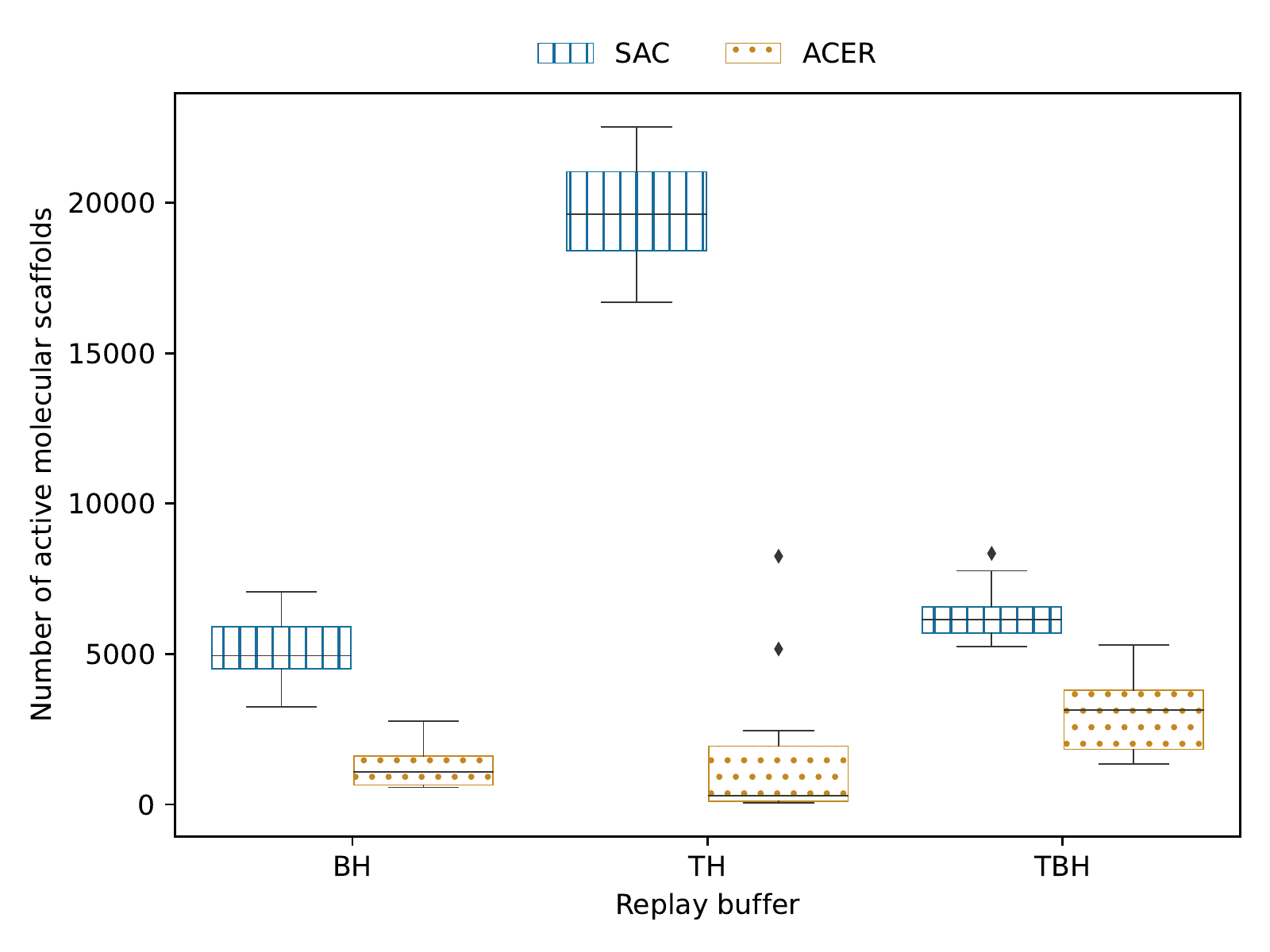}\label{fig:mol_scaff_nofilter_offpolicy}}
    \hfill
     \subfloat[Number of active topological scaffolds]{\includegraphics[width=0.49\textwidth]{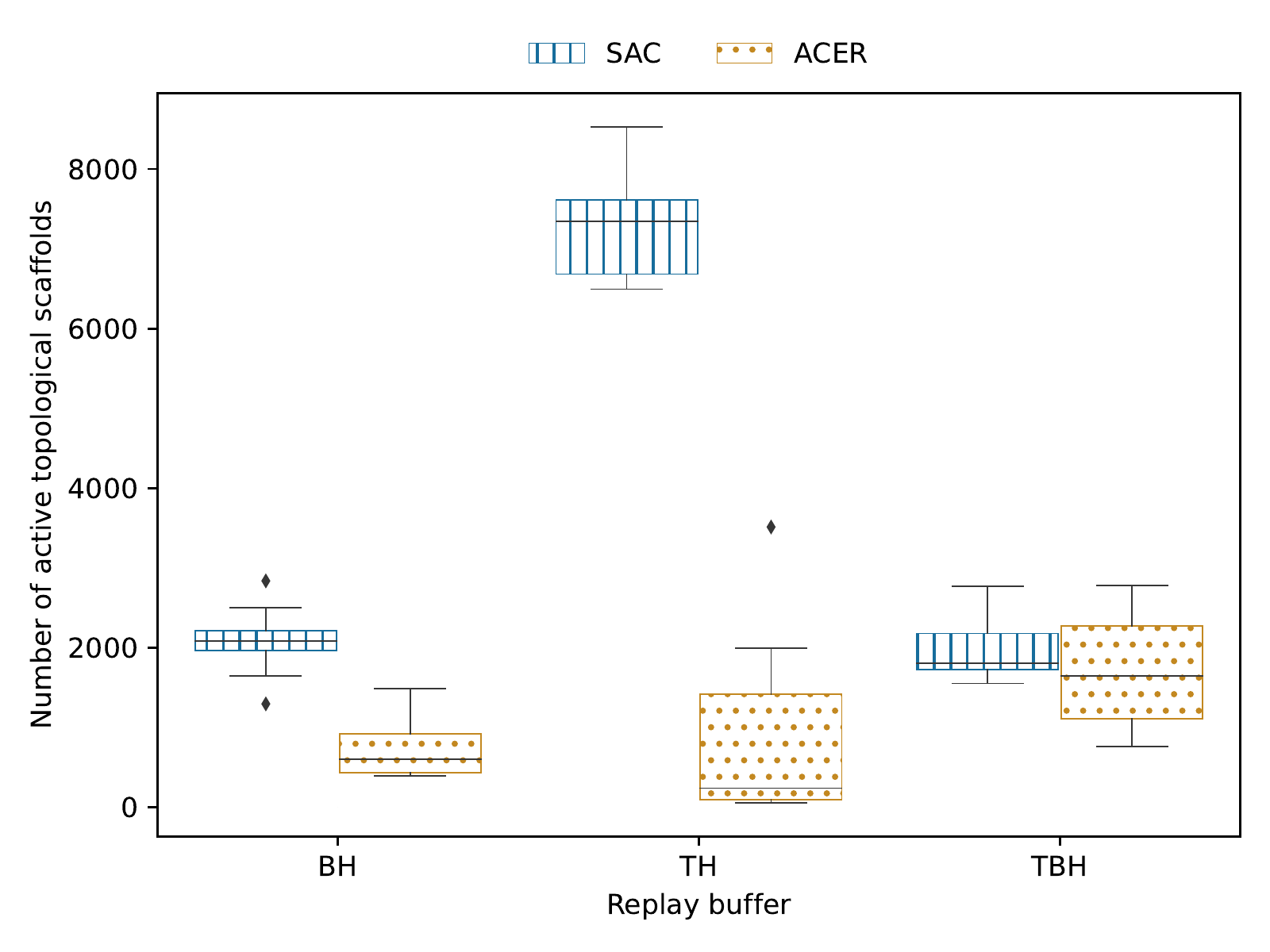}\label{fig:topo_scaff_nofilter_offpolicy}}
    \caption{Box plots of the number of unique active molecules and active scaffolds for the off-policy algorithms ACER and SAC (higher is better) when utilizing no diversity filter. It shows mean and standard deviation over 11 runs for each policy and replay buffer. 128 SMILES strings are sampled in each episode, with a budget of 2000 episodes in total.}
    \label{fig:results_nofilter_offpolicy}
\end{figure}

\figref{fig:results_nofilter_offpolicy} displays box plots of the number of active molecules and scaffolds, over 11 repeated runs for each combination of off-policy algorithms and replay buffers. Note that invalid SMILES are displayed again with a reward of $0$ in this figure but are given a reward of $-1$ during training. No diversity filter is used, i.e., the generation of similar molecules is not penalized between episodes. It is observed that SAC using \textit{Top history} generates the largest number of (unique) active molecules and scaffolds, on par with the best on-policy without a diversity filter, Regularized MLE with \textit{Bin current}. For all replay buffers, SAC seems to generate a significantly larger number of active molecules and scaffolds compared to ACER, except for topological scaffolds when using \textit{Top-bottom} history where they display similar performance.

\begin{figure}[h]
     \centering
     \subfloat[SAC]{
    \includegraphics[width=0.49\textwidth]{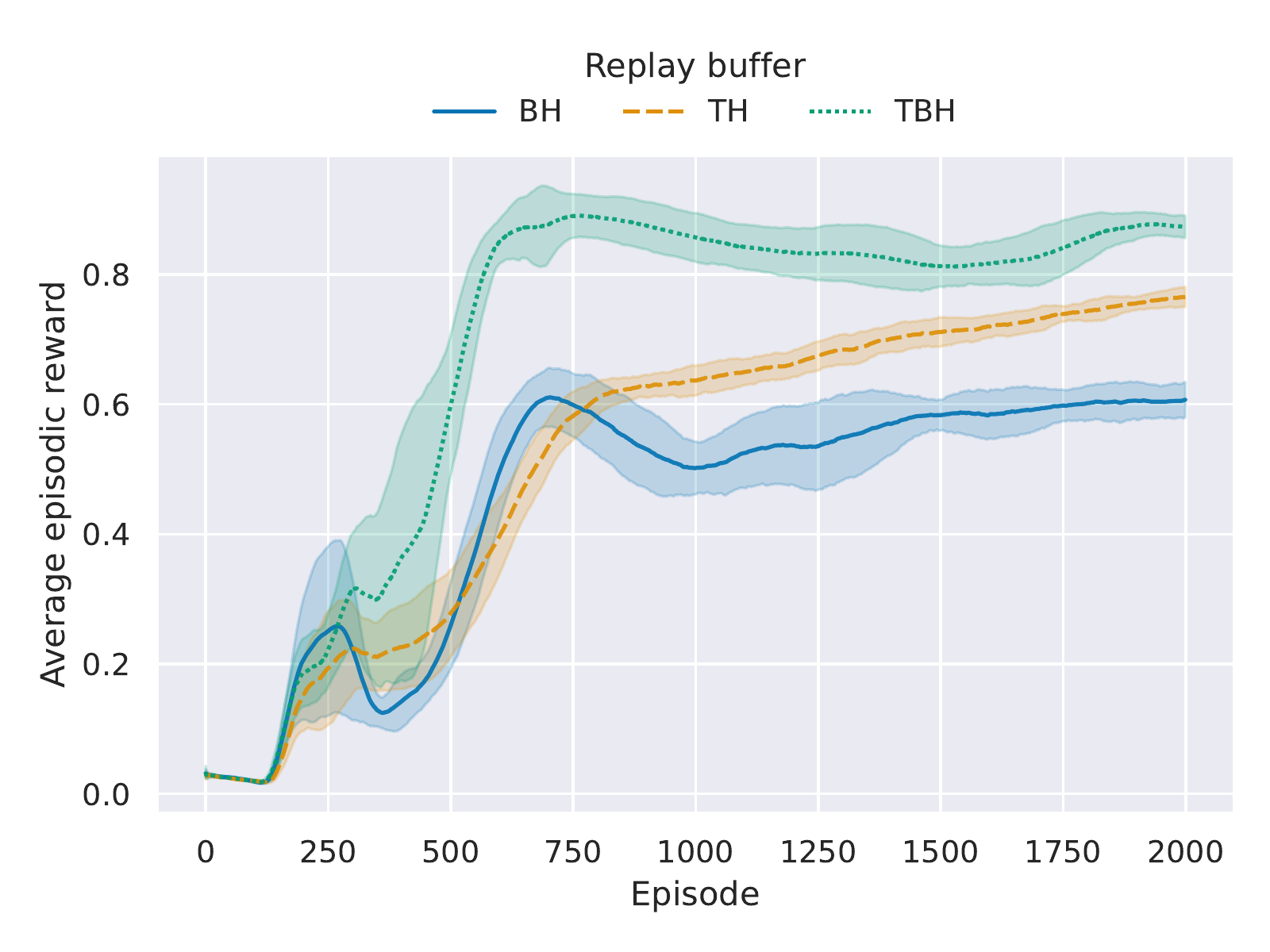}\label{fig:reward_sac_nofilter}}
    \hfill
    \subfloat[ACER]{\includegraphics[width=0.49\textwidth]{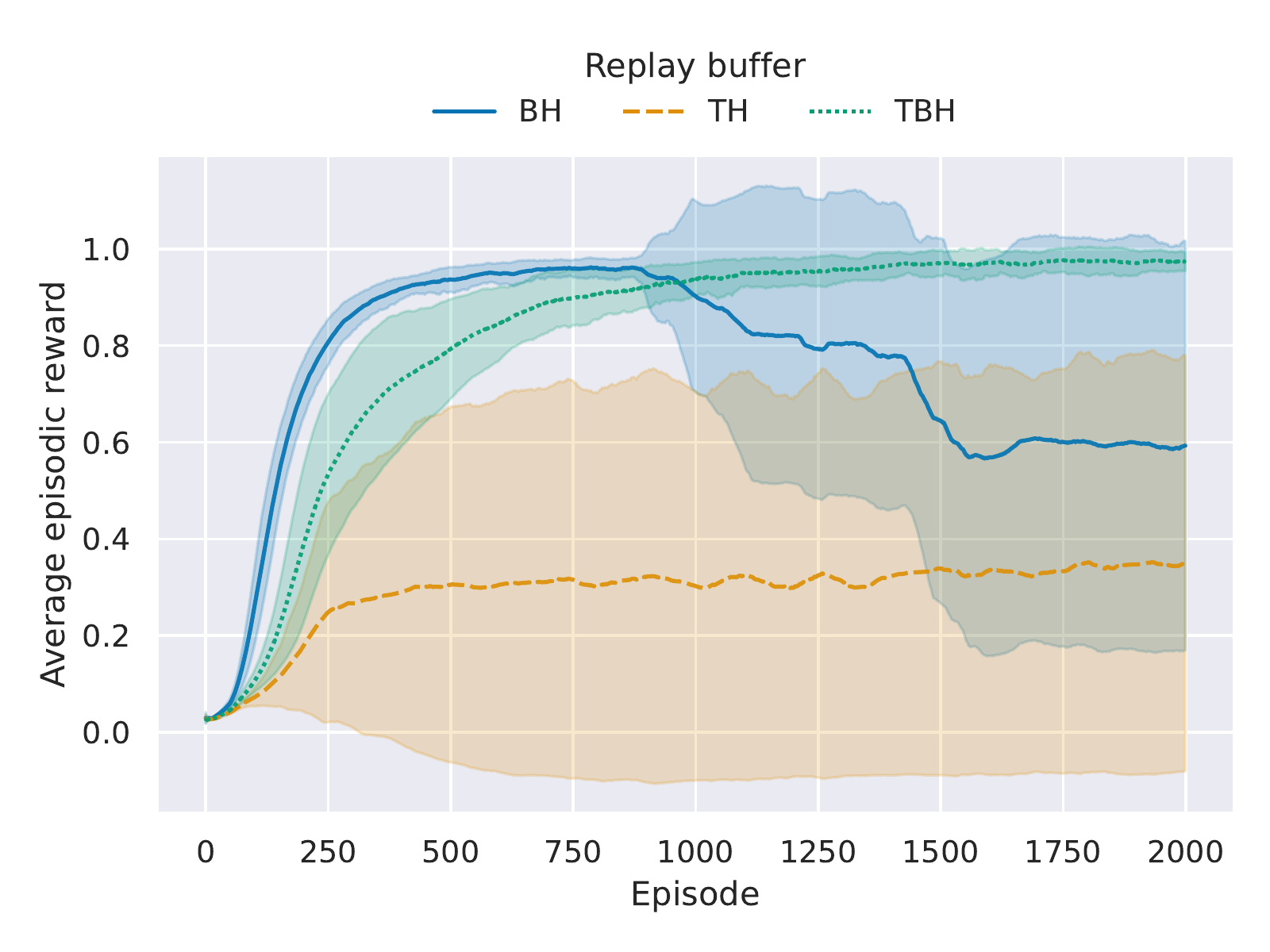}\label{fig:reward_acer_nofilter}}
    \caption{Average episodic reward computed over a batch of sequences, for the off-policy algorithms SAC and ACER (higher is better) when utilizing no diversity filter. It shows mean and standard deviation of moving average, of window size 50, over 11 runs for each policy and replay buffer. 128 SMILES strings are sampled in each episode, with a budget of 2000 episodes in total. Note that invalid SMILES are displayed with a reward of $0$ in this figure, but are given a reward of $-1$ during training.}
    \label{fig:reward_nofilter_offpolicy}
\end{figure}

\figref{fig:reward_nofilter_offpolicy} displays means and standard deviations, over 11 repeated runs, of the average episodic rewards of sampled molecules for all combinations of off-policy algorithms and replay buffers using no diversity filter. For SAC, using \textit{Top-Bottom history} gives an average episodic reward of over 0.8; while \textit{Bin history} converges to an average episodic reward of around 0.6 and \textit{Top history} keeps improving, almost reaching an average episodic reward of 0.8. For ACER, \textit{Top-Bottom history} reaches an average episodic reward close to 1. When using \textit{Bin history}, a significantly faster improvement in the first 500 episodes, compared to \textit{Top-Bottom history}, is observed but then diverges. A possible explanation for this is that it finds a mode with higher entropy, but that gives a lower score, it accidentally forgets too much of the pre-trained model. For \textit{Top history}, ACER shows a large variability and does not consistently reach an average episodic reward above 0.4. 

\section{Discussion}\label{sec:discussion}
In this section, we discuss the \textit{de novo} drug design of investigated on- and off-policy algorithms and replay buffers. 
Both the use of a diversity filter and the use of no diversity filter are discussed. We consider a diversity filter that penalizes molecules with identical molecular scaffolds. As a result, there is a natural increase in diversity regarding molecular scaffolds, but not necessarily topological scaffolds since different molecular scaffolds can have the same topological scaffold. When using no diversity filter, a large number of molecular scaffolds seems to lead to a large number of topological scaffolds. However, this is not as evident when using the diversity filter. A reason for this can be that there is a substantially larger number of molecular scaffolds generated when using a diversity filter which naturally leads to more topological scaffolds being the same.

Molecules are generated using a model pre-trained on a data set derived from the ChEMBL database which will teach the agent to generate ChEMBL-like molecules. The DRD2 data set, which is used to create the scoring function, comprises data from both the ChEMBL and PubChem databases \cite{wang2017pubchem}. Hence, this experimental setup only considers small molecules and the agent is not anticipated to have the ability to produce molecules that differ significantly from those present in the ChEMBL database. Previous work suggests that there are close to \num{500000} molecular scaffolds in total in the ChEMBL database \cite{liang2019bioactivity}. On the other hand, all scaffolds do not necessarily contain molecules active toward DRD2. The budget of 2000 episodes used in this work also limits the search for new scaffolds. Typically, we are able to identify scaffolds consisting of SMILES with a score of 1 and hence successfully traverse the reward landscape.

\subsection{On-policy Algorithms}
This section discusses the results of the on-policy algorithms and replay buffers; both with and without a diversity filter. 

\subsubsection{With Diversity Filter}
Both A2C and PPO generate more active molecules but fewer active scaffolds, compared to Regularized MLE. One possible explanation for this could be that A2C and PPO stay for a longer time close to a penalized scaffold before moving to the next scaffold. Regularized MLE uses a fixed pre-trained actor in the loss and can, therefore, easily jump between scaffolds without forgetting how to generate valid SMILES strings. This comes at the cost of its likelihood staying close to that of the pre-trained model. However, we do not observe that this notably limits its performance. We observe that on-policy algorithms can enhance diversity by not only relying on high-scoring molecules, particularly regarding the diversity of topological scaffolds.

A2C and PPO achieve high average episodic rewards by either using all on-policy data or a subset comprising both high- and low-scoring molecules. For some combinations of algorithm and replay mechanism is it advantageous to use historical samples instead of only current samples. 
However, there seems to be no significant added performance gain in using experiences from previous iterations, particularly compared to using all current samples. One possible explanation for this could be that the diversity filter can impose a significant change in the reward landscape after each episode by initiating penalization of frequently generated scaffolds and therefore possibly making previously acquired knowledge outdated, which may need to be accounted for. The investigated diversity filter uses a discrete threshold to determine when molecules of a scaffold should obtain a zero reward. A diversity filter that in a stepwise or continuous manner penalizes molecules from the same scaffold could have other effects on the reward landscape which possibly could be more suitable for some replay buffers.

\subsubsection{Without Diversity Filter}
As expected, the number of actives molecules and scaffolds is significantly lower compared to when using a diversity filter. There is an evident relationship between the number of active molecules, molecular scaffolds, and topological scaffolds.  This is not necessarily evident when utilizing diversity filter. 

When using no diversity filter, Regularized MLE generates a substantially larger number of active molecules and scaffolds, compared to A2C and PPO, especially when used in combination with the replay buffers \textit{Bin current}, \textit{Bin history} and \textit{All current}. There is no significant difference in using a replay mechanism utilizing historical or current samples, except for the bin-based replay mechanisms in combination with Regularized MLE. \textit{Bin current} in combination with Regularized MLE displays the overall largest number of active molecules and scaffolds while displaying among the lowest average episodic reward. This replay buffer seems to be able to enhance the exploration of Regularized MLE, possibly because it does not only exploit high-scoring molecules. Hence, it is evidently important to use a diverse set, in terms of reward, for learning without a diversity filter. 

Why Regularized MLE performs better, in general, is likely because it is heavily regularized to stay close to the pre-trained model in terms of likelihood and, therefore, inherently can jump between scaffolds without using the initial knowledge. Without a diversity filter, PPO generates the least number of active molecules and scaffolds, while still displaying a high average episodic reward. This must be because of an early mode collapse, leading to PPO generating more or less only the same molecules through all episodes.

\subsection{Off-policy Algorithms}
In this section, the results of the off-policy algorithms and replay buffers are discussed; both with and without a diversity filter. 

\subsubsection{With Diversity Filter}
For the best off-policy algorithm and replay buffer combination, i.e., ACER with \textit{Bin history}, the number of active molecules is on par with the best on-policy combinations. On the other hand, several off-policy results struggle to reach an episodic reward significantly above 0.4, which is the minimum reward for a generated SMILES string to be stored in the diversity filter. Only ACER with \textit{Bin history} can consistently reach an average episodic reward above 0.7. With the help of the diversity filter, this leads to the highest number of active molecules and scaffolds. Hence, the slower convergence rate can possibly be explained by a more elaborate exploration phase. However, ACER shows a large variability. Even though SAC does generally achieve lower average episodic rewards, it is more stable and shows a larger median of active molecules and scaffolds compared to ACER utilizing \textit{Top history} and \textit{Top-bottom history}. It appears that the average episodic reward of SAC with \textit{Bin history} has not converged yet, indicating that it is still in the learning phase. It is possible that this specific combination requires over 2000 episodes to conduct adequate exploration.

\subsubsection{Without Diversity Filter}
For the off-policy algorithms with no diversity filter, it is observed that SAC using \textit{Top history} generates the largest number of unique active molecules and scaffolds. However, it does not reach the highest average episodic reward in this setting, which is achieved by ACER. ACER with \textit{Top-bottom} history converges to the highest average episodic reward, among the off-policy algorithms without diversity filter, but is still outperformed by SAC for the generation of active molecules and scaffolds. This is because it generates many duplicates of the same high-scoring molecules. Hence, SAC will generate more unique molecules when no diversity filter is used and, consequently, improve the exploration. This behavior yields an enhancement in the diversity of the active molecules, especially when the top-scoring molecules are used for off-policy updates. Overall, this highlights the positive impact an appropriate replay buffer can have on the generation of diverse molecules. It also emphasizes the positive effect that the usage of a diversity filter has and that inherent exploration of the algorithms is not necessarily enough to generate a structurally diverse set of active molecules.

\section{Conclusions}\label{sec:conclusions}
We explore on- and off-policy RL algorithms for SMILES-based molecular \textit{de novo} drug design using recurrent neural networks. The investigation has focused on how well the algorithms sample structurally diverse and high-rewarding molecules for different replay buffers. This has been done by studying their behaviors both with and without using a diversity filter that penalizes the generation of similar molecules between episodes. 

For on-policy algorithms, we observe that it is often favorable to use all generated molecules from the current batch for learning. Regularized MLE utilizing the full batch for learning, in combination with a diversity filter, leads to the overall best performance in terms of both the reward and diversity. However, it is possible to obtain similar performance by learning from fewer samples if the training data includes at least both high-rewarding and low-rewarding data points. For these on-policy algorithms applied with no diversity filter, it is also important to use intermediate-rewarding samples, either from the current batch of sampled molecules or previously sampled molecules.

There is a potential performance gain in using off-policy algorithms with a suitable replay buffer. When using no diversity filter, we observe that SAC provides good exploration, leading to a more structurally diverse generation. Hence, when no diversity filter is used, the policy must keep its randomness to avoid mode collapse. Interestingly, when using a diversity filter ACER yields better performance, displaying the potential to be on par with Regularized MLE or even better. 

We release the source code of the methods as an open-source framework, to enable further exploration of reinforcement learning algorithms and replay buffers mechanisms. 

\section*{Acknowledgments}
This work was partially supported by the Wallenberg Artificial Intelligence, Autonomous Systems, and Software Program (WASP), funded by the Knut and Alice Wallenberg Foundation, Sweden.

\begin{appendices}

\section{Hyperparameters}\label{app:hyper}

\subsection{Regularized Maximum Likelihood Estimation}

\begin{table}[h]
\begin{center}
\caption{Regularized MLE Hyperparameters}\label{table:reg_mle}%
\begin{tabular}{ c | c  }
\toprule
Parameter & Value \\
\midrule
Optimizer & Adam \cite{kingma2014adam}\\
 $\sigma$ & 128 \\
 margin threshold & 50 \\
 learning rate & \num{0.0001} \\
 gradient steps per update & 1\\
 \botrule
\end{tabular}
\end{center}
\end{table}

\subsection{Proximal Policy Opimization}
\begin{table}[h]
\begin{center}
\caption{PPO Hyperparameters}\label{table:ppo}%
\begin{tabular}{ c | c  }
\toprule
Parameter & Value \\
\midrule
Optimizer & Adam \cite{kingma2014adam}\\
 clipping range $(\epsilon)$ & 0.2 \\
 number of updates per episode & 4\\
 discount $(\gamma)$ & 0.99\\
 number of mini-batches & 4 \\
 norm for gradient clipping & L2 norm\\
 maximum gradient norm & 0.5\\
 learning rate & \num{0.0001} \\
gradient steps per update & 1\\
 \botrule
\end{tabular}
\end{center}
\end{table}

\subsection{Advantage Actor-Critic}

\begin{table}[h]
\begin{center}
\caption{A2C Hyperparameters}\label{table:a2c}%
\begin{tabular}{ c | c  }
\toprule
Parameter & Value \\
\midrule
Optimizer & Adam \cite{kingma2014adam}\\
 discount $(\gamma)$ & 0.99\\
 norm for gradient clipping & L2 norm\\
 maximum gradient norm & 0.5\\
 learning rate actor & \num{0.0001} \\
learning rate critic & \num{0.0001}\\
gradient steps per update & 1\\
 \botrule
\end{tabular}
\end{center}
\end{table}

\subsection{Actor-Critic with Experience Replay}

\begin{table}[h]
\begin{center}
\caption{ACER Hyperparameters}\label{table:acer}%
\begin{tabular}{ c | c  }
\toprule
Parameter & Value \\
\midrule
Optimizer & Adam \cite{kingma2014adam}\\
replay rate $(\lambda)$ & 4\\
 discount $(\gamma)$ & 0.99\\
 target smoothing coefficient $(\tau)$ & 0.95\\
 entropy weight & 0.001 \\ 
 trust factor variance reduction $(\delta)$ & 1\\
 Trust region clipping range $(c)$ & 10 \\
 norm for gradient clipping & L2 norm\\
 maximum gradient norm & 0.5\\
 learning rate & \num{0.0001} \\
 number of off-policy replay samples $(k)$ & 128\\
gradient steps per update & 1\\
number of initial episodes without update & 10\\
 \botrule
\end{tabular}
\end{center}
\end{table}

\subsection{Soft Actor-Critic}
\begin{table}[h]
\begin{center}
\caption{SAC Hyperparameters}\label{table:sac}%
\begin{tabular}{ c | c  }
\toprule
Parameter & Value \\
\midrule
Optimizer & Adam \cite{kingma2014adam}\\
number of off-policy updates & 4\\
discount $(\gamma)$ & 1\\
smoothing coefficient $(\tau)$ & 0.99\\
initial temperature $(\alpha_0)$ & 0.001\\
norm for gradient clipping & L2 norm\\
maximum gradient norm & 0.5\\
entropy target $(\Bar{H})$ & 0.3 \\
learning rate actor $(\lambda_\pi)$ & \num{0.0001} \\
learning rate critic $(\lambda_Q)$ & \num{0.0001}\\
learning rate temperature $(\lambda_\alpha)$ & \num{0.0001}\\
average policy weight & 0.5 \\
number of off-policy replay samples $(k)$ & 64\\
gradient steps per update & 1\\
number of initial episodes without update $(K_{\mathrm{init}})$ & 10\\
clipping range $(c)$ & 0.5\\
 \botrule
\end{tabular}
\end{center}
\end{table}

\section{Technical Details}\label{app:tech}
For all models, training was done using Python 3.8.15 and PyTorch 1.13.1. Computations in this work were performed on a Linux cluster using Nvidia Tesla K80 and Nvidia V100 graphic cards utilizing CUDA 11.4. The experiments were run in parallel on around 50 graphic cards, where each run of an experiment was restricted to one graphic card at runtime. 

For the scoring function and diversity filter, reinvent-scoring 0.0.73 is used with reinvent-chemistry 0.0.51. To train the DRD2 predictive model, which is used as scoring function, scikit-learn 1.2.0 is used. To compute fingerprints and scaffolds, RDKit 2022.9.3 is used. 

\section{Visual Comparison of Generated SMILES}\label{app:smiles}
Top 10 generated SMILES of unique topological scaffolds of one representative run of each combination. 

\subsection{On-policy Algorithms}
\subsubsection{With Diversity Filter}
\begin{figure}[h]
     \centering
     \subfloat[All current (AC)]{
    \includegraphics[width=0.48\textwidth]{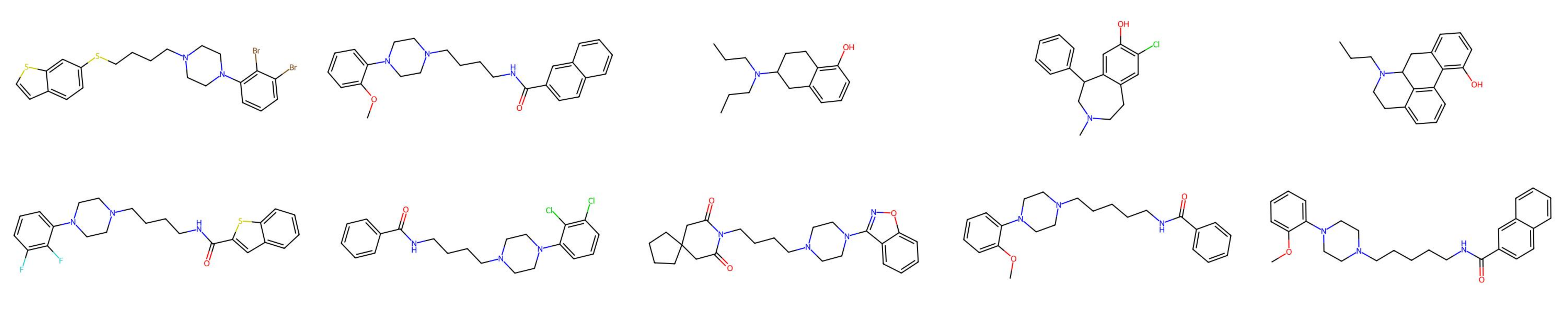}\label{fig:a2c_ac}}
     \hfill
    \subfloat[Bin current (BC)]{\includegraphics[width=0.48\textwidth]{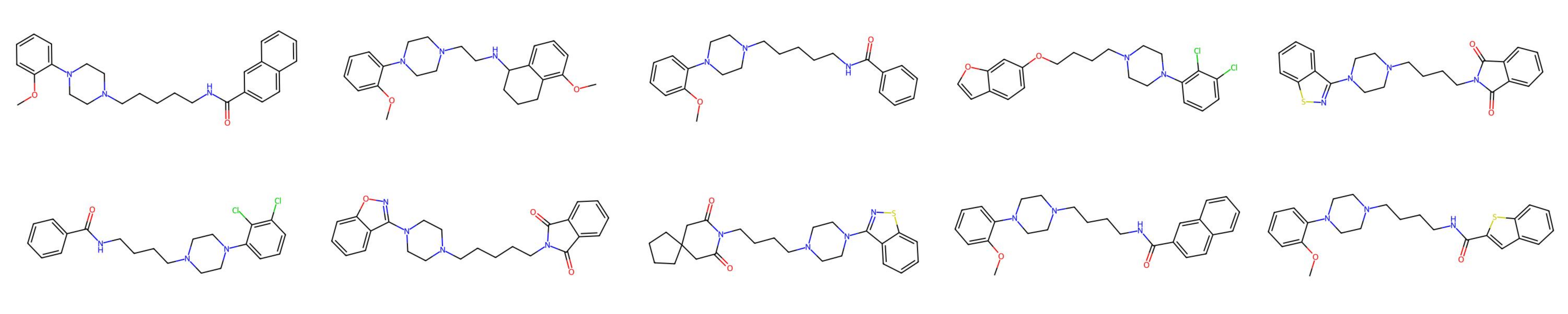}\label{fig:a2c_bc}}
     \hfill
     \subfloat[Bin history (BH)]{\includegraphics[width=0.48\textwidth]{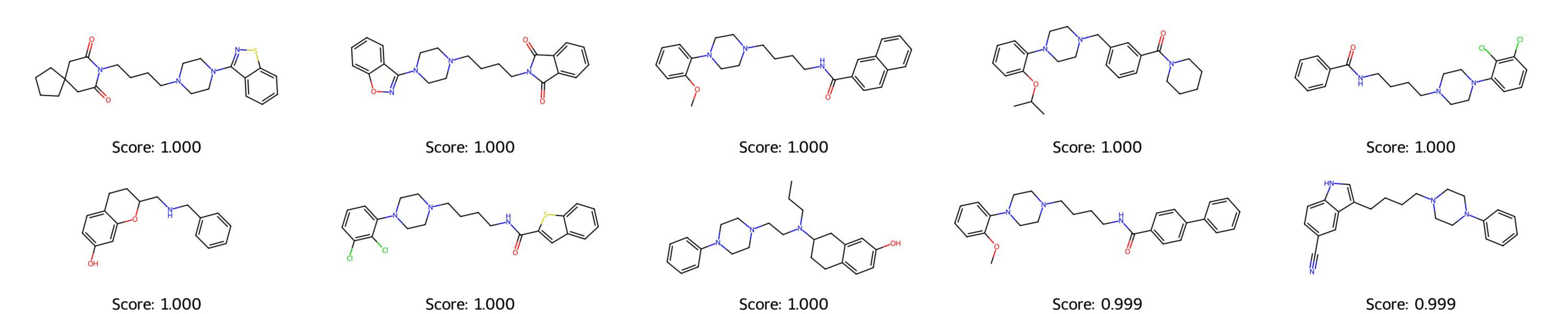}\label{fig:a2c_bh} }
     \hfill
     \subfloat[Top-bottom current (TBC)]{\includegraphics[width=0.48\textwidth]{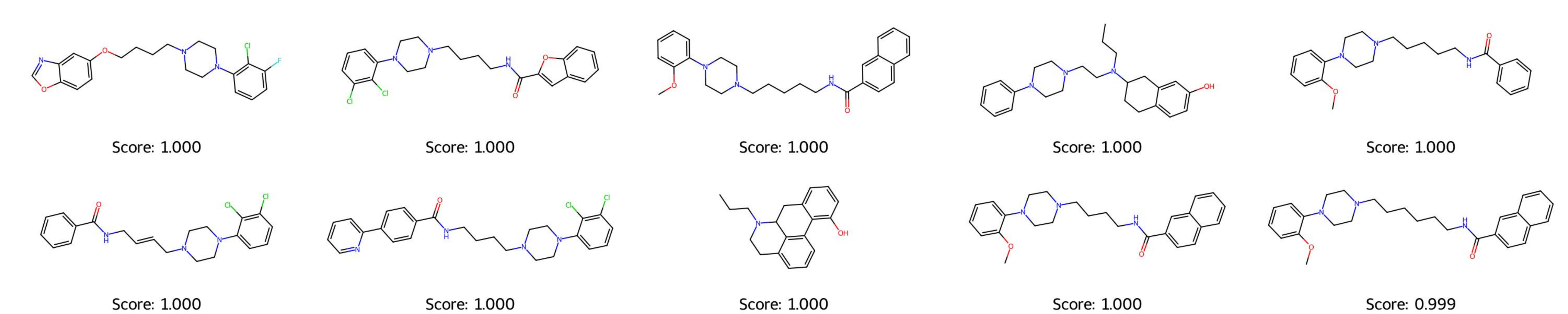}\label{fig:a2c_tbc} }
     \hfill
     \subfloat[Top-bottom history (TBH)]{\includegraphics[width=0.48\textwidth]{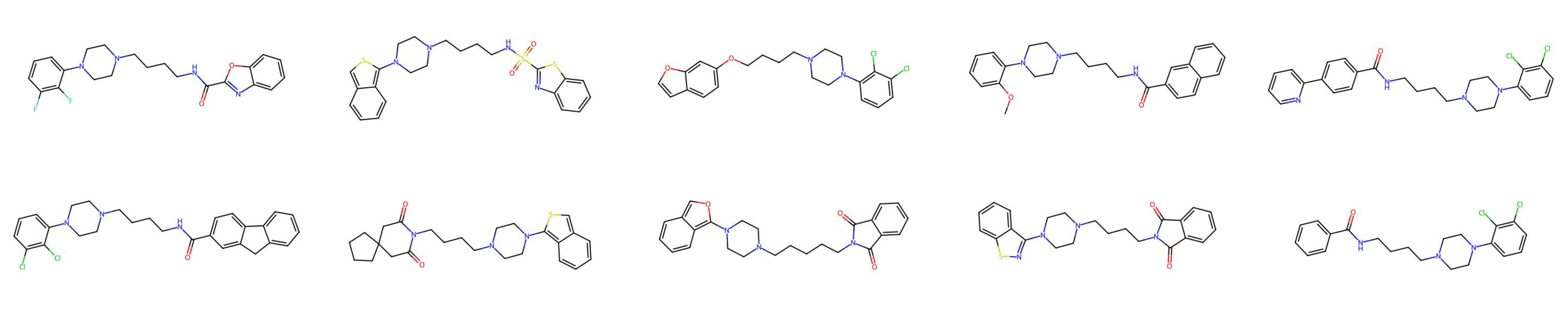}\label{fig:a2c_tbh} }
     \hfill
     \subfloat[Top current (TC)]{\includegraphics[width=0.48\textwidth]{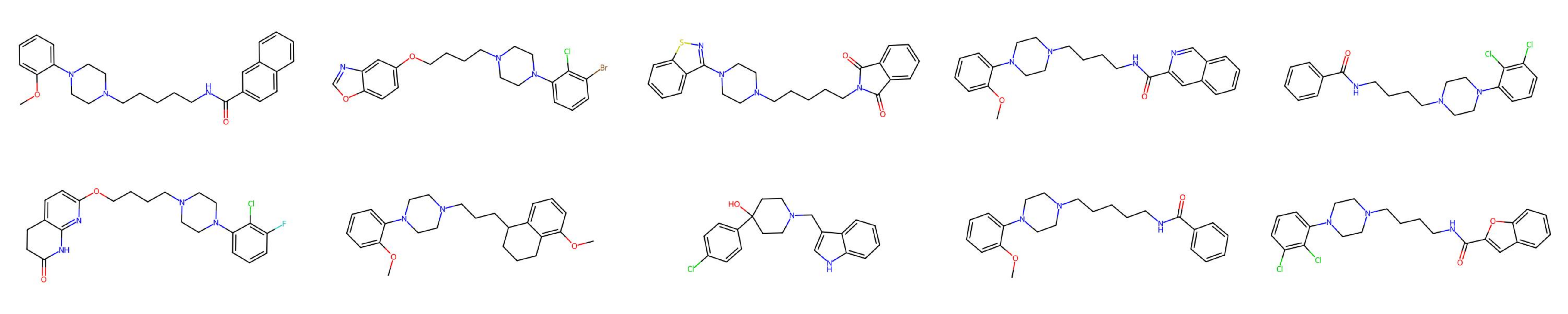}\label{fig:a2c_tc} }
     \hfill
     \subfloat[Top history (TH)]{\includegraphics[width=0.48\textwidth]{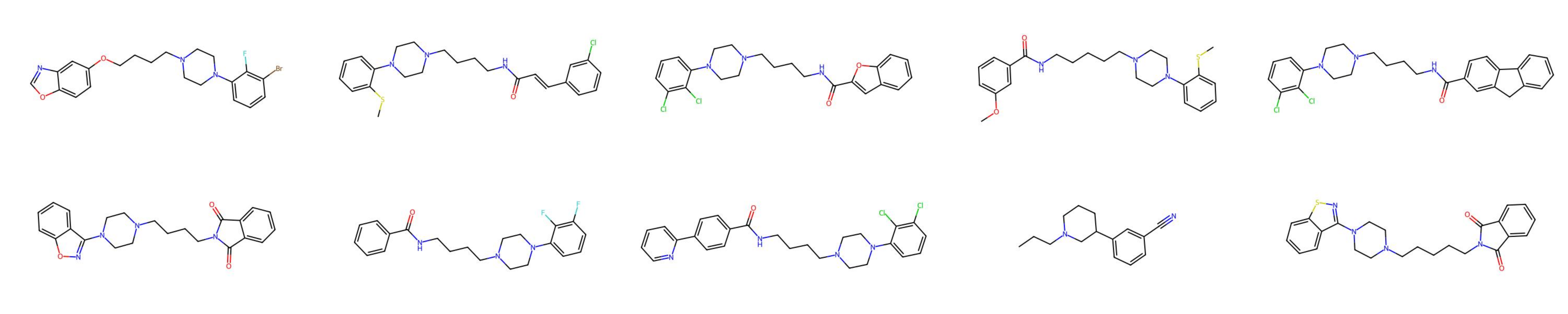}\label{fig:a2c_th} }
    \caption{A2C with diversity filter penalizing the generation of SMILES with the same molecular scaffold. If no score is displayed, all scores are at least 0.9995. Otherwise, scores are rounded upwards to 3 decimals.}
    \label{fig:smiles_a2c}
\end{figure}

\begin{figure}[h]
     \centering
     \subfloat[All current (AC)]{
    \includegraphics[width=0.48\textwidth]{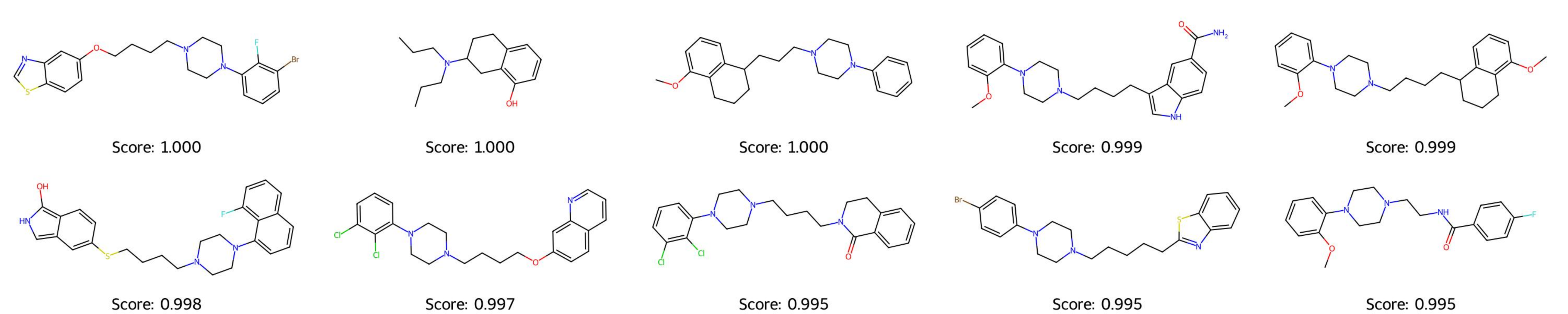}\label{fig:ppo_ac}}
     \hfill
    \subfloat[Bin current (BC)]{\includegraphics[width=0.48\textwidth]{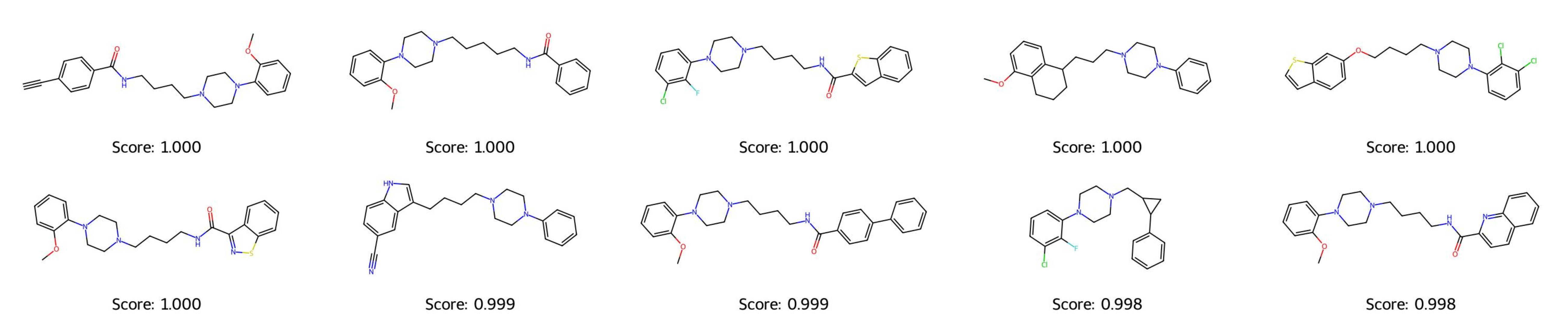}\label{fig:ppo_bc}}
     \hfill
     \subfloat[Bin history (BH)]{\includegraphics[width=0.48\textwidth]{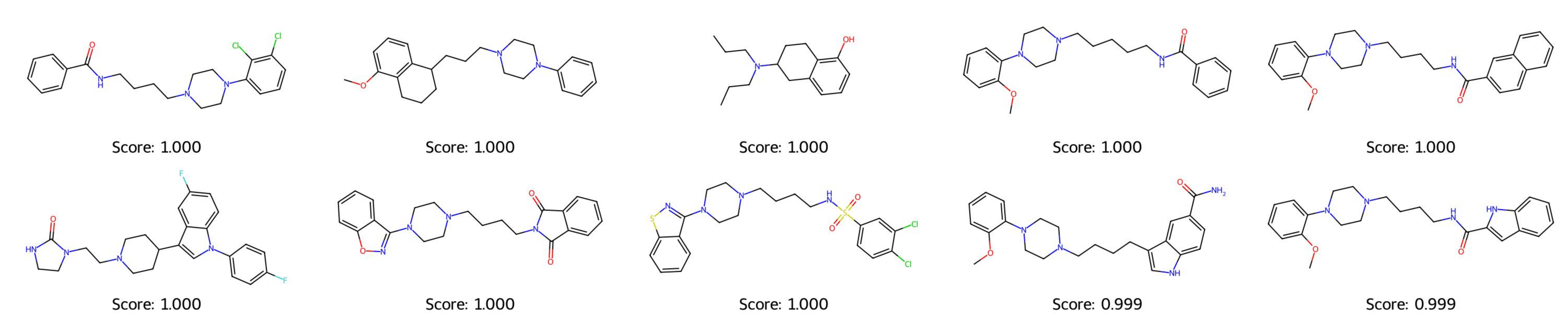}\label{fig:ppo_bh} }
     \hfill
     \subfloat[Top-bottom current (TBC)]{\includegraphics[width=0.48\textwidth]{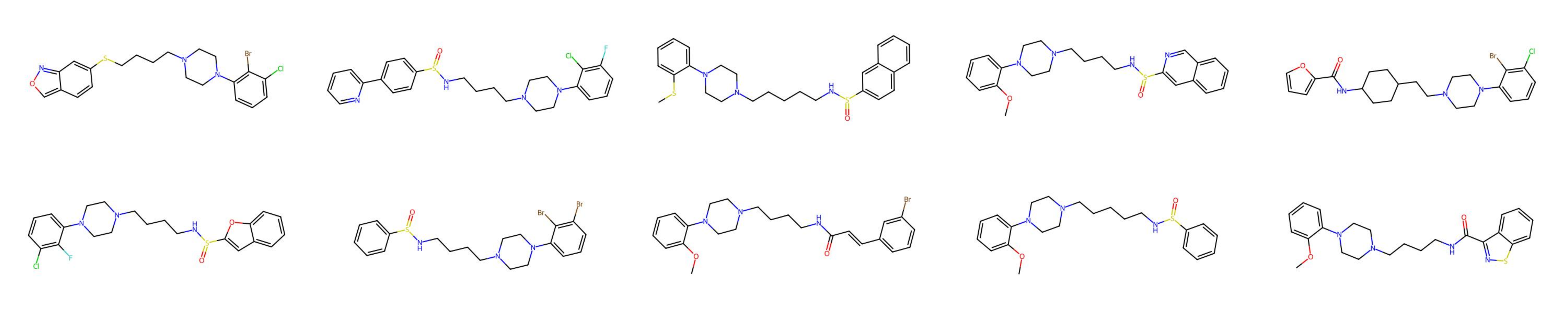}\label{fig:ppo_tbc} }
     \hfill
     \subfloat[Top-bottom history (TBH)]{\includegraphics[width=0.48\textwidth]{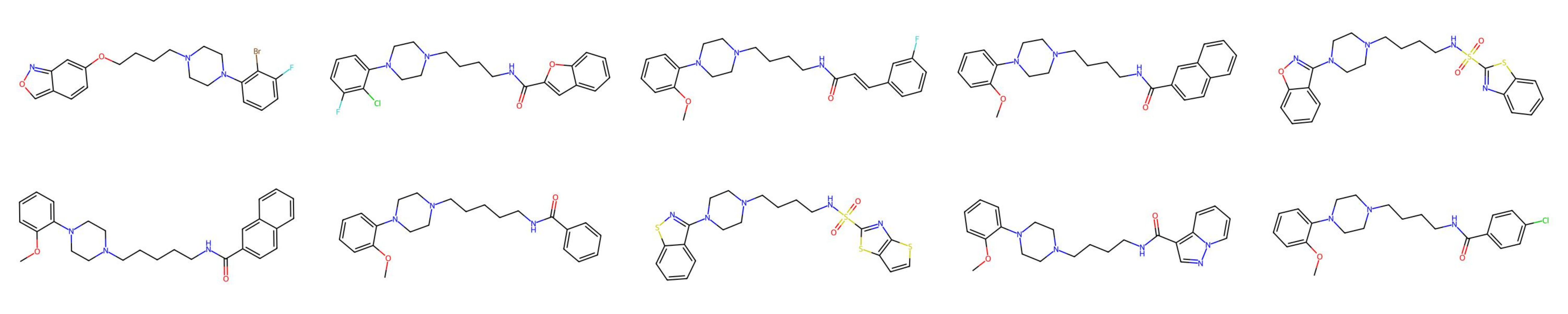}\label{fig:ppo_tbh} }
     \hfill
     \subfloat[Top current (TC)]{\includegraphics[width=0.48\textwidth]{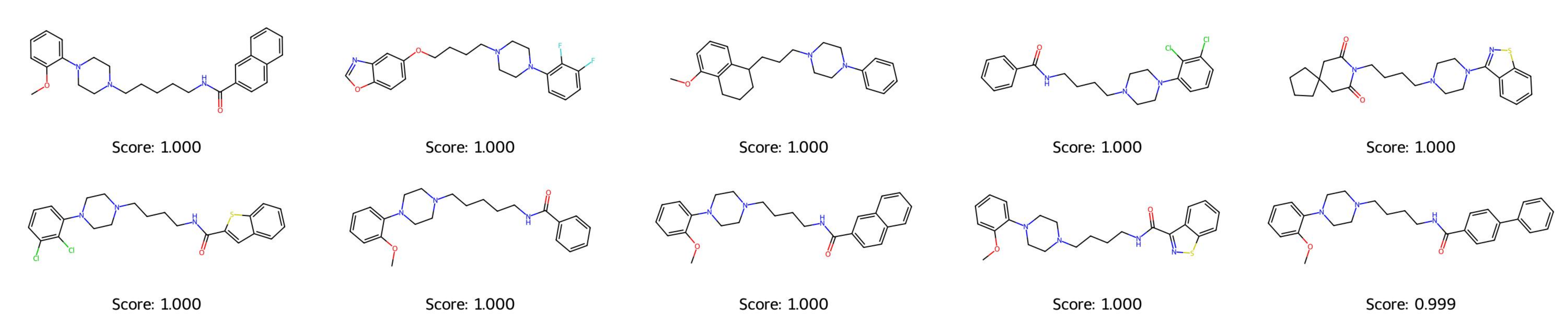}\label{fig:ppo_tc} }
     \hfill
     \subfloat[Top history (TH)]{\includegraphics[width=0.48\textwidth]{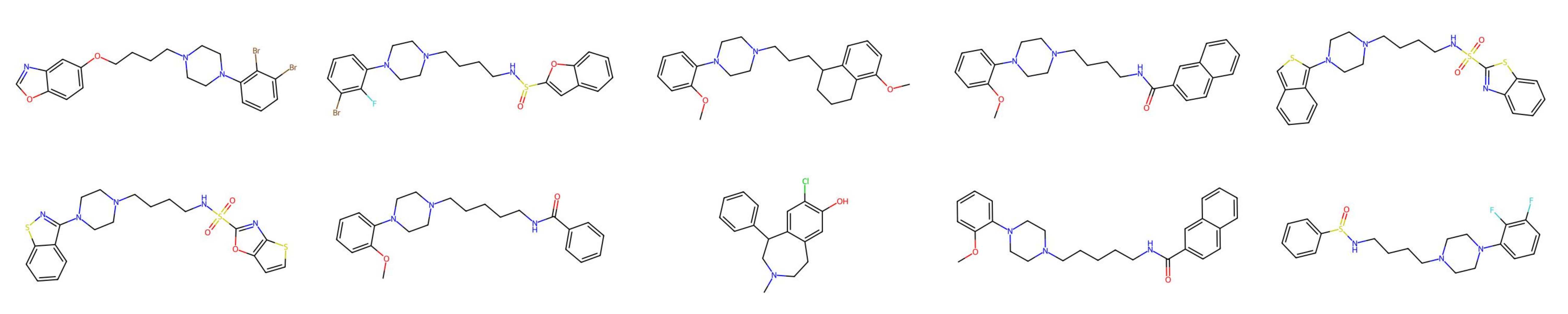}\label{fig:ppo_th} }
    \caption{PPO with diversity filter penalizing the generation of SMILES with the same molecular scaffold. If no score is displayed, all scores are at least 0.9995. Otherwise, scores are rounded upwards to 3 decimals.}
    \label{fig:smiles_ppo}
\end{figure}

\begin{figure}[h]
     \centering
     \subfloat[All current (AC)]{
    \includegraphics[width=0.48\textwidth]{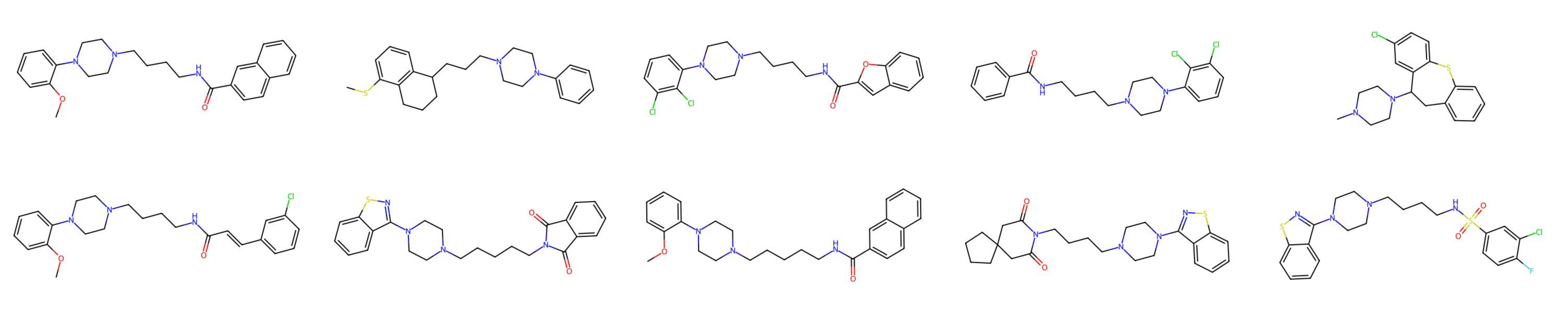}\label{fig:reinvent_ac}}
     \hfill
    \subfloat[Bin current (BC)]{\includegraphics[width=0.48\textwidth]{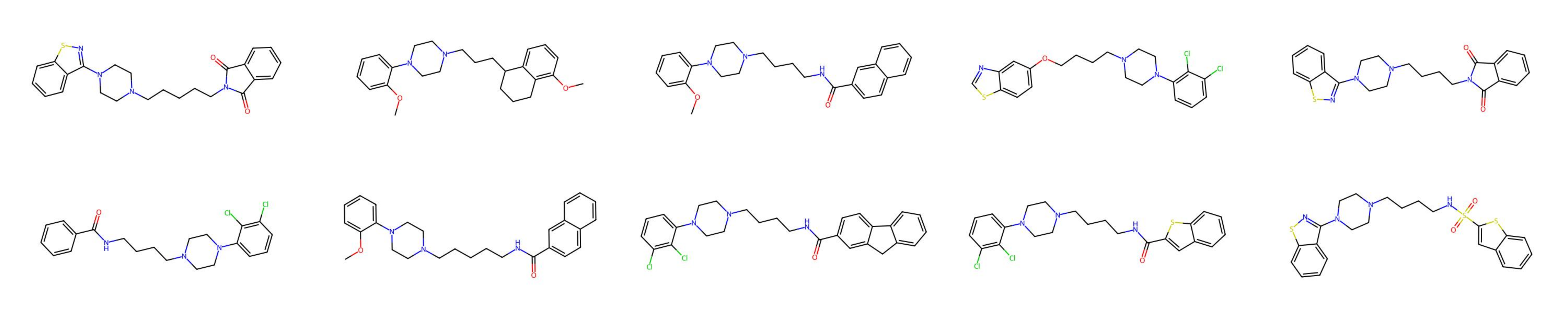}\label{fig:reinvent_bc}}
     \hfill
     \subfloat[Bin history (BH)]{\includegraphics[width=0.48\textwidth]{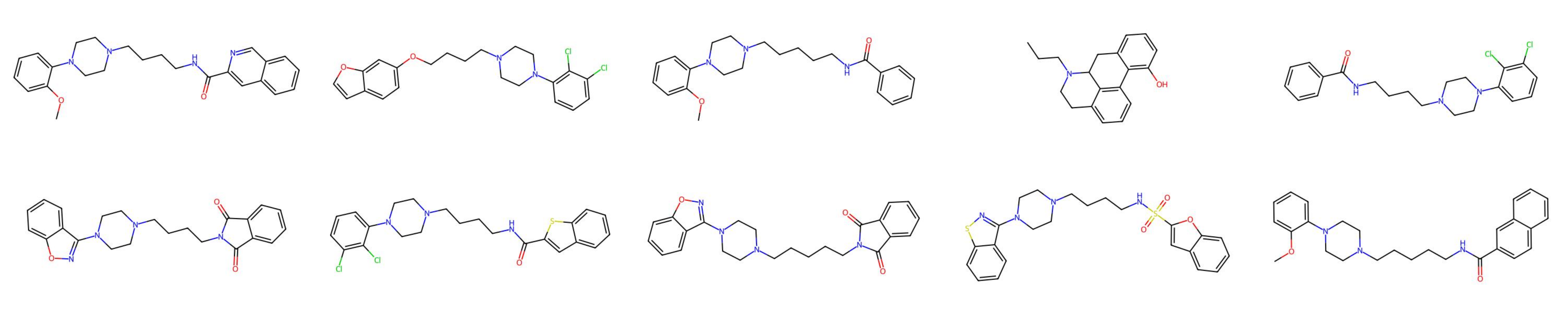}\label{fig:reinvent_bh} }
     \hfill
     \subfloat[Top-bottom current (TBC)]{\includegraphics[width=0.48\textwidth]{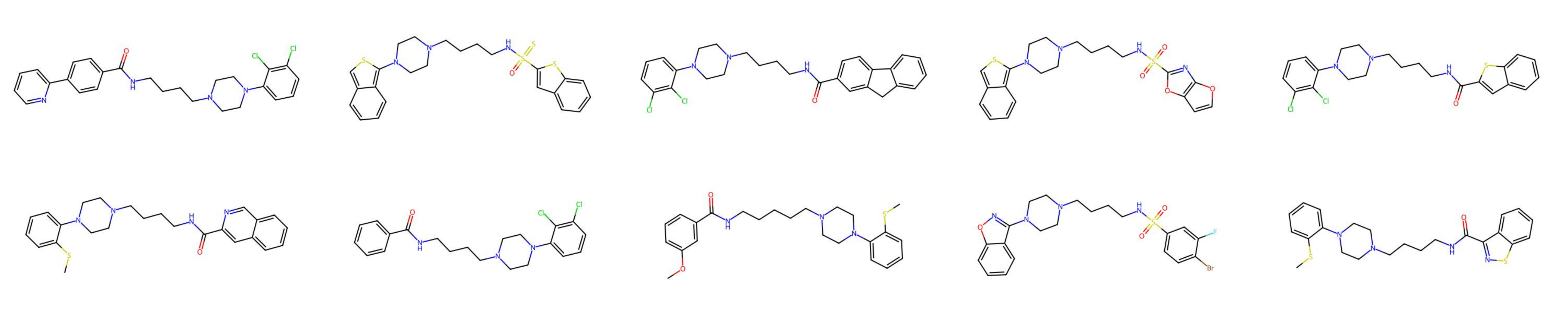}\label{fig:reinvent_tbc} }
     \hfill
     \subfloat[Top-bottom history (TBH)]{\includegraphics[width=0.48\textwidth]{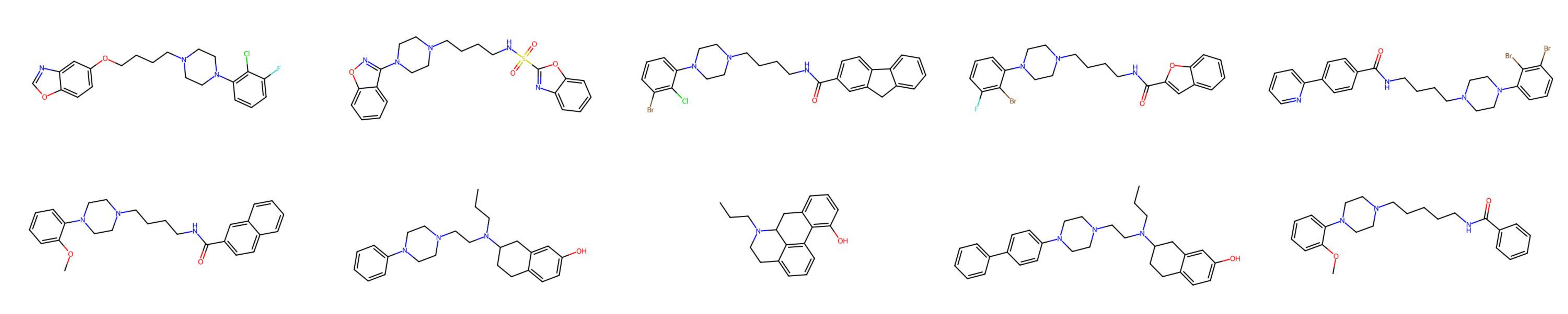}\label{fig:reinvent_tbh} }
     \hfill
     \subfloat[Top current (TC)]{\includegraphics[width=0.48\textwidth]{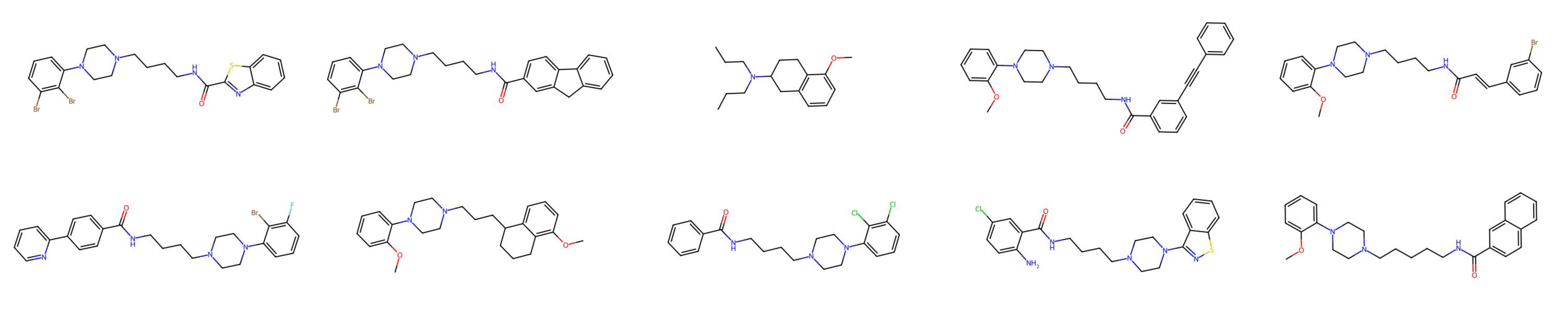}\label{fig:reinvent_tc} }
     \hfill
     \subfloat[Top history (TH)]{\includegraphics[width=0.48\textwidth]{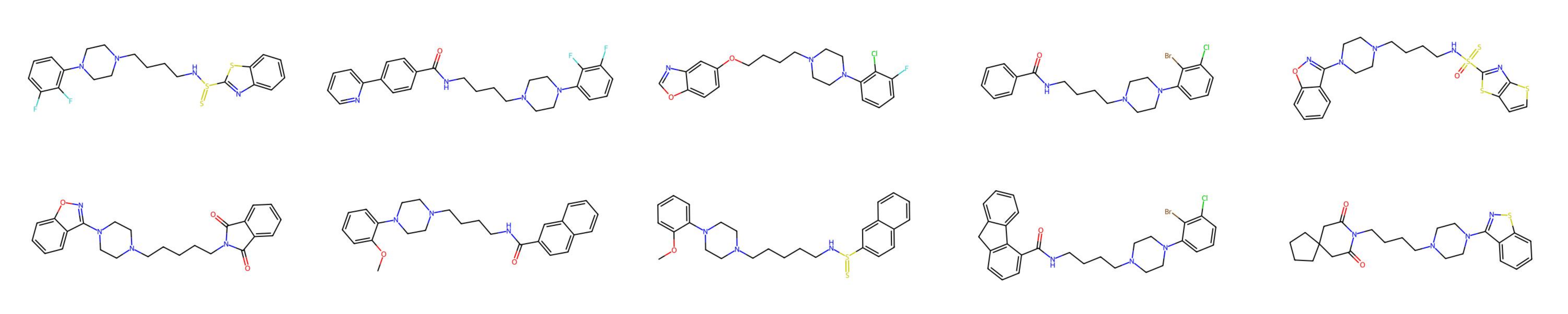}\label{fig:reinvent_th} }
    \caption{Regularized MLE with diversity filter penalizing the generation of SMILES with the same molecular scaffold. If no score is displayed, all scores are at least 0.9995. Otherwise, scores are rounded upwards to 3 decimals.}
    \label{fig:smiles_reinvent}
\end{figure}

\subsection{Without Diversity Filter}

\begin{figure}[h]
     \centering
     \subfloat[All current (AC)]{
    \includegraphics[width=0.45\textwidth]{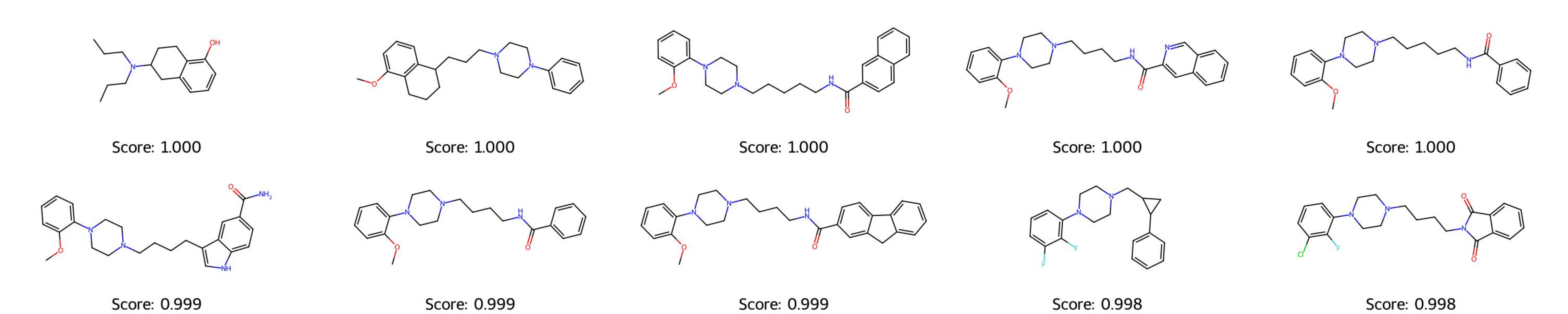}\label{fig:a2c_ac_nofilter}}
     \hfill
    \subfloat[Bin current (BC)]{\includegraphics[width=0.48\textwidth]{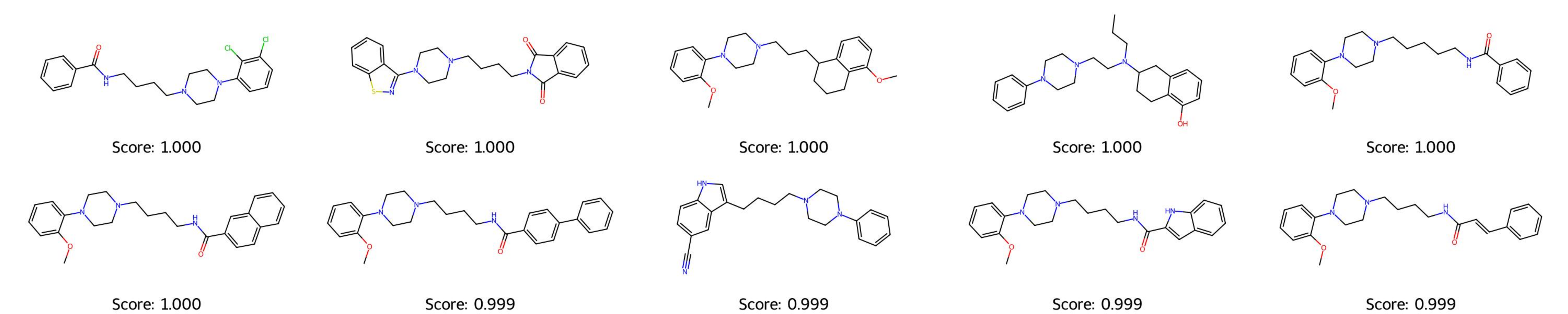}\label{fig:a2c_bc_nofilter}}
     \hfill
     \subfloat[Bin history (BH)]{\includegraphics[width=0.48\textwidth]{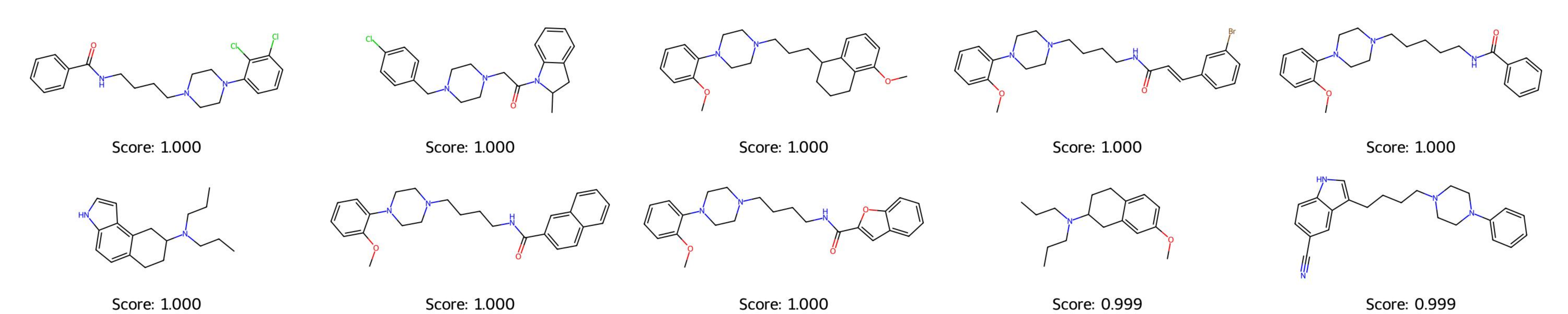}\label{fig:a2c_bh_nofilter} }
     \hfill
     \subfloat[Top-bottom current (TBC)]{\includegraphics[width=0.48\textwidth]{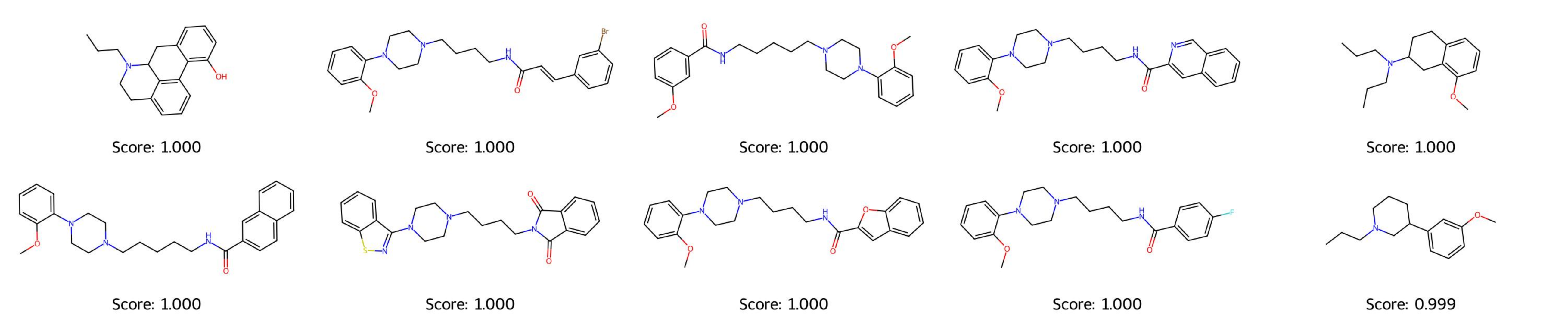}\label{fig:a2c_tbc_nofilter} }
     \hfill
     \subfloat[Top-bottom history (TBH)]{\includegraphics[width=0.48\textwidth]{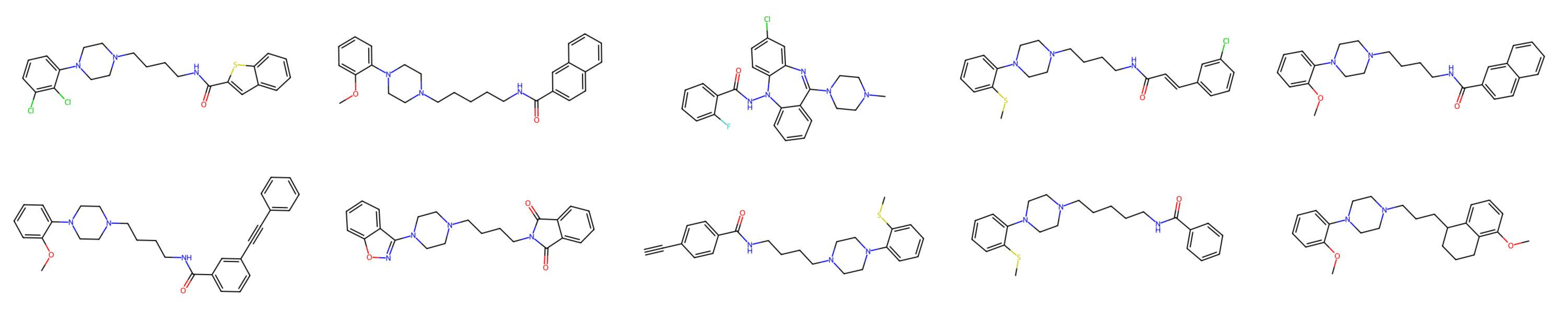}\label{fig:a2c_tbh_nofilter} }
     \hfill
     \subfloat[Top current (TC)]{\includegraphics[width=0.48\textwidth]{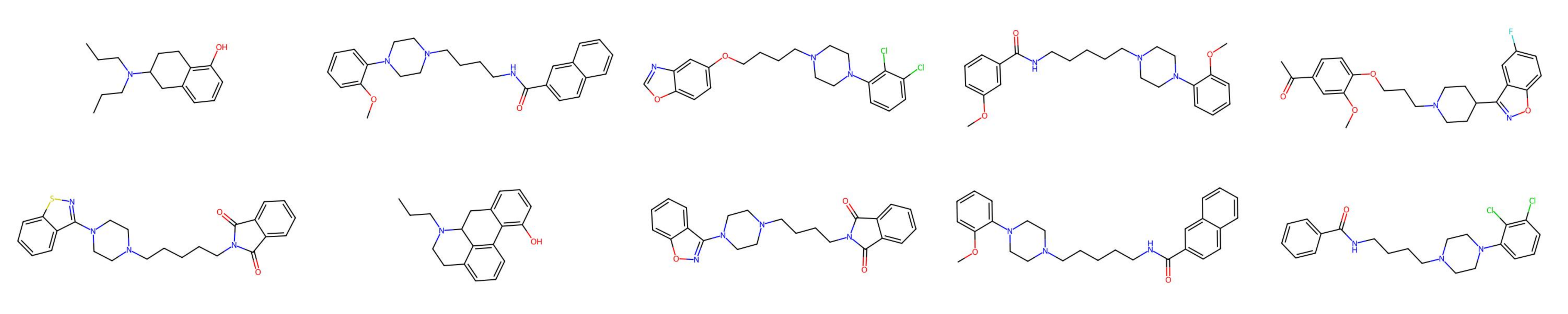}\label{fig:a2c_tc_nofilter} }
     \hfill
     \subfloat[Top history (TH)]{\includegraphics[width=0.48\textwidth]{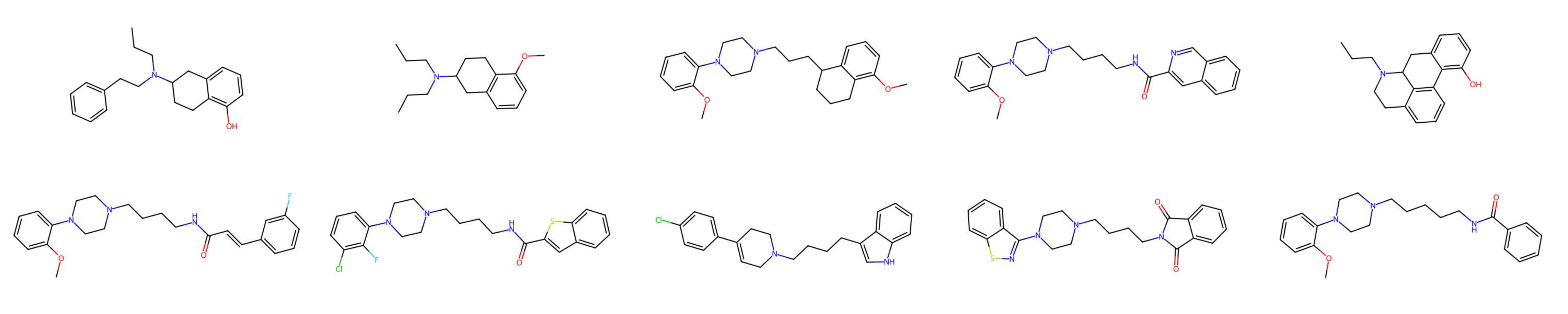}\label{fig:a2c_th_nofilter} }
    \caption{A2C without diversity filter. If no score is displayed, all scores are at least 0.9995.}
    \label{fig:smiles_a2c_nofilter}
\end{figure}

\begin{figure}[h]
     \centering
     \subfloat[All current (AC)]{
    \includegraphics[width=0.45\textwidth]{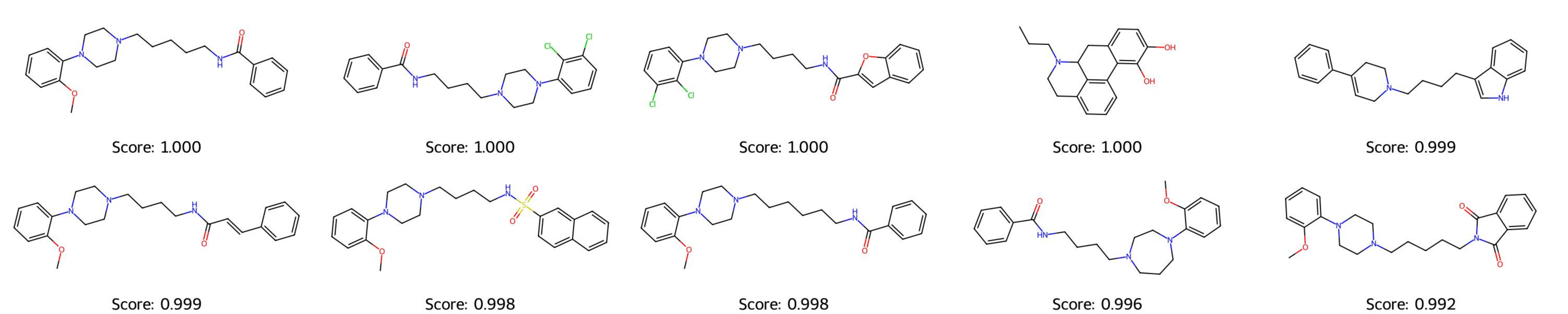}\label{fig:ppo_ac_nofilter}}
     \hfill
    \subfloat[Bin current (BC)]{\includegraphics[width=0.48\textwidth]{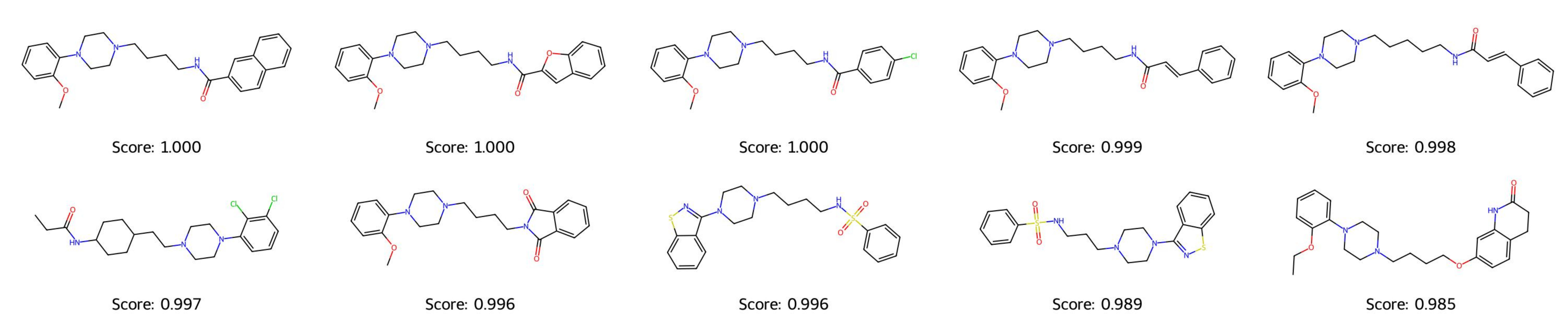}\label{fig:ppo_bc_nofilter}}
     \hfill
     \subfloat[Bin history (BH)]{\includegraphics[width=0.48\textwidth]{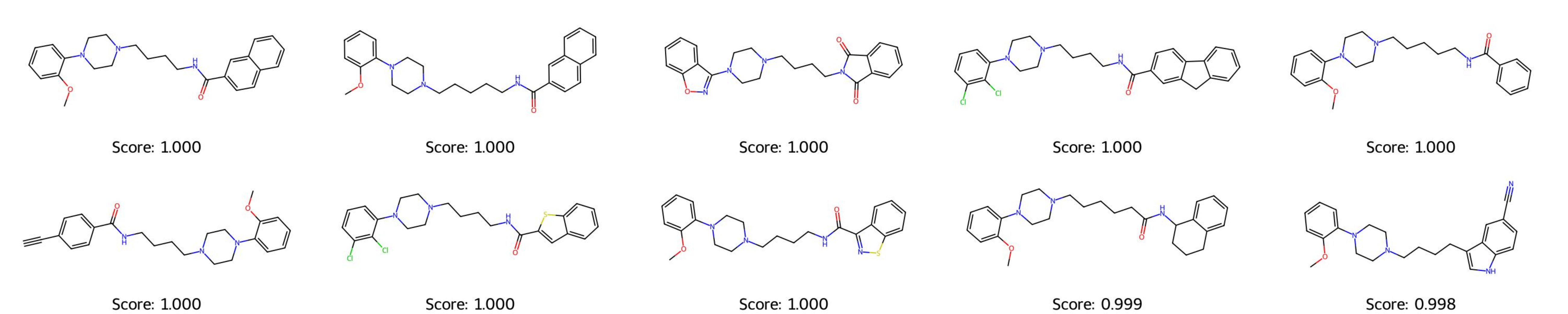}\label{fig:ppo_bh_nofilter} }
     \hfill
     \subfloat[Top-bottom current (TBC)]{\includegraphics[width=0.48\textwidth]{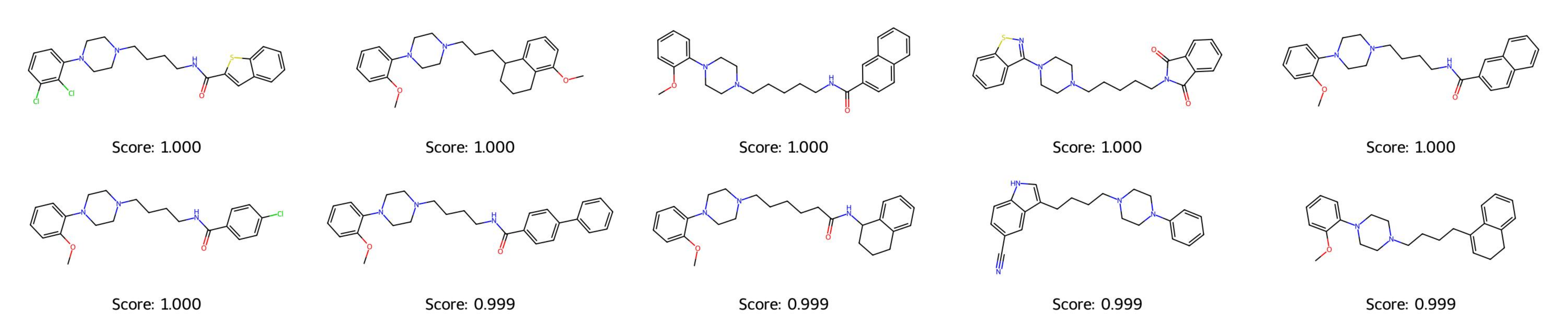}\label{fig:ppo_tbc_nofilter} }
     \hfill
     \subfloat[Top-bottom history (TBH)]{\includegraphics[width=0.48\textwidth]{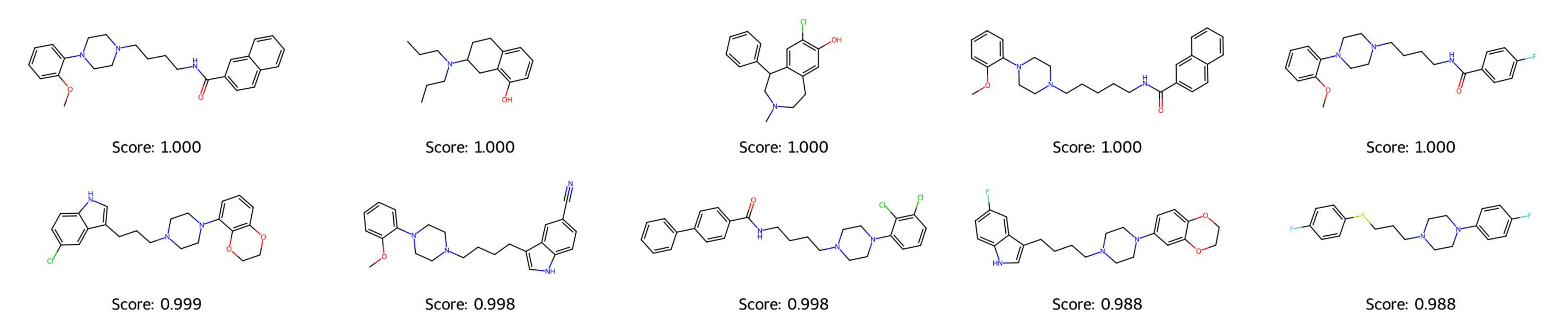}\label{fig:ppo_tbh_nofilter} }
     \hfill
     \subfloat[Top current (TC)]{\includegraphics[width=0.48\textwidth]{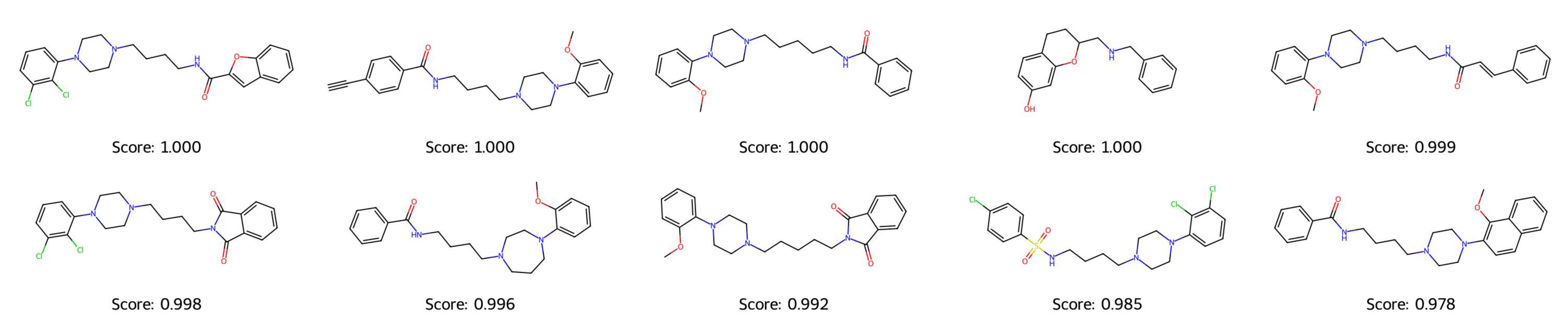}\label{fig:ppo_tc_nofilter} }
     \hfill
     \subfloat[Top history (TH)]{\includegraphics[width=0.48\textwidth]{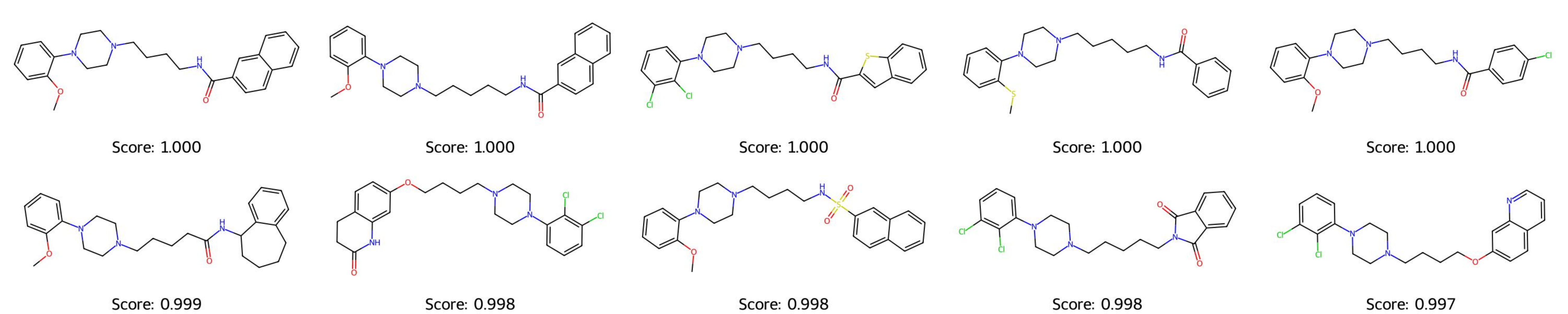}\label{fig:ppo_th_nofilter} }
    \caption{PPO without diversity filter. If no score is displayed, all scores are at least 0.9995.}
    \label{fig:smiles_ppo_nofilter}
\end{figure}

\begin{figure}[h]
     \centering
     \subfloat[All current (AC)]{
    \includegraphics[width=0.45\textwidth]{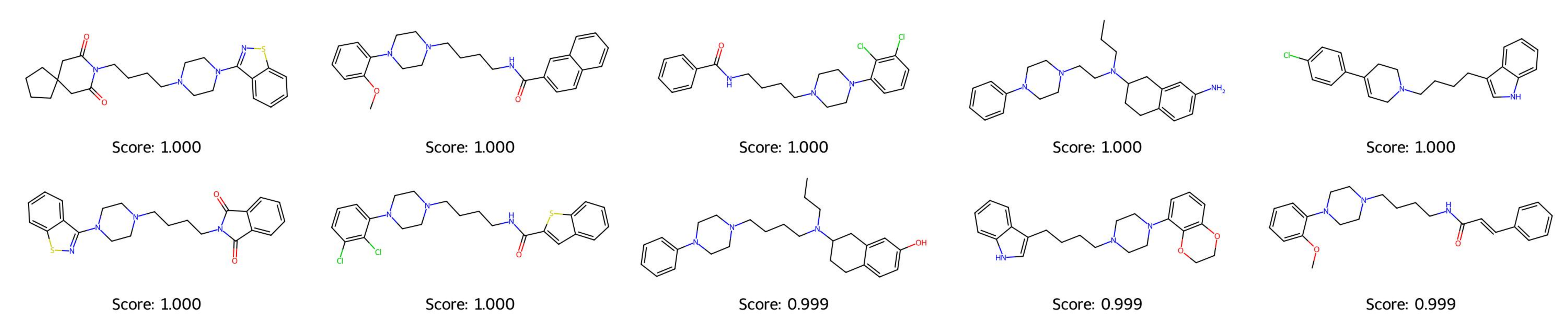}\label{fig:reinvent_ac_nofilter}}
     \hfill
    \subfloat[Bin current (BC)]{\includegraphics[width=0.48\textwidth]{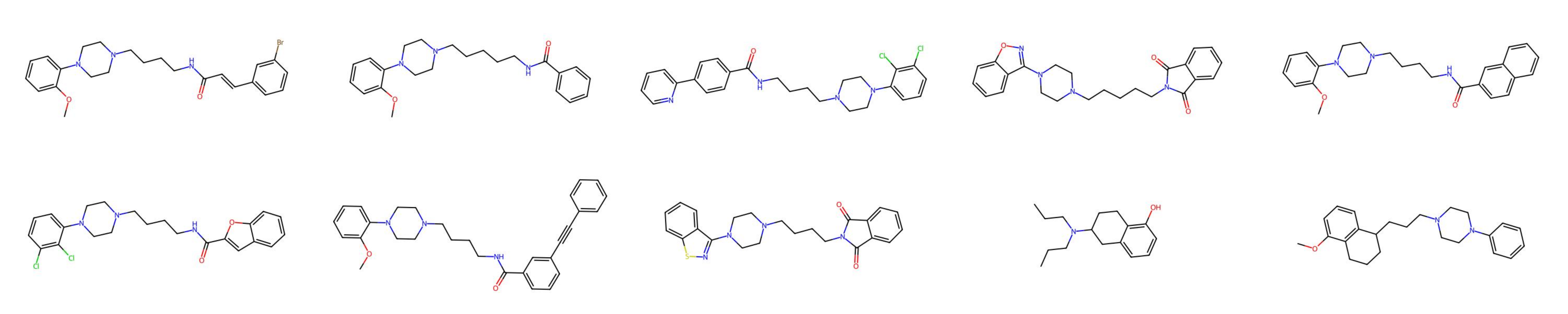}\label{fig:reinvent_bc_nofilter}}
     \hfill
     \subfloat[Bin history (BH)]{\includegraphics[width=0.48\textwidth]{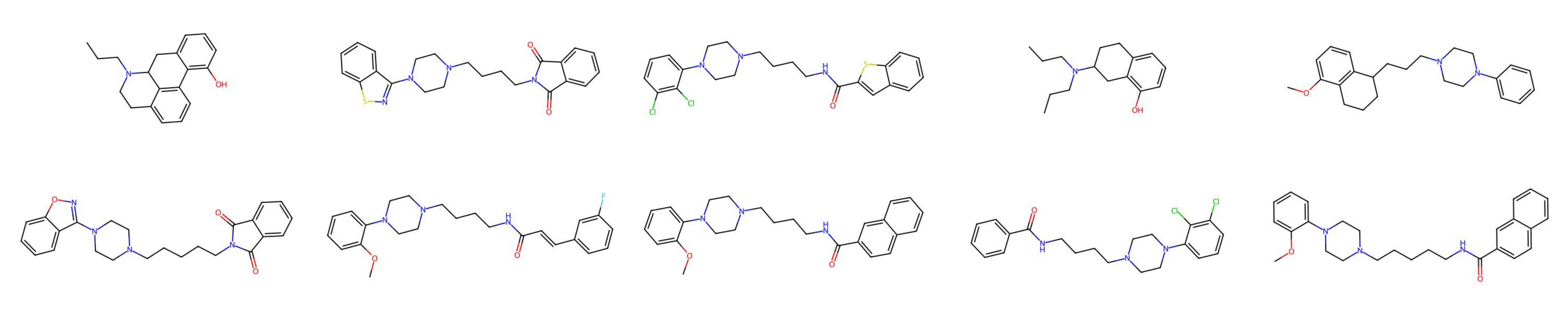}\label{fig:reinvent_bh_nofilter} }
     \hfill
     \subfloat[Top-bottom current (TBC)]{\includegraphics[width=0.48\textwidth]{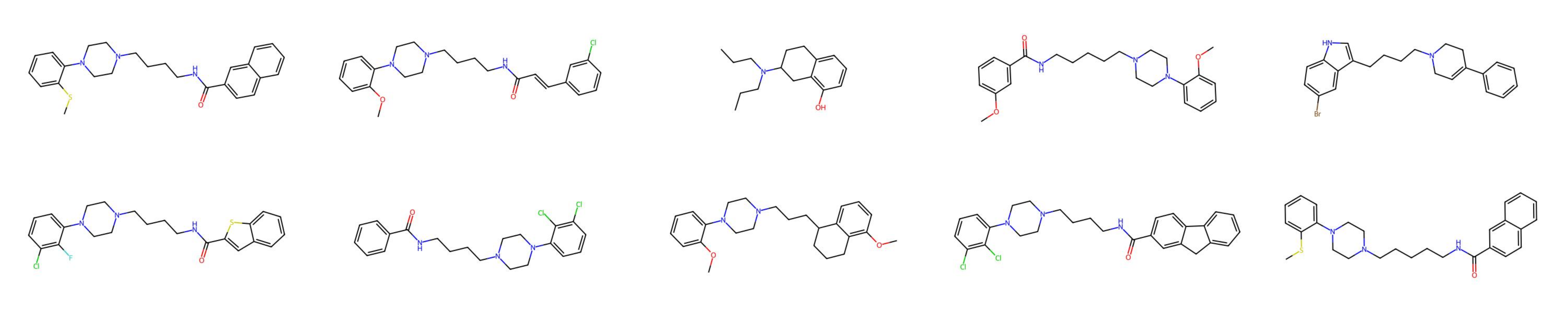}\label{fig:reinvent_tbc_nofilter} }
     \hfill
     \subfloat[Top-bottom history (TBH)]{\includegraphics[width=0.48\textwidth]{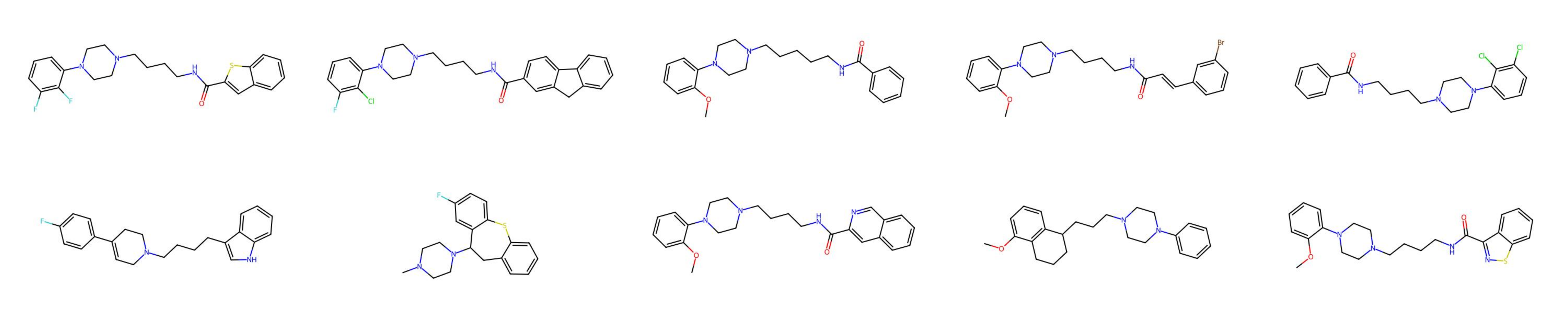}\label{fig:reinvent_tbh_nofilter} }
     \hfill
     \subfloat[Top current (TC)]{\includegraphics[width=0.48\textwidth]{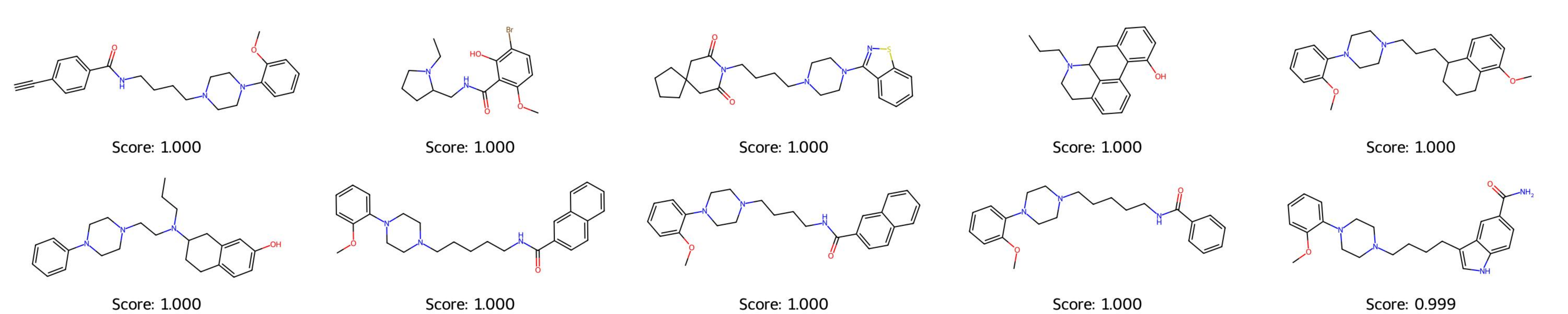}\label{fig:reinvent_tc_nofilter} }
     \hfill
     \subfloat[Top history (TH)]{\includegraphics[width=0.48\textwidth]{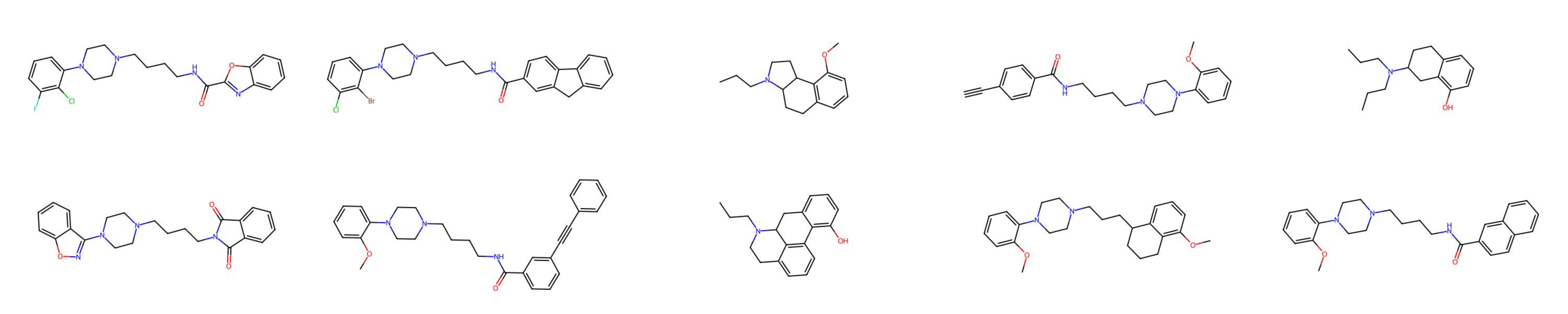}\label{fig:reinvent_th_nofilter} }
    \caption{Regularized MLE without diversity filter. If no score is displayed, all scores are at least 0.9995.}
    \label{fig:smiles_reinvent_nofilter}
\end{figure}

\subsection{Off-policy Algorithms}
\subsubsection{With Diversity Filter}

\begin{figure}[h]
     \centering
     \subfloat[Bin history (BH)]{\includegraphics[width=0.48\textwidth]{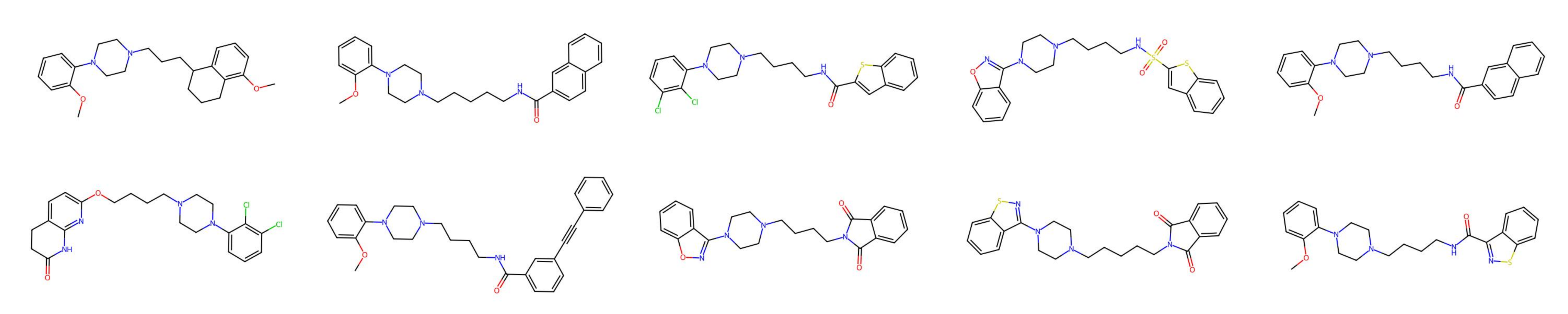}\label{fig:acer_bh} }
     \hfill
     \subfloat[Top-bottom history (TBH)]{\includegraphics[width=0.48\textwidth]{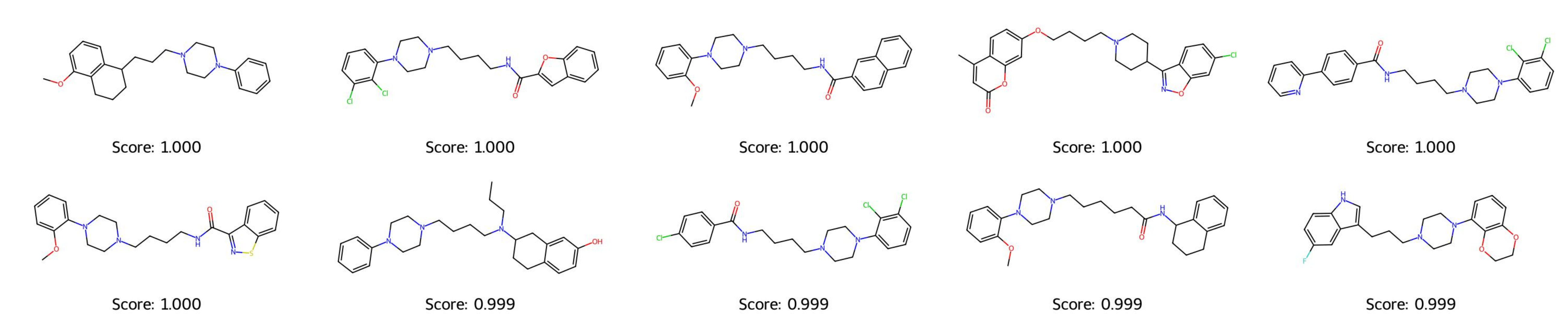}\label{fig:acer_tbh} }
     \hfill
     \subfloat[Top history (TH)]{\includegraphics[width=0.48\textwidth]{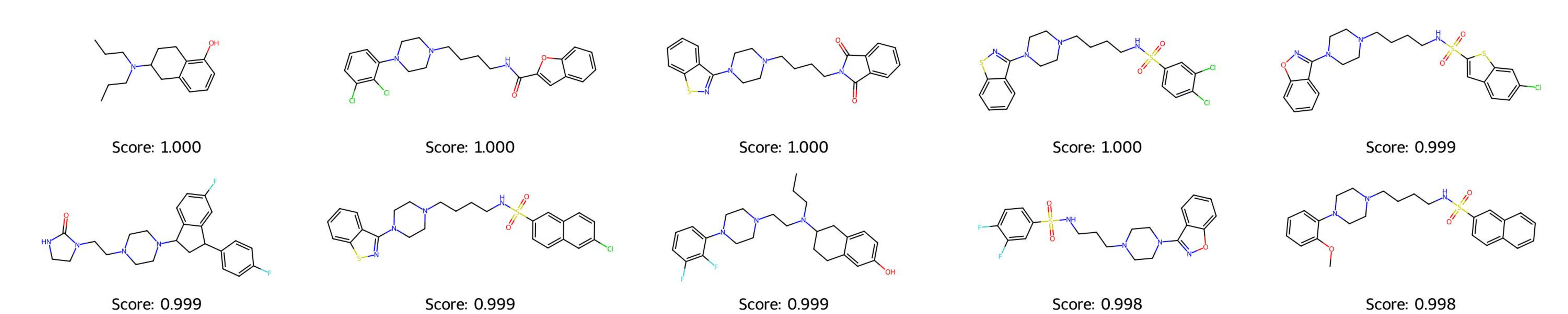}\label{fig:acer_th} }
    \caption{ACER with diversity filter penalizing the generation of SMILES with the same molecular scaffold. If no score is displayed, all scores are at least 0.9995. Otherwise, scores are rounded upwards to 3 decimals.}
    \label{fig:smiles_acer}
\end{figure}

\begin{figure}[h]
     \centering
     \subfloat[Bin history (BH)]{\includegraphics[width=0.48\textwidth]{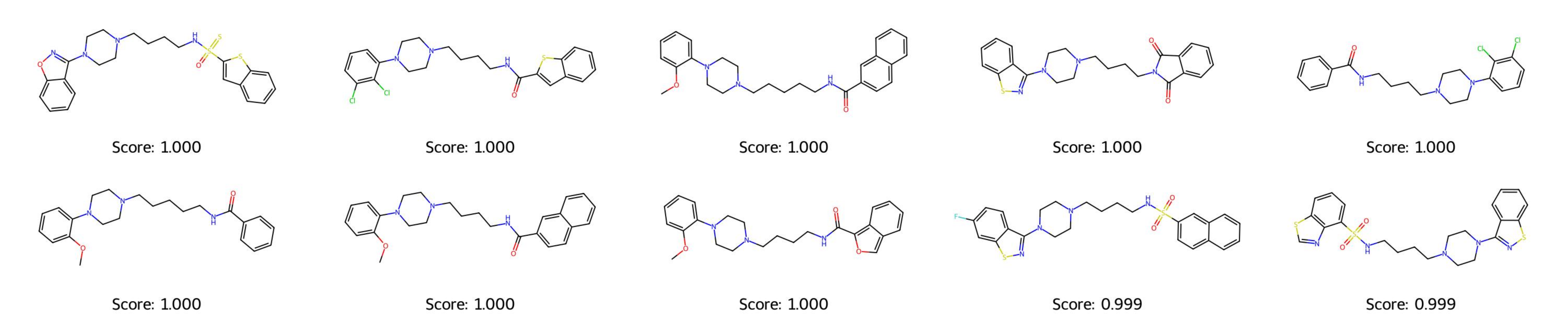}\label{fig:sac_bh} }
     \hfill
     \subfloat[Top-bottom history (TBH)]{\includegraphics[width=0.48\textwidth]{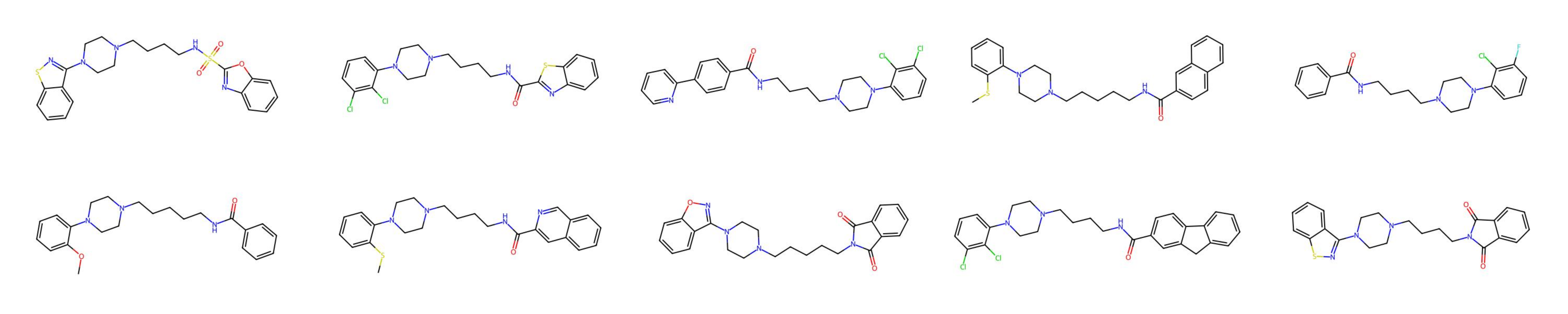}\label{fig:sac_tbh} }
     \hfill
     \subfloat[Top history (TH)]{\includegraphics[width=0.48\textwidth]{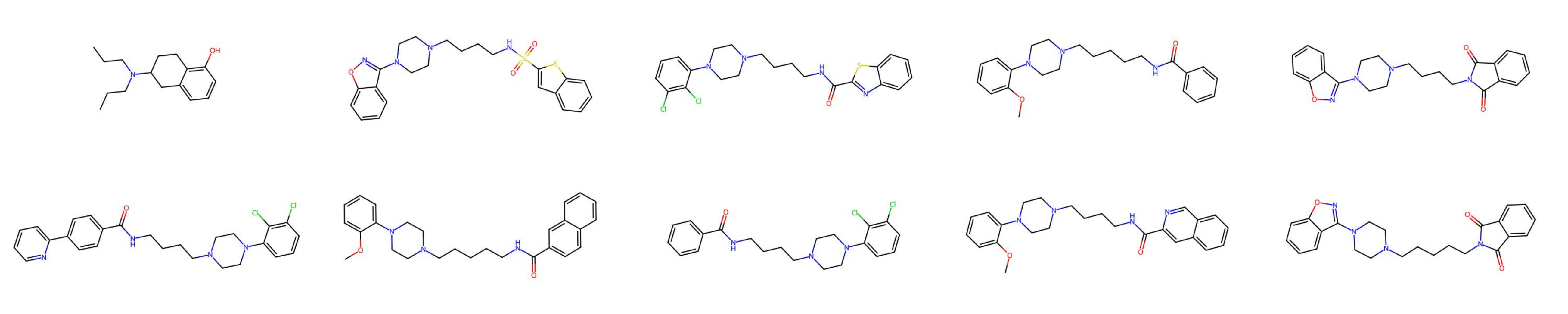}\label{fig:sac_th} }
    \caption{SAC with diversity filter penalizing the generation of SMILES with the same molecular scaffold. If no score is displayed, all scores are at least 0.9995. Otherwise, scores are rounded upwards to 3 decimals.}
    \label{fig:smiles_sac}
\end{figure}

\subsection{Without Diversity Filter}

\begin{figure}[h]
     \centering
     \subfloat[Bin history (BH)]{\includegraphics[width=0.48\textwidth]{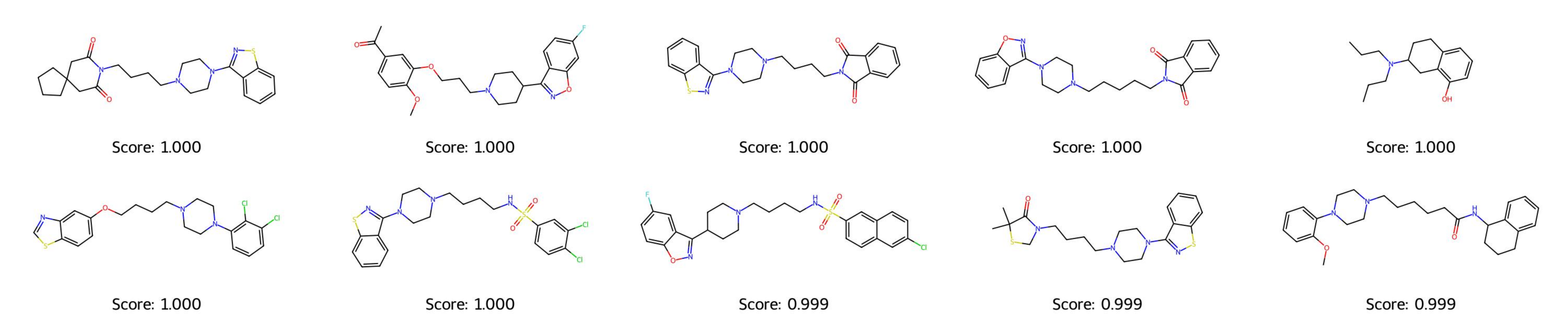}\label{fig:acer_bh_nofilter} }
     \hfill
     \subfloat[Top-bottom history (TBH)]{\includegraphics[width=0.48\textwidth]{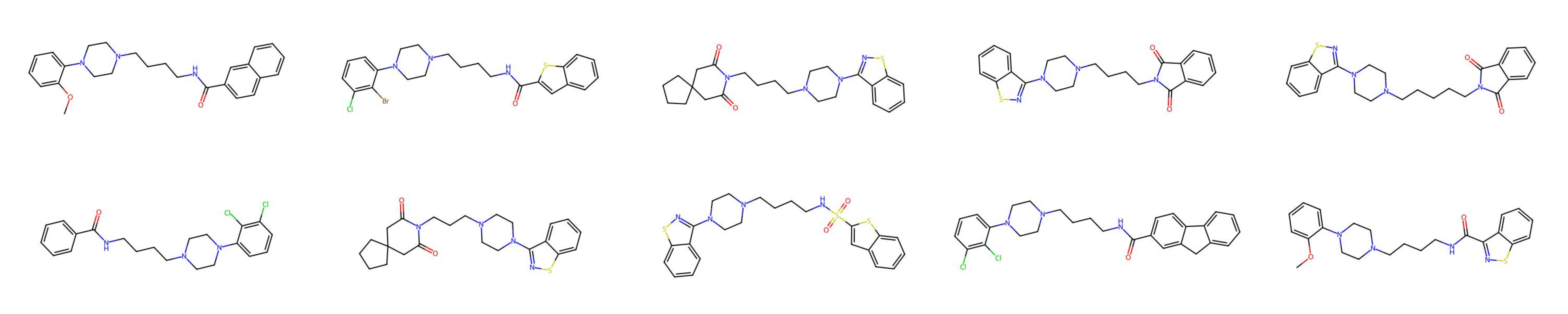}\label{fig:acer_tbh_nofilter} }
     \hfill
     \subfloat[Top history (TH)]{\includegraphics[width=0.48\textwidth]{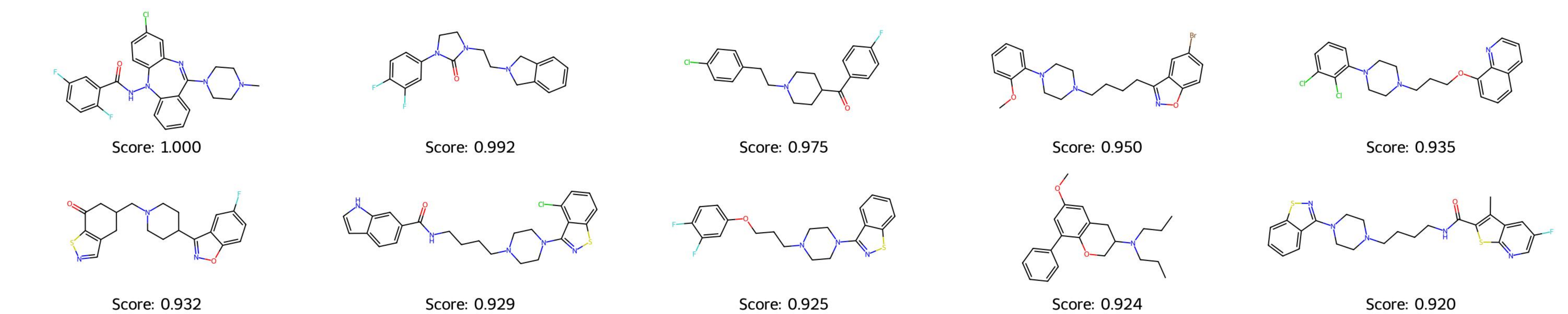}\label{fig:acer_th_nofilter} }
    \caption{ACER without diversity filter. If no score is displayed, all scores are at least 0.9995.}
    \label{fig:smiles_acer_nofilter}
\end{figure}

\begin{figure}[h]
     \centering
     \subfloat[Bin history (BH)]{\includegraphics[width=0.48\textwidth]{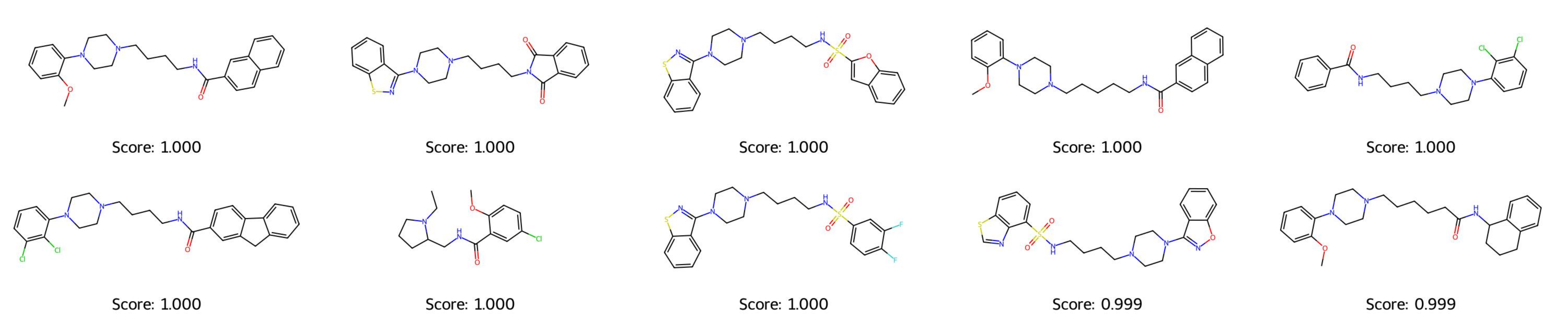}\label{fig:sac_bh_nofilter} }
     \hfill
     \subfloat[Top-bottom history (TBH)]{\includegraphics[width=0.48\textwidth]{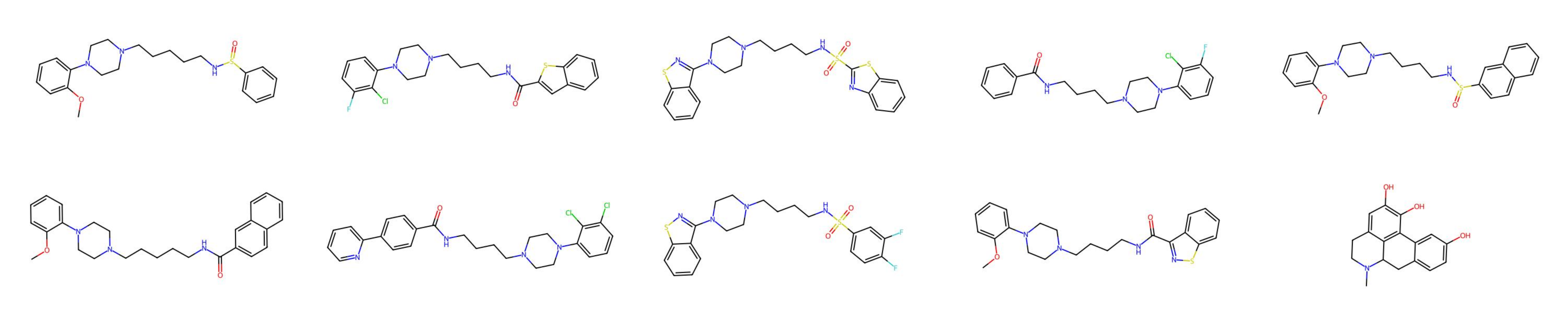}\label{fig:sac_tbh_nofilter} }
     \hfill
     \subfloat[Top history (TH)]{\includegraphics[width=0.48\textwidth]{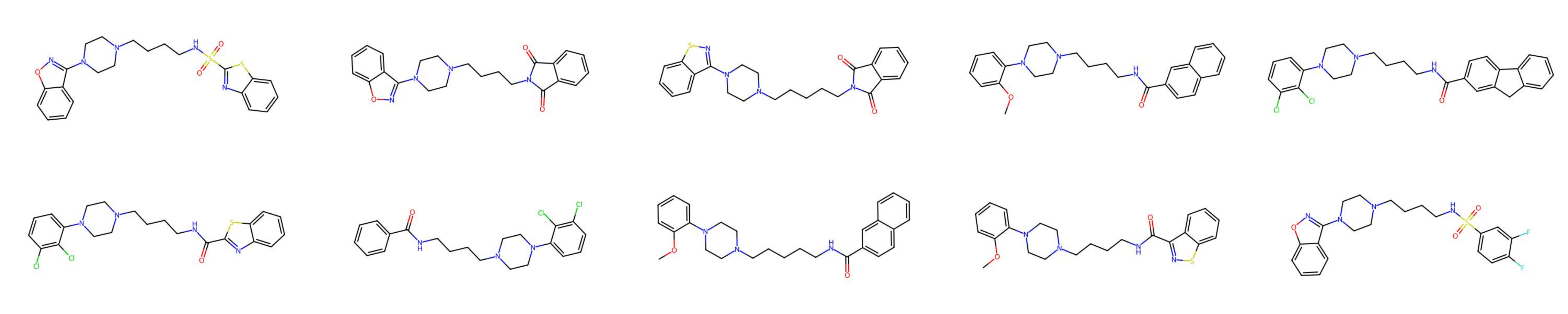}\label{fig:sac_th_nofilter} }
    \caption{SAC without diversity filter. If no score is displayed, all scores are at least 0.9995.}
    \label{fig:smiles_sac_nofilter}
\end{figure}

\end{appendices}

\bibliography{sn-bibliography}

\end{document}